\documentclass{jfm}
\usepackage{graphicx}
\usepackage{epstopdf, epsfig}
\usepackage{amsfonts,amssymb,amsmath,mathbbol}   
\usepackage{caption}
\usepackage{subcaption}

\usepackage{xcolor} 

\def\v{{\rm{\bf v}}}
\def\e{{\rm{\bf e}}}

\def\n{{\rm \bf n}}
\def\t{{\rm \bf t}}

\def\I{{\rm \bf I}}

\usepackage{color} 

\shorttitle{Freezing droplet} 
\shortauthor{Kavuri et al.} 
\title{Evaporation-driven coalescence of two droplets undergoing freezing}

\author
 {
 Sivanandan Kavuri,\aff{1}
 George Karapetsas,\aff{2}
 Chander Shekhar Sharma\aff{3}
  \and
  Kirti Chandra Sahu\aff{1}\corresp{\email{gkarapetsas@auth.gr; ksahu@che.iith.ac.in}}
  }

\affiliation
{
\aff{1}
Department of Chemical Engineering, Indian Institute of Technology Hyderabad, Sangareddy 502 284, Telangana, India
\aff{2}
Department of Chemical Engineering, Aristotle University of Thessaloniki, Thessaloniki 54124, Greece
\aff{3}
Department of Mechanical Engineering, Indian Institute of Technology Ropar, 140001 Rupnagar, India
}

\begin{document}

\maketitle

\begin{abstract}
We examine the evaporation-induced coalescence of two droplets undergoing freezing by conducting numerical simulations employing the lubrication approximation. When two sessile drops undergo freezing in close vicinity over a substrate, they interact with each other through the gaseous phase and the simultaneous presence of evaporation/condensation. In an unsaturated environment, the evaporation flux over the two volatile sessile drops is asymmetric, with lower evaporation in the region between the two drops. This asymmetry in the evaporation flux generates an asymmetric curvature in each drop, which results in a capillary flow that drives the drops closer to each other, eventually leading to their coalescence. This capillary flow, driven by evaporation, competes with the upward movement of the freezing front, depending on the relative humidity in the surrounding environment. We found that higher relative humidity reduces the evaporative flux, delaying capillary flow and impeding coalescence by restricting contact line motion. For a constant relative humidity, the substrate temperature governs the coalescence phenomenon, and resulting condensation can accelerate this process. Interestingly, lower substrate temperatures are observed to facilitate faster propagation of the freezing front, which, in turn, restricts coalescence.
\end{abstract}

\begin{keywords}
Freezing, sessile droplets, frost halo, evaporation, lubrication approximation
\end{keywords}

\section{Introduction} \label{sec:intro}

The solidification of water droplets holds substantial importance in various fields, including the food industry \citep{george1993freezing,james2015review}, energy storage \citep{sharma2022numerical}, and freeze-drying processes \citep{franks1998freeze,assegehegn2019importance}. However, in applications such as aircraft operation \citep{gent2000aircraft,cao2015aircraft,cao2018aircraft}, marine vessels \citep{zhou2022research}, food storage facilities \citep{zhu2021biomimetic}, and wind energy generation \citep{fakorede2016ice,kraj2010phases}, the solidification of water droplets can yield negative consequences.

The freezing process of a sessile droplet on a cold substrate unfolds in two discernible phases. During the early recalescence phase, an ice-crystal scaffold is rapidly formed, accompanied by a rise in temperature within the remaining liquid \citep{jung2012frost,hu2010icing}. This phase is characterized by the release of energy as liquid transitions into ice, leading to a temperature increase. A slower second phase ensues after the recalescence phase, solidifying the remaining liquid through isothermal freezing. As this occurs, additional heat is released as the freezing front advances toward the apex of the droplet. The vapour condenses on the substrate around the droplet, forming frost halos \citep{jung2012frost,kavuri2023freezing}. Numerous researchers have investigated the freezing of stationary water drops and uncovered intriguing physics, such as freezing front propagation, cusp formation, volume expansion, and frost halo formation \citep{angell1983supercooled,ajaev2004effect,hu2010icing,anderson1996case,Tembely2019,marin2014universality}. The evolution of the water-ice freezing front, including its behaviour in the recalescence phase, was examined by \cite{meng2020dynamic} theoretically. \cite{nauenberg2016theory} and \cite{zhang2016freezing} studied the evolution of the freezing front during the freezing of a sessile water drop on different surfaces through theory and experiments. \cite{zhang2018simulation,zhang2019shape} conducted experiments to investigate the volume expansion and shape variations of various volumes of freezing water drops on hydrophilic and hydrophobic surfaces and compared these findings with a numerical model they developed. \cite{marin2014universality} explored the universality of tip singularity for a droplet deposited on a cold substrate. In contrast, \cite{starostin2022universality,starostin2023effects} investigated the freezing of water drops on metallic surfaces and surfaces lubricated with silicone oil, including analysis of the effect of asymmetric cooling of sessile droplets on the orientation of the freezing tip. Other studies also focused on the effect of surface roughness \citep{fuller2024analysis}, curvature \citep{liu2021supercooled,jin2017impact,zhang2018experimental,ju2018impact}, and wettability \citep{pan2019experimental,peng2020study} on the freezing of the drops. Many studies also focused on the freezing delay phenomenon \citep{guo2021icing,shi2022freezing,jung2011superhydrophobic,boinovich2014effect,hao2014freezing} of the sessile water droplets as it helps in designing better ice-phobic surfaces.

Additionally, several researchers have explored the freezing dynamics of sessile droplets using various numerical simulation techniques. These include approaches based on Navier-Stokes equations with front-tracking \citep{vu2015numerical,vu2018numerical}, level-set and volume-of-fluid \citep{blake2015simulating,tembely2018numerical}, and lattice Boltzmann methods \citep{perez2021investigations,wang2022new}. \cite{zadravzil2006droplet} and \cite{Tembely2019} employed the lubrication approximation method, which resolved shear stress singularities at the solid-liquid interface using a precursor layer, to investigate the freezing of sessile drops. \cite{sebilleau2021air} conducted experimental and theoretical studies on the influence of humidity on freezing fronts and cusp formation. In contrast to earlier numerical studies that frequently disregarded evaporation, our prior research focusing on the freezing of sessile droplets and the formation of frost halos, employing the lubrication approximation, examined the influence of evaporation on these phenomena \citep{kavuri2023freezing}. Our findings reveal a direct association between the formation of frost halos and the inherent evaporation process during the early stages of freezing.

While all the studies mentioned above concentrated on the freezing dynamics of individual drops on a surface, realistic scenarios typically involve multiple drops in close proximity. In such situations, these drops can undergo freezing and dynamically interact with one another during the process. The droplet interaction during the freezing process is caused by significant vaporization effects, leading to different frost propagation mechanisms. \cite{yancheshme2020mechanisms} suggests four different mechanisms for frost propagation: ice-bridge formation,  cascade freezing, frost halos and droplet explosion. \cite{graeber2018cascade} found that multiple droplets interact with each other through strong vapourization, and this vapour front leads to cascade freezing. \cite{nath2017review} discusses frost halos, inter-droplet bridging and dry zones occurring during condensation frosting. \cite{castillo2021quantifying} found that the droplet droplet interactions lead to asymmetric solidification. For volatile fluids, the interaction through vapourization is much more significant. \cite{Moosman1980,Ajaev2001,Craster2009c,Karapetsas2016,williams2021} explored the evaporation of sessile drops on a hot substrate using the lubrication approach. The spreading of a sessile drop due to Marangoni flow has also been studied using lubrication theory \citep{karapetsas2014thermocapillary,karapetsas2013effect}. Additionally, several researchers have experimentally investigated the evaporation of sessile drops \citep{hari2024,hari2022counter,katre2020evaporation,Karapetsas2012}. \citet{sadafi2019vapor} and \citet{wen2019vapor} conducted experimental studies on the evaporation-driven coalescence of single-component volatile droplets on high-energy substrates at room temperature and ambient humidity. Similar behaviors were observed across various organic fluids, including n-hexane, n-pentane, cyclohexane, ethyl acetate, HFE7000, HFE7100, HFE7200, and polypropylene glycol \citep{sadafi2019vapor, wen2019vapor, man2017vapor}. In addition, \citet{man2017vapor} investigated the theoretical aspects of attraction, repulsion, and chasing behaviors of evaporating droplets, demonstrating that droplet motion can result from asymmetries in the evaporation flux, even in the absence of the Marangoni effect. In contrast to these single-component systems, the behavior of binary droplets \citep{cira2015vapour, man2017vapor} is influenced by the Marangoni effect, leading to more complex interactions driven by differences in evaporation rates and surface tensions of the components. Furthermore, \citet{sadafi2019vapor} found that single-component volatile droplets can coalesce due to substrate-mediated forces, thermal Marangoni forces, and evaporation-induced effects. A similar study, albeit for droplets on soft substrates, has been recently presented by \cite{Karapetsas2024}, where it was shown that, besides the thermal Marangoni and evaporation-induced effects, the droplets might also interact through the deforming elastic substrate.

Due to its practical relevance, the interaction of sessile droplets, even in the absence of evaporation and freezing, has been extensively studied by several researchers \citep{sui2013inertial,eddi2013influence,varma2021coalescence,kapur2007morphology,paulsen2011viscous,hernandez2012symmetric,xia2019universality}. Additionally, investigations into the coalescence of drops on a substrate and the coalescence of volatile drops at room temperatures have been conducted \citep{sadafi2019vapor,Karapetsas2024}. However, despite experimental studies on the freezing of multiple drops \citep{graeber2018cascade,jung2012frost,castillo2021quantifying,yancheshme2020mechanisms}, the freezing behaviour of volatile drops has not been theoretically explored.

To the best of our knowledge, the present study is the first attempt to theoretically explore the interaction between two volatile drops undergoing freezing, employing the lubrication approximation while considering evaporation and condensation. Our findings suggest that when the freezing front propagation exhibits limited advancement and fails to restrict the contact line of the drops, rapid coalescence occurs between the two volatile sessile drops closely placed on a substrate. We explore the coalescence mechanism of volatile drops undergoing freezing and also study the effect of relative humidity and the initial separation between the two drops on the coalescence dynamics.

The remainder of this paper is structured as follows. In \S \ref{sec:model}, we present the problem formulation, governing equations, scaling considerations, and the numerical approach utilized within the framework of the lubrication approximation. The results are discussed in \S \ref{sec:dis}. It also presents the coalescence mechanism of volatile drops undergoing freezing, accompanied by an extensive parametric study. Finally, concluding remarks are provided in \S \ref{sec:Conc}.

\section{Problem formulation} \label{sec:model}

\begin{figure}
\centering
\includegraphics[width=0.75\textwidth]{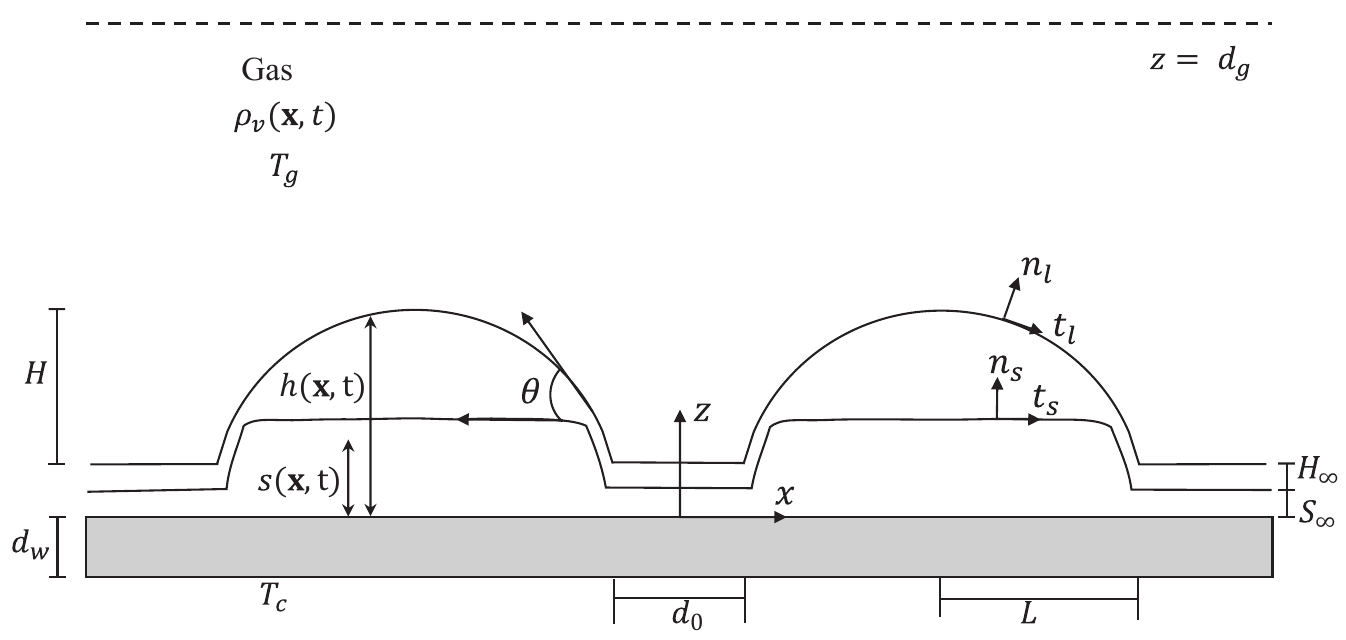}
\caption{Schematic representation of two sessile droplets undergoing freezing on a solid substrate. Here, $S_\infty$, $H_\infty$ and $d_0$ are the initial thickness of microscopic ice-layer, the thickness of the precursor layer, and the distance between the two drops, respectively. $T_{g}$ and $T_{c}$ are the temperatures of the ambient and the bottom of the substrate.}
\label{fig:geom}
\end{figure}

The freezing process of two thin sessile liquid droplets separated by an initial distance of $d_0$ placed on a cold, horizontal solid substrate is investigated numerically using the lubrication approximation. Figure \ref{fig:geom} depicts a schematic diagram of two droplets separated by a initial distance $d_0$, during freezing, along with various parameters used in our modelling. The bottom of the substrate with thickness $(d_w)$, thermal conductivity $(\lambda_w)$ and specific heat $(C_{pw})$ is maintained at a constant temperature, $T_c$. We assume that, at $t=0$, the early recalescence phase has already taken place with the liquid having reached the melting temperature. At this point, the slower solidification step driven by the heat conduction ensues and a very thin layer of ice of uniform thickness, $S_\infty$, has formed along the solid substrate. The height and half-width of the droplets are denoted by $H$ and $L$, respectively. The liquid is assumed to be incompressible and Newtonian, with constant density $(\rho_l)$, specific heat capacity $(C_{pl})$, thermal conductivity $(\lambda_l)$ and viscosity $(\mu_l)$. The surface tension of the liquid-gas interface $(\gamma_{lg})$ is assumed to be constant. The frozen solid phase has constant density $(\rho_s)$, specific heat capacity $(C_{ps})$ and thermal conductivity $(\lambda_s)$. The droplet is bounded from above by an inviscid gas. A Cartesian coordinate system $(x,z)$ with its origin at the centre of the droplet on the solid substrate is employed in our study as shown in figure \ref{fig:geom}. Here, $z=s(x,t)$ and $z=h(x,t)$ represent the liquid-ice and liquid-gas interfaces, respectively. In the present work, we consider the drop to be very thin $(H \ll L)$. Thus, the aspect ratio of the droplet, $\epsilon=H/L$, is assumed to be very small. This assumption permits us to use the lubrication theory, employed below, to derive a set of evolution equations that govern the freezing dynamics of the sessile droplet considering the liquid, ice and gas phases. However, earlier studies demonstrated the validity of lubrication model for contact angles up to $60^\circ$ \citep{charitatos2020thin,Tembely2019}.

\subsection{Dimensional governing equations}\label{sec:gov_eqn}

\subsubsection{Liquid phase}

The dynamics in the liquid phase is governed by the mass, momentum and energy conservation equations, which are given by
\begin{equation}\label{eq:mom}
\rho_l \left( \frac{\partial \v}{\partial t} + \v \cdot \nabla \v \right) = - \nabla p + \mu_l \nabla^2 \v,
\end{equation}
\begin{equation}\label{eq:cont}
\nabla \cdot \v = 0,
\end{equation}
\begin{equation}\label{eq:energy_liquid}
\rho_l C_{pl} \left( \frac{\partial T_l}{\partial t} + \v \cdot \nabla T_l \right) = \lambda_l \nabla^2 T_l,
\end{equation}
where $\v$, $p$ and $T_l$ denote the velocity, pressure and temperature in the liquid phase, respectively; $\nabla$ represents the gradient operator. At the free surface ($z=h(x,t)$), the liquid velocity, $\v$, and the velocity of the liquid-gas interface, $\v_{lg}$, are related as
\begin{equation}
\v = \v_{lg} + (J_v/\rho_l) \; \n_l,
\end{equation}
\begin{equation}
\n_l = (-h_{x}e_x - h_{y}e_y + e_z)/\sqrt{(h_{x}^2+h_{y}^2+1)},
\end{equation}
where $J_v$ denotes the evaporative flux and $\n_l$ is the outward unit normal on the interface. However, the tangential components of both velocities, $\v_{\tau} = \v - (\v \cdot \n_l) \n_l = \v_{lg} - (\v_{lg} \cdot \n_l) \n_l$, are the same. Moreover, at $z=h(x,t)$, the velocity field satisfies the local mass, force and energy balance in the liquid and gas phases \citep{Karapetsas2016}. Thus,
\begin{equation}
\rho_l (\v -\v_{lg}) \cdot \n_l = \rho_g (\v_g -\v_{lg}) \cdot \n_l,
\end{equation}
\begin{equation}
J_v (\v -\v_g) - \n_l \cdot \left[-p \I + \mu_l \left( \nabla \v + \nabla \v ^T \right) \right]  = p_g \n_l - \gamma_{lg} \kappa_l \n_l - \Pi \n_l ,
\end{equation}
\begin{equation}
J_v L_v + \lambda_l \nabla T_l \cdot \n_l - \lambda_g \nabla T_g \cdot \n_l = 0,
\label{energy1}
\end{equation}
where $\rho_g$, $\lambda_g$, $\v_g$ and $T_g$ denote the density, thermal conductivity, velocity field, and gas phase temperature, respectively. Here, $\I$ represents the identity tensor, $T_{lg}$ indicates the temperature at the liquid-gas interface, $L_v$ is the specific internal latent heat of vaporization, $\kappa_{lg} = - \nabla_{s,l} \cdot \n_l$ denotes the mean curvature of the free surface, and $\nabla_{s,l} = (I-\n_l\n_l) \cdot \nabla$ represents the surface gradient operator. The disjoining pressure ($\Pi$) that accounts for the van der Waals interaction is defined as
\begin{equation} \label{eq:disj_press}
\Pi = A \left[ \left( \frac{B}{h-s} \right)^n - \left( \frac{B}{h-s} \right)^m \right],
\end{equation}
where $A= A_{Ham}/B^n \geq 0$ is a constant that describes the magnitude of the energy of the intermolecular interactions between the liquid-gas and liquid-ice interfaces, and $B$ denotes the precursor layer thickness. Here, $n > m > 1$ and $A_{Ham}$ denotes the dimensional Hamaker constant.

At the liquid-ice interface $(z=s(x,t))$, the velocity is given by
\begin{equation} \label{eq:mass_bc_solid_ND}
\v =  \v_{ls} - (J_s/\rho_l) \; \n_s,
\end{equation}
\begin{equation} 
\n_l = (-s_{x}e_x - s_{y}e_y + e_z)/\sqrt{(s_{x}^2+s_{y}^2+1)},
\end{equation}
where $J_s$ denotes the freezing mass flux and $\v_{ls}$ the velocity of the liquid-ice interface. As we impose the no-slip condition at the liquid-ice interface, the tangential component of the velocity is given by
\begin{equation} \label{eq:ice_no_slip}
\v \cdot \t_s = 0.
\end{equation}
Here, $\n_s$ and $\t_s$ are the outward unit normal and unit tangential vectors on the liquid-ice interface, respectively.

\subsubsection{Solid (ice) phase} \label{solid_ice_phase}

The energy conservation equation in the solid (ice) phase is given by
\begin{equation}\label{eq:energy_ice}
\rho_s C_{ps} \frac{\partial T_s}{\partial t} = \lambda_s \nabla^2 T_s,
\end{equation}
where $T_s$ denotes the temperature in the solid phase.

At the solid substrate ($z=0$), we impose continuity of temperature
\begin{equation}
T_s = T_w, \label{eq2p12}
\end{equation}
where $T_w$ is the temperature of the substrate at $z=0$.

At the freezing front ($z=s(x,t)$), the boundary condition for the temperature is expressed as
\begin{equation} \label{eq:temp_bc_liquid_solid}
T_s = T_l = T_f
\end{equation}
We also assume that equilibrium temperature at the freezing front, $T_f$, is the same as the melting temperature, $T_m$.

At $z=s(x,t)$, the conservation of mass and energy between the liquid and solid phases leads to
\begin{equation} \label{eq:mass_bc_solid}
J_s = \rho_l (\v_{ls} - \v) \cdot \n_s = \rho_s (\v_{ls} - \v_s) \cdot \n_s,
\end{equation}
\begin{equation} \label{eq:energy_bc_solid}
\rho_s (\v_{ls} - \v_s) \cdot \n_s H_s - \lambda_s \nabla T_s \cdot \n_s = \rho_l (\v_{ls} - \v) \cdot \n_s H_l - \lambda_l \nabla T_l \cdot \n_s,
\end{equation}
where $\v_s$ denotes the velocity of the ice phase; $H_s$ and $H_l$ denote the enthalpy of the ice and liquid, respectively. By combining eqs. (\ref{eq:mass_bc_solid}) and (\ref{eq:energy_bc_solid}) and assuming that $\v_s=0$, we get
\begin{equation} \label{eq:energy_bc_solid_final}  
J_s \Delta H_{sl} - \lambda_s \nabla T_s \cdot \n_s + \lambda_l \nabla T_l \cdot \n_s = 0,
\end{equation}
where $\Delta H_{sl} = H_s - H_l$ denotes the enthalpy jump at the liquid-ice interface. Considering that $H_s = C_{ps} (T_f - T_m) + L_f(T_m)$ and $H_l = C_{pl} (T_f - T_m)$, we evaluate 
\begin{equation} \label{eq:DHsl}  
\Delta H_{sl} = (C_{ps} - C_{pl}) (T_f - T_m) + L_f,
\end{equation}
where $L_f$ denotes the latent heat of fusion considering the melting temperature, $T_m$ as the reference temperature. As will be shown below, eq. (\ref{eq:energy_bc_solid_final}) can be used to evaluate the position of the freezing front, $s(x,t)$.

\begin{table}
\centering
\begin{tabular}{ccc}
Property   &     Notation       & Value  \\  \hline
Density of the liquid phase   & $\rho_{l}$ &  1000 Kg m$^{-3}$ \\
Density of the frozen solid phase   & $\rho_{s}$ & 900 Kg m$^{-3}$  \\
Density of the gas phase   & $\rho_{g}$ &  $5 \times 10^{-3}$ Kg m$^{-3}$ \\
Viscosity of the liquid phase   & $\mu_{l}$ &  $10^{-3}$ Pa$\cdot$s\\
Viscosity of the gas phase   & $\mu_{g}$ &  $1.81 \times 10^{-5}$ Pa$\cdot$s \\
Melting temperature  & $T_m$ &  273 K \\
Ambient temperature	 & $T_g$ &  273 K \\
Temperature at the bottom of the substrate   & $T_c$ & 263-272.7 K  \\ 
Thickness of the substrate   & $d_w$ & $3 \times 10^{-3}$ m \\  
Thermal conductivity of the substrate   & $\lambda_w$ &  0.33 W m$^{-1}$ K$^{-1}$\\ 
Thermal conductivity of the liquid phase   & $\lambda_l$ &  0.57 W m$^{-1}$ K$^{-1}$ \\ 
Thermal conductivity of the frozen solid phase   & $\lambda_s$ &  2.21 W m$^{-1}$ K$^{-1}$ \\
Thermal conductivity of the gas phase   & $\lambda_g$ &  0.02 W m$^{-1}$ K$^{-1}$ \\ 
Specific heat capacity of the substrate   & $C_{pw}$ & 300 - 3000 J Kg$^{-1}$K$^{-1}$ \\ 
Specific heat capacity of the liquid phase   & $C_{pl}$ &  4220 J Kg$^{-1}$K$^{-1}$\\
Specific heat capacity of the frozen solid phase   & $C_{ps}$ &  2050 J Kg$^{-1}$K$^{-1}$ \\
Specific heat capacity of the gas phase   & $C_{pg}$ & 4220 J Kg$^{-1}$K$^{-1}$  \\
Surface tension of the liquid-gas interface at $T_m$  & $\gamma_{lg}$ &  0.07 N m$^{-1}$ \\
Surface tension of the liquid-ice interface    & $\gamma_{ls}$ &  0.02 N m$^{-1}$ \\
Latent  heat  of  evaporation & $L_{v}$ &  $2.45\times10^6$ J Kg$^{-1}$ \\
Latent  heat  of  fusion   & $L_{f}$ &  $3.35\times10^5$ J Kg$^{-1}$\\
Diffusion coefficient of the vapour in the gas phase   & $D_m$ &  $1.89 \times 10^{-5}$ m$^2$/s\\
Universal  gas  constant   & $R_g$ & 8.314 J K$^{-1}$ mol$^{-1}$ \\
Accommodation  coefficient & $\alpha$ &  $\approx 1$ \\
Relative humidity & ${RH}$ &  $0 -1$ \\
\end{tabular}
\vspace{2mm}
\caption{Typical values of the physical parameters considered in our simulations. These properties are for water-air system and Polymethyl methacrylate (PMMA) substrate.}
\label{T:wecr} 
\end{table}

\subsubsection{Gas phase}

In order to account for situations when a droplet freezes in an environment with varying humidity, we consider that the gas phase is inviscid and consists of both air and vapour. The gas phase velocity ($\v_{g}$) is assumed to be varied linearly between the liquid-gas interface and far away from the droplet, such that $\v_g \cdot \t_l=\v_{lg} \cdot \t_l$ at $z=h$ and $\v_{g}=0$ at $z=D_g$. The dynamics in the gas phase is governed by
\begin{equation} \label{eq:Lap_vap_conc}
\frac{\partial \rho_v}{\partial t} + \v_g \cdot \nabla \rho_v = D_{m}\nabla^2\rho_v,
\end{equation}
where $\rho_v$ is the concentration of the vapor in the gas phase and $D_{m}$ represents the diffusion coefficient of the vapour in the gas phase. We consider the case of a well-mixed gas phase where proper ventilation maintains constant vapour concentration at some distance from the droplet $(\rho_{vi})$. Thus, at ${z=D_g}$, $\rho_v = \rho_{vi}$.

The energy conservation equation in the gas phase is given by
\begin{equation}\label{eq:energy_gas}
\rho_g C_{pg} \frac{\partial T_g}{\partial t} = \lambda_g \nabla^2 T_g,
\end{equation}
where $T_g$ denotes the temperature in the gas phase.
At the liquid-gas interface ($z=h(x,t)$), we impose continuity of temperature
\begin{equation}
T_g = T_l,\label{eq2p20}
\end{equation}
where $T_l$ is the temperature of the drop at $z=h(x,t)$. Eqs. (\ref{eq2p12}) and (\ref{eq2p20}) imply that the contact resistance between ice and substrate, as well as the interfacial resistance between liquid and gas, are neglected.

The liquid in the droplet and the vapour in the gas phase are coupled as a consequence of the evaporation and condensation at the liquid-gas interface. In the gas phase, the vapour mass flux is related to the departure from the uniform vapour density, which is given by
\begin{equation} \label{eq:ev_flux}
J_v = - D_m \left( \n_l \cdot \nabla \rho_v \right) \; \mbox{at} \; z = h(x,t).
\end{equation} 

Additionally, the kinetic theory leads to a linear constitutive relation between the mass and the departure from equilibrium at the interface, which is known as the Hertz–Knudsen relationship \citep{Prosperetti1984, Sultan2005, Karapetsas2016}. This is given by
\begin{equation} \label{eq:Hertz_Knudsen}
J_v = \alpha \left( \frac{R_g T_{m}}{2 \pi M} \right)^{1/2} \left( \rho_{ve} (T_{lg}) - \rho_v \right),
\end{equation} 
where $R_g$ is the universal gas constant, $M$ is the molecular weight and $\alpha$ is the accommodation coefficient (close to unity). The equilibrium vapour concentration, $\rho_{ve} (T_{lg})$, can be obtained by employing a linear temperature-dependent equation of state:
\begin{equation}
\rho_{ve} (T_{lg}) = \rho_{ve} (T_m) \left[ 1 + \frac{M(p - p_{g})}{\rho_{l}R_gT_{m}} + \frac{ML_{v}}{R_{g}T_{m}}\left(\frac{T_{lg}}{T_{m}} - 1\right)\right].
\end{equation} 
Here, $L_{v}$ is the latent heat of evaporation. The combination of eqs. (\ref{eq:ev_flux}) and (\ref{eq:Hertz_Knudsen}) leads to
\begin{equation} \label{eq:int_vap_conc}
D_{m} \left( \n_l \cdot \nabla \rho_v \right) = - \alpha \left( \frac{R_g T_{m}}{2 \pi M} \right)^{1/2} \left(  \rho_{ve} (T_{lg}) - \rho_v  \right),
\end{equation} 
which can be used to evaluate the local interfacial vapour concentration, $\rho_v$. We use eq. (\ref{eq:int_vap_conc}) as the boundary condition at $z=h(x,t)$ to solve eq. (\ref{eq:Lap_vap_conc}).

\subsubsection{Solid substrate}

The energy conservation equation in the solid substrate is given by
\begin{equation}\label{eq:energy_solid}
\rho_w C_{pw} \frac{\partial T_w}{\partial t} = \lambda_w \nabla^2 T_w,
\end{equation}
where $\rho_{w}$ represents the density of the substrate. The above equation is subjected to the continuity of thermal flux at the ice-substrate interface ($z=0$): 
\begin{equation}
\lambda_s \frac{\partial T_s}{\partial z}  = \lambda_w \frac{\partial T_w}{\partial z},
\end{equation}
and at the bottom of the substrate, ${z=-d_w}$,
\begin{equation}
T_w = T_c.
\end{equation}

The properties of fluids and range of physical conditions considered in our study are listed in Table \ref{T:wecr}. 

\subsection{Scaling} \label{sec:scaling}
The governing equations and boundary conditions are nondimensionalized using the following scalings (wherein tilde denotes the dimensionless variable):
\begin{equation}
\begin{gathered}
( x, z, h, D_{w}, D_{g} ) = L ( \tilde{x}, \epsilon \tilde{z}, \epsilon \tilde{h} , \epsilon \tilde{D_{w}}, \epsilon \tilde{D_{g}}), ~ t = \frac{L}{U} \tilde{t}, ~ ( u, w ) = U ( \tilde{u}, \epsilon \tilde{w} ),\\ (u_g, w_g ) = U ( \tilde{u_{g}}, \epsilon \tilde{w_{g}} ), ~ (p,\Pi) = \frac{\mu_l U L}{H^2} (\tilde{p},\tilde{\Pi}), ~ T_i = \Delta T \; \tilde{T}_i  + T_c \; (i=l,s,w), \\ 
 ~ J_s = \epsilon \rho_s U \; \tilde{J}_s, ~ J_v = \epsilon \rho_{ve}(T_g) U \tilde{J}_v, ~ \rho_v = \rho_{ve}(T_g) \; \tilde{\rho}_v, \\ \nabla = \frac{1}{L} \tilde{\nabla}, ~ \nabla_{s,i} = \frac{1}{L} \tilde{\nabla}_{s,i} \; (i=l,s),
\end{gathered}
\end{equation}
where $\Delta T = T_m - T_c$, $\widetilde{\nabla} = \e_x\widetilde{\partial}_x +  \e_z \epsilon^{-1}\widetilde{\partial}_z$ and $\widetilde{\nabla}_{s,i} = (I-\n\n) \cdot \widetilde{\nabla} \; (i=l,s)$. The velocity scale, $U = \epsilon^3 \gamma_{lg} / \mu_l$,  such that $Ca/\epsilon^2 = 1$. Here, $Ca=\mu_l U / (\epsilon \gamma_{lg})$ denotes the capillary number. Henceforth, the tilde notation is suppressed, and ${\partial / \partial x}$, ${\partial / \partial z}$ and ${\partial / \partial t}$ are represented by the subscripts $x$, $z$ and $t$, respectively. By employing these scalings and incorporating the lubrication approximation ($\epsilon \ll 1$), we obtain the following dimensionless governing equations and boundary conditions in the liquid, ice and gas phases. 

\subsubsection{Liquid phase}
The dimensionless governing equations in the liquid phase are given by
\begin{equation} \label{eq:xmom_scaled}
\partial^2_z u = \partial_x p,
\end{equation}
\begin{equation} \label{eq:zmom_scaled}
\partial_z p = 0,
\end{equation}
\begin{equation} \label{eq:cont_scaled}
\partial_x u + \partial_z w= 0,
\end{equation}
\begin{equation} \label{eq:Tl_scaled}
\partial^2_z T_l = 0.
\end{equation}
The boundary conditions at the liquid-gas interface ($z=h(x,t)$) are
\begin{equation} \label{eq:ph_scaled}
p  = - \kappa_{lg}  - \Pi,
\end{equation}
\begin{equation} \label{eq:tstress_scaled}
\partial_z u  = 0,
\end{equation}
\begin{equation} \label{eq:ebc_scaled}
\partial_z T_l  = - \chi J_v +  \Lambda_g \partial_z T_l,
\end{equation}
\begin{equation} \label{eq:kin_scaled}
\partial_t h + u \partial_x h - w  = -D_v J_v.
\end{equation}
Here, $D_v = \rho_{ve}(T_g)/ \rho_l$ represents the density ratio, $\Lambda_g = \lambda_g/\lambda_l$ represents the thermal conductivity ratio of the gas phase to the liquid phase, and $\chi = \epsilon \rho_{ve}(T_g) U L_v H / (\lambda_l \Delta T)$ denotes the scaled latent heat of vaporization.

The boundary conditions at the liquid-ice interface (at $z=s(x,t)$) are  
\begin{equation} \label{eq:noslip_ls_scaled}
u  = 0 \; \; {\rm and} \;\; T_l = T_f,
\end{equation}
where $T_f=T_m = 1$.

The dimensionless disjoining pressure is given by 
\begin{equation}
\Pi = {A_{n} \epsilon^{-2}} \left[ \left( \frac{\beta}{h-s} \right)^n - \left( \frac{\beta}{h-s} \right)^m \right],
\label{eq:dimensionless_disj_press}
\end{equation}
where $\beta$ is of the same order as the equilibrium precursor layer thickness \citep{pham2019imbibition} and $A_{n} = H A / \gamma_{lgo}$ denotes the dimensionless Hamaker constant. The interaction between the repulsive and attractive components of eq. (\ref{eq:dimensionless_disj_press}) dictates the value of the equilibrium contact angle $\theta_{eq}$ \citep{Schwartz1998,zadravzil2006droplet,espin2015droplet,pham2019imbibition,Tembely2019}. This can be approximated \citep{pham2019imbibition}
\begin{equation} \label{eq:CA}
\theta_{eq} \approx \sqrt{\beta A_{n}}.
\end{equation}

The full mean curvatures at the liquid-gas and liquid-ice interfaces, retaining higher order contributions, are given by 
\begin{equation}
\kappa_{lg} = \frac{h_{xx}}{(1+\epsilon^2 h_x^2)^{\frac{3}{2}}} ~ {\rm and} ~
\kappa_{sl} = \frac{s_{xx}}{(1+\epsilon^2 s_x^2)^{\frac{3}{2}}}.
\end{equation}

\subsubsection{Solid (ice) phase}

Under the lubrication approximation ($\epsilon \ll 1$), the dimensionless energy conservation equations for the ice phase is given by
\begin{equation}
\partial^2_z T_s = 0.
\end{equation}
The energy balance at the liquid-ice interface ($z=s(x,t)$) gives
\begin{equation} \label{Stefan_scaled}
Ste \left( \Lambda_s \partial_z T_s - \partial_z T_l \right) = J_s,
\end{equation}
where $\Lambda_s = \lambda_s / \lambda_l$ is the thermal conductivity ratio and $Ste = \lambda_l \Delta T / ( \epsilon \rho_s U L_f H )$ denotes the Stefan number. At $z=s(x,t)$, the conservation of mass (eq. \ref{eq:mass_bc_solid_ND}) gives
\begin{equation} \label{eq:Js_scaled}
\partial_t s  - w = D_s J_s, ~ {\rm where} ~ J_s = \partial_t s.
\end{equation}
where $D_s=\rho_s/\rho_l$ is the density ratio of the frozen phase to the liquid phase.

At the ice-solid interface ($z=0$), we impose
\begin{equation}
T_s = T_w.
\end{equation}

\subsubsection{Gas phase}

The dimensionless conservation equation for the vapor concentration becomes
\begin{equation}\label{eq:rho_v_scaled}
Pe_v({\partial_t\rho_v}+u_g\partial_x\rho_v+w_g\partial_z\rho_v) = \partial^2_z\rho_v +\epsilon^2\partial^2_x\rho_v.
\end{equation}
At the liquid-gas interface ($z=h(x,t)$), eq. (\ref{eq:int_vap_conc})
reduces to
\begin{equation}\label{bc:1_mass_bal}
\frac{Pe_v}{K} \left[ \rho_{veR}(1 + \Delta p + \Psi \left (T_{l}-1 \right)) - \rho_v \right]= - \partial_z \rho_v, 
\end{equation}
while the constitutive equation for the evaporation flux gives
\begin{equation}\label{bc:k}
K J_v = \rho_{veR}(1 + \Delta p + \Psi \left (T_{l}-1 \right)) - \rho_v.
\end{equation}
The boundary condition far from the droplet ($z = D_g$) is given by
\begin{equation} 
\rho_v = RH,\label{eq:vap_conc_hum_scaled}
\end{equation}
where $RH$ denotes the relative humidity. The various dimensionless numbers appearing in eqs. (\ref{eq:rho_v_scaled} - \ref{bc:k}) are
\begin{equation}
\begin{gathered}
Pe_v = \frac{\epsilon U H}{D_m}, ~ K = \frac{\epsilon U}{\alpha } \left( \frac{2 \pi M}{R_g T_{m}} \right)^{1/2}, ~ \Delta = \frac{\mu_{l}ULM}{H^{2}\rho_{l}R_gT_m}, \\  \Psi = \frac{L_{v}M\Delta T}{R_{g}T_{m}^2}, ~\rho_{veR} = {{\rho_{ve}(T_m)} \over \rho_{ve}(T_g)}, ~ {\rm and} ~ RH ={\rho_{vi} \over \rho_{ve}(T_g)}.
\end{gathered}
\label{eq:nondimK}
\end{equation}

Under the lubrication approximation ($\epsilon \ll 1$), the dimensionless energy conservation equations for the gas phase is given by
\begin{equation}
\partial^2_z T_g = 0.
\end{equation}
The boundary condition far from the droplet ($z = D_g$) is given by
\begin{equation} \label{eq:Tg_top}
T_g = T_v,
\end{equation}
At the liquid-gas interface ($z=h(x,t)$)
\begin{equation} \label{eq:Tg_liq-gas}
T_g = T_l.
\end{equation}

\subsubsection{Solid substrate}

The scaled energy conservation equation for the solid substrate is given by
\begin{equation}
\partial^2_z T_w = 0.
\end{equation}
The boundary condition at the solid-ice interface ($z=0$) is
\begin{equation}
\Lambda_s \partial_z T_s = \Lambda_w \partial_z T_w,
\end{equation}
where $\Lambda_w = \lambda_w / \lambda_l$ denotes the thermal conductivity ratio. At the bottom of the substrate (at $z=-D_w$), we impose
\begin{equation}
T_w  = 0.
\end{equation}

\subsection{Evolution equations}

By integrating eqs. (\ref{eq:xmom_scaled}) and (\ref{eq:zmom_scaled}) with respect to $z$ and using eqs. (\ref{eq:tstress_scaled}), (\ref{eq:noslip_ls_scaled}) and (\ref{eq:ph_scaled}), we get 
\begin{equation}
u = \frac{\partial_x p}{2}   (z^2 - s^2) - h \partial_x p (z-s),
\end{equation}
\begin{equation}
p = - \kappa_{lg}  - \Pi.
\end{equation}

By integrating eq. (\ref{eq:cont_scaled}) and using eq. (\ref{eq:kin_scaled}), we get the following evolution equation: 
\begin{equation}
\partial_t h - \partial_t s = - \partial_x q_l - D_v J_v - D_s J_s,
\end{equation}
where
\begin{equation}
q_l = \frac{\partial_x p}{2} \left (\frac{h^3}{3} - s^2 h + \frac{2s^3}{3} \right) - h \partial_x p  \left (\frac{h^2}{2}-s h  + \frac{s^2}{2} \right).
\end{equation}

Similarly, by integrating eq. (\ref{eq:Tl_scaled}) and using eqs. (\ref{eq:ebc_scaled}) and (\ref{eq:noslip_ls_scaled}), we get 
\begin{equation}
T_l = \left(-\chi J_v+\Lambda_g\left(\frac{T_{v}-T_{l}\big|_{h}}{D_g-h}\right)\right) (z-s) + T_f. \label{eq262}
\end{equation}
The temperature distribution in the ice phase is governed by
\begin{equation}
T_s = \frac{T_f}{D_w+s \Lambda_w/\Lambda_s} \left( D_w + z  \frac{\Lambda_w}{\Lambda_s} \right).
\end{equation}
Using the above expression along with eq. (\ref{eq:Js_scaled}) and introducing them into eq. (\ref{Stefan_scaled}), we get
\begin{equation}\label{eq:Freezing_rate}
\partial_t s = Ste \left(  \frac{\Lambda_w T_f}{D_w+s \Lambda_w/\Lambda_s} + \chi J_v -  \Lambda_g\left(\frac{T_{v}-T_{l}\big|_{h}}{D_g-h}\right)\right). 
\end{equation}

The temperature profile in the solid substrate is given by 
\begin{equation}
T_w = \frac{T_f}{D_w + s \; \Lambda_w/\Lambda_s}(z+D_w).
\end{equation}

\subsubsection{Gas phase - K\'{a}rm\'{a}n-Pohlhausen approximation}

In order to retain the advection terms in the vapour concentration balance equation, we apply the K\'{a}rm\'{a}n-Pohlhausen integral approximation and define the integrated form of $\rho_v$, which is given by
\begin{equation}
\int_{h}^{D_g} \rho_v dz = f. \label{eq:f_int}
\end{equation}
In order to be able to evaluate eq. (\ref{eq:f_int}), we need to prescribe the form of $\rho_v$ as a function of the vertical coordinate. To this end, we assume that $\rho_v$
can be approximated by a polynomial of the form 
\begin{equation}
    \rho_v = c_1z^2 + c_2z+c_3.
\end{equation}
By substituting the corresponding polynomial in eq. (\ref{eq:f_int}) and applying the appropriate boundary conditions, i.e. eqs. (\ref{bc:1_mass_bal}) and (\ref{eq:vap_conc_hum_scaled}), it is possible to evaluate the polynomial constants and eventually derive the following expressions for the constants $c_1$, $c_2$ and $c_3$.
\begin{subeqnarray}
    c_1 & = & \frac{f-\frac{Pe_v J_v}{2}(D_g-h)^2-RH(D_g-h)}{\frac{2}{3}(h-D_g)^3},\\[3pt]
    c_2 & = & -Pe_ vJ_v - 2h\left[\frac{f-\frac{Pe_v J_v}{2}(D_g-h)^2-RH(D_g-h)}{\frac{2}{3}(h-D_g)^3}\right],\\[3pt]
    c_3 & = & RH-{D_g}^2\left[\frac{f-\frac{Pe_vJ_v}{2}(D_g-h)^2-RH(D_g-h)}{\frac{2}{3}(h-D_g)^3}\right] + Pe_v J_vD_g \nonumber\\
    && + 2h D_g\left[\frac{f-\frac{Pe_v J_v}{2}(D_g-h)^2-RH(D_g-h)}{\frac{2}{3}(h-D_g)^3}\right].
\end{subeqnarray}
Then, by integrating eq. (\ref{eq:rho_v_scaled}) and using the boundary conditions, we get the following integrated form of the concentration equation:
\begin{eqnarray}
     Pe_v\left[\frac{\partial f}{\partial t} + \frac{\partial g_v}{\partial x}  - \rho_{v}\big|_{h}(D_v J_v)\right]  &=& \nonumber \\
     \frac{\partial \rho_v}{\partial z}\big|_{D_g} - \frac{\partial \rho_v}{\partial z}\big|_{h} &+& \epsilon^2 \left[\frac{\partial }{\partial x}\left(\frac{\partial f}{\partial x} + \rho_v\big|_{h}h_x\right) + \frac{\partial \rho_v}{\partial x}\big|_{h}h_x\right],
\end{eqnarray}
where 
\begin{equation} \label{eq:g_int}
 \int_{h}^{D_g} u_g\rho_v dz = g_v.
\end{equation}
To evaluate eq. (\ref{eq:g_int}), we need to prescribe the $x$-component of the velocity profile in the gas phase, which can be approximated by considering the following linear profile
\begin{equation}\label{eq:u_g_scaled}
u_g = a z + b.
\end{equation}
The constants $a$ and $b$ can be evaluated by simply considering that, at the liquid-gas interface ($z=h(x,t)$), the velocity of the gas is equal to the velocity of liquid
\begin{equation}\label{eq:vap_velocity_bc1}
u_{g} = u,~ {\rm and} ~ w_{g} = w,
\end{equation}
and at the far-field ($z = D_g$),
\begin{equation}\label{eq:vap_velocity_bc2}
u_{g} = 0,~ {\rm and} ~ w_{g} = 0.
\end{equation}
Thus,
\begin{subeqnarray}
a & = & \frac{\frac{\partial_x p}{2}  (h^2 - s^2) - h \partial_x p (h-s)}{(h-D_g)},\\[3pt]
b & = & -D_g\left[\frac{\frac{\partial_x p}{2}   (h^2 - s^2) - h \partial_x p  (h-s)}{(h-D_g)}\right].
\end{subeqnarray}

\subsection{Initial and boundary conditions - Numerical procedure}

In our modelling, the droplet is deposited on a thin precursor layer that resides on top of a thin layer of ice. By selecting the value of the dimensionless Hammaker constant $A_{n}$ appropriately, we ensure that the droplets achieve an equilibrium contact angle with the substrate. This choice governs the equilibrium contact angle by balancing the repulsive and attractive components of the disjoining pressure interaction. The following initial conditions are imposed on the domain.
\begin{eqnarray}
& h(x,t=0) = \max(h_{\infty}+s_{\infty}-x^{2}-x(d_{0}+2)-\frac{d_{0}^2}{4}-d_{0},h_{\infty}+s_{\infty}),~ \nonumber\\
& f(x,t=0) = RH(D_{g}-h(x,t=0)).
\end{eqnarray}
The dimensionless equilibrium precursor layer thickness ($h_{\infty}=(h-s_{\infty})= H_{\infty}/{H}$) far from the droplet can be estimated by considering that in this region, the fluid is flat with zero mean curvature and sufficiently thin such that the attractive van der Waals forces suppress evaporation. Therefore, by taking the constitutive equation for the evaporation flux and setting it to zero, the dimensionless equilibrium precursor thickness $h_{\infty}$ can be evaluated by solving the following nonlinear equation
\begin{equation}
\rho_{veR}\left(1 + \Delta\left(-{\epsilon^{-2}A_{n}} \left[ \left( \frac{\beta}{h_\infty} \right)^n - \left( \frac{\beta}{h_\infty} \right)^m \right]\right) + \Psi\left(\frac{\frac{\Lambda_g}{\Lambda_w}\frac{D_w}{D_g}{T_v}}{1 + \frac{\Lambda_g}{\Lambda_w}\frac{D_w}{D_g}}-1\right)\right) - RH= 0.
\label{pre_model}
\end{equation}
In all our simulations, we have taken $\beta=0.01$. The initial thickness of the thin ice layer is taken to be $s_{\infty}=10^{-3}$, but we have verified that our findings remain unchanged when considering, e.g. one or two orders of magnitude smaller values of $s_{\infty}$. To prevent the precursor layer from freezing, we adopt a similar approach to that of \cite{zadravzil2006droplet} and introduce the thickness-dependent Stefan number $(Ste (x))$, which is given by the following expression.
\begin{equation}
Ste(x) = \frac{1}{2}(1+\tanh[4\times10^{3}((h-s)-1.4\beta)])Ste.  \label{pre_model1}
\end{equation}
We also assume that when the thickness of the ice layer is very low ($s_{\infty}=10^{-3}$), the temperature of the liquid-gas interface and the ice-liquid interface are the same as the top of the substrate by using a thickness-dependent boundary condition at the liquid-ice interface ($z=s(x,t)$) and the liquid-gas interface ($z=h(x,t)$). This is given by.
\begin{equation}
T_s = F(s)\left(T_{f}-\frac{\frac{\Lambda_g}{\Lambda_w}\frac{D_w}{D_g}{T_v}}{1 + \frac{\Lambda_g}{\Lambda_w}\frac{D_w}{D_g}}\right) + \frac{\frac{\Lambda_g}{\Lambda_w}\frac{D_w}{D_g}{T_v}}{1 + \frac{\Lambda_g}{\Lambda_w}\frac{D_w}{D_g}},
\end{equation}
where 
\begin{equation}
F(s)=\frac{1}{2}(1+\tanh[4\times10^{3}(s-1.5 \beta)]).  
\end{equation}
\begin{equation}
T_{l}\big|_{h} = F(s) \left(\frac{\left(-\chi J_v+\left(\frac{\Lambda_g T_{v}}{D_g-h}\right)\right)(h-s) + {T_f}}{1+\frac{\Lambda_g}{D_{g} - h}(h-s)}-\frac{\frac{\Lambda_g}{\Lambda_w}\frac{D_w}{D_g} {T_v}}{1 + \frac{\Lambda_g}{\Lambda_w}\frac{D_w}{D_g}}\right)
+ \frac{\frac{\Lambda_g}{\Lambda_w}\frac{D_w}{D_g} {T_v}}{1 + \frac{\Lambda_g}{\Lambda_w}\frac{D_w}{D_g}}
\end{equation}
Finally, we impose the following set of boundary conditions
\begin{equation}
h_{x}(0,t) = h_{xxx}(0,t) = h_{x}(x_{\infty},t) = h_{xxx}(x_{\infty},t)= 0,
\end{equation}
\begin{equation}
s_{x}(0,t) = s_{xxx}(0,t) = s_{x}(x_{\infty},t) = s_{xxx}(x_{\infty},t)= 0,
\end{equation}
\begin{equation}
h(x_{\infty},t)-s(x_{\infty},t) = h_{\infty}, ~ s(x_{\infty},t) = s_{\infty}, 
\end{equation}
\begin{equation}
f_x(0,t) = 0,
\end{equation}
\begin{equation}
f(x_{\infty},t) = RH(D_{g}-s(x_{\infty},t)-h_{\infty}).
\end{equation}
where $x_{\infty}$ represents the end of the domain.

The dimensionless governing equations are discretized using the Galerkin Finite Element Method, with weak forms derived for each equation. The solutions are iteratively obtained using the Newton-Raphson scheme, advancing in time via an implicit Euler method with an adaptive time step. The time step adapts based on the maximum residual errors from the previous step, a characteristic feature of the adaptive implicit Euler method. The LAPACK linear algebra package is utilized, with the iterative program written in FORTRAN. We validate our model by comparing the tip angle at the end of freezing with the angle observed in the experimental study by \citet{marin2014universality} for a typical set of parameters, as shown in figure \ref{fig:Tip_angle}. Further validation was conducted by simulating a scenario previously examined by \citet{kavuri2023freezing} and \citet{zadravzil2006droplet} in figure \ref{fig:freez_comp} that illustrates the temporal evolution of the droplet shape, $h$ (solid line), and the freezing front, $s$ (dot-dashed line), for a drop placed on a cold substrate. Our simulations employ a one-dimensional mesh along the $x$-direction, covering a computational domain of twelve dimensionless units with 9601 grid points. The optimal computational domain size and grid density were determined through a thorough investigation of domain size effects and a grid convergence test. Appendix \ref{sec:Val} presents the details about the validation of the present numerical model (figures \ref{fig:Tip_angle} and \ref{fig:freez_comp}) and the grid convergence test (figure \ref{fig:Grid_study}). For more information, refer to \cite{kavuri2023freezing, wang2024role, williams2021}.

\section{Results and Discussion} \label{sec:dis}

\begin{figure}
\centering
\includegraphics[width=0.95\textwidth]{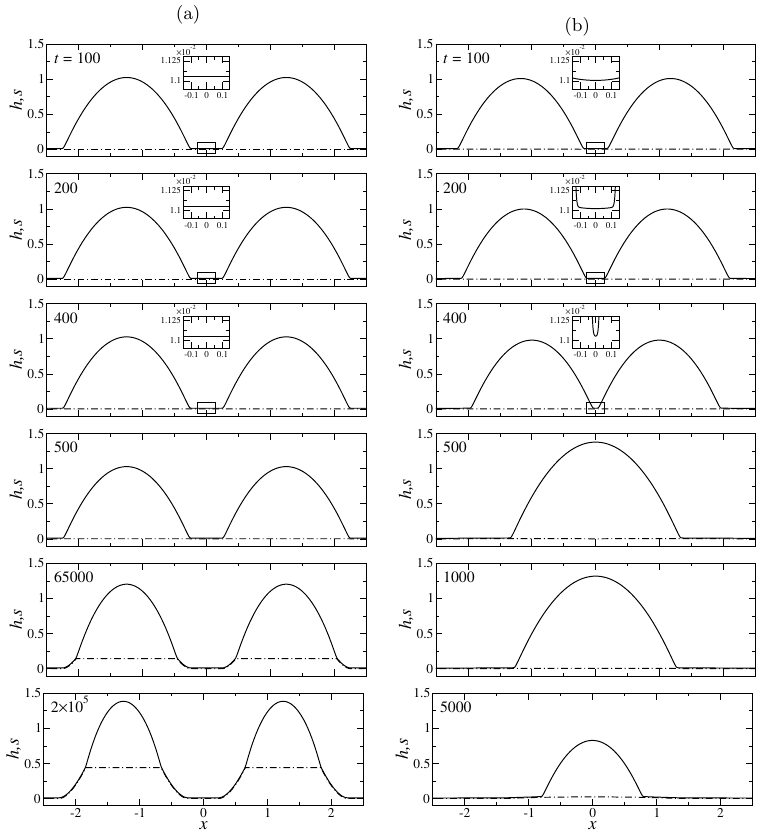}
\caption{Temporal evolution of the shape of droplets, $h$ (solid line) and the freezing front, $s$ (dot-dashed line) for two droplets placed on a cold substrate with initial separation distance $d_0 = 0.5$. (a) Without evaporation ($RH = 1.0$, $\chi = 0$, $\Delta = 0$, $\Psi = 0$, and $Pe_{v} = 0$) and (b) with evaporation ($RH = 0.9$, $\chi = 1.6$, $\Delta = 10^{-4}$, $\Psi = 0.02$, and $Pe_{v} = 1$). The remaining dimensionless parameters in both the systems are $Ste = 2.53\times10^{-5}$, $T_{v} = 1.0$, $A_{n} = 17.0$, $D_v = 10^{-3}$, $D_{g} = 2.0$, $D_{s} = 0.9$, $\Lambda_{S} = 3.89$, $\Lambda_{W} = 0.33$, $\Lambda_{g} = 0.041$, $K = 8\times10^{-4}$, $D_{w} = 15.0$, $\epsilon=0.2$, and $\rho_{veR} = 1.0$.}
\label{fig:coalescence_demo}
\end{figure}

We investigate the evaporation-driven coalescence phenomenon of two volatile droplets placed in close vicinity on a cold substrate undergoing freezing. In order to demonstrate this phenomenon, we begin the presentation by considering two scenarios. In the first scenario (depicted in figure \ref{fig:coalescence_demo}(a)), we examine a system comprising two droplets initially separated by a distance of $d_0 = 0.5$ on a cold substrate undergoing freezing. In this case, we neglect the evaporation by setting $RH = 1.0$, $\chi = 0$, $\Delta = 0$, $\Psi = 0$, and $Pe_{v} = 0$. In the second scenario (illustrated in figure \ref{fig:coalescence_demo}(b)), we analyze the same system but with evaporation taken into account by setting $RH = 0.9$, $\chi = 1.6$, $\Delta = 10^{-4}$, $\Psi = 0.02$, and $Pe_{v} = 1$. The remaining dimensionless parameters ($Ste = 2.53\times10^{-5}$, $T_{v} = 1.0$, $A_{n} = 17.0$, $D_v = 10^{-3}$, $D_{g} = 2.0$, $D_{s} = 0.9$, $\Lambda_{S} = 3.89$, $\Lambda_{W} = 0.33$, $\Lambda_{g} = 0.041$, $K = 8\times10^{-4}$, $D_{w} = 15.0$, $\epsilon=0.2$, and $\rho_{veR} = 1.0$) are kept the same in both the systems. The dimensionless parameters considered in figure \ref{fig:coalescence_demo}(b) are termed as `base' parameters. It is to be noted that the volatility of the liquid considered in our study is significantly higher than that of water. The value of the dimensionless parameter characterizing the volatility ($D_v$) is set at $10^{-3}$, which corresponds to highly volatile liquids, e.g. butane, pentane, ammonia, propyl benzene, benzol, and toluene. Figure \ref{fig:coalescence_demo}(a) and (b) depict the temporal evolution of the shape of the droplet $(h)$ and the freezing front $(s)$. It can be seen in figure \ref{fig:coalescence_demo}(a) that for the system, when the evaporation is neglected, the contact line of the droplets remains fixed while the freezing front propagates upward at later times, keeping the volume of liquid the same. In contrast, in the system undergoing evaporation, the droplets migrate closer at early times, leading to their coalescence at $t=400$. Subsequently, the droplets merge to form a single droplet, whose size decreases due to the associated evaporation.  

\subsection{Mechanism} \label{sec:mechanism}

\begin{figure}
\centering
\hspace{0.75cm}{\large (a)}   \hspace{5.5cm}  {\large (b)} \\
\includegraphics[width=0.45\textwidth]{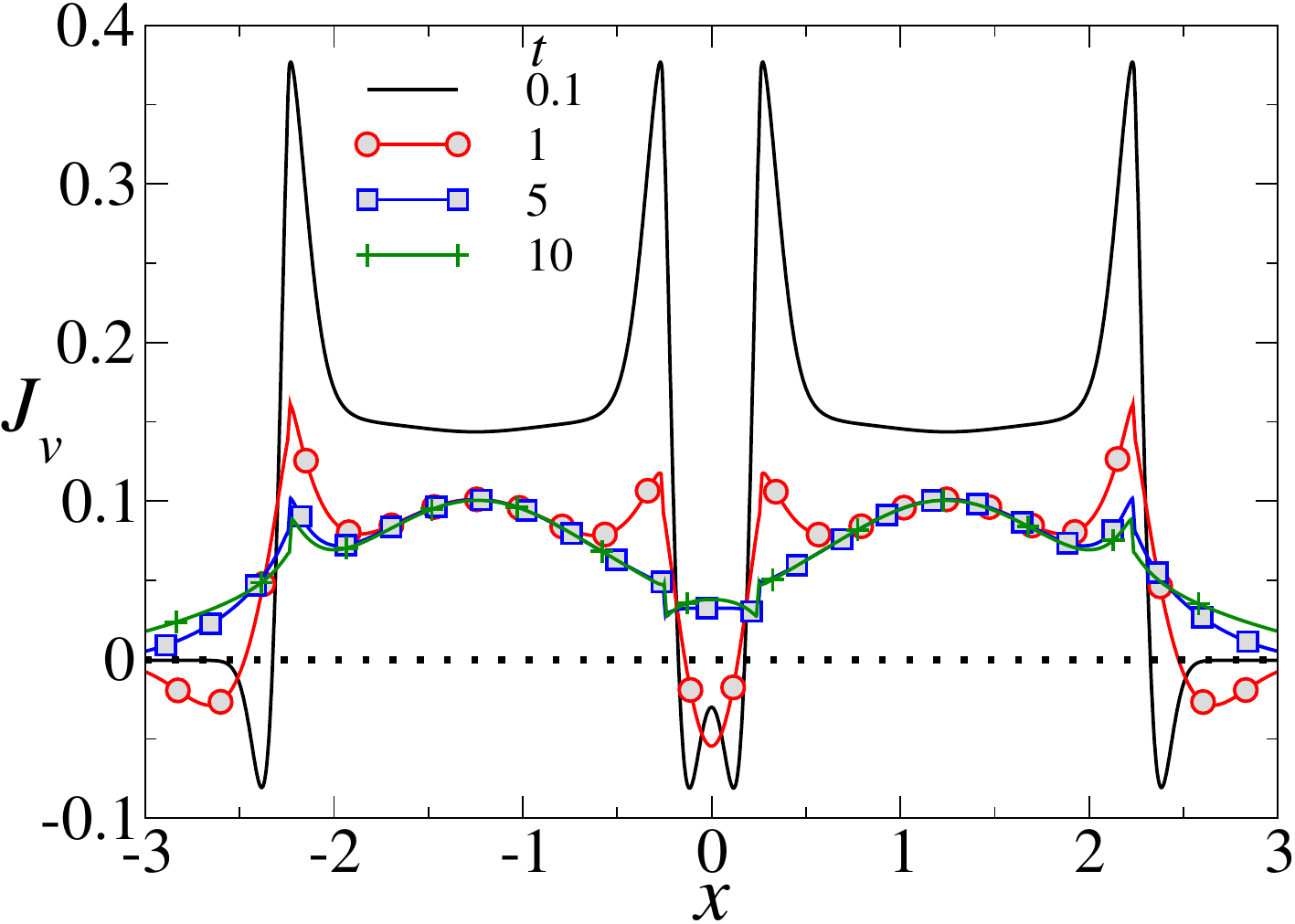}
\includegraphics[width=0.45\textwidth]{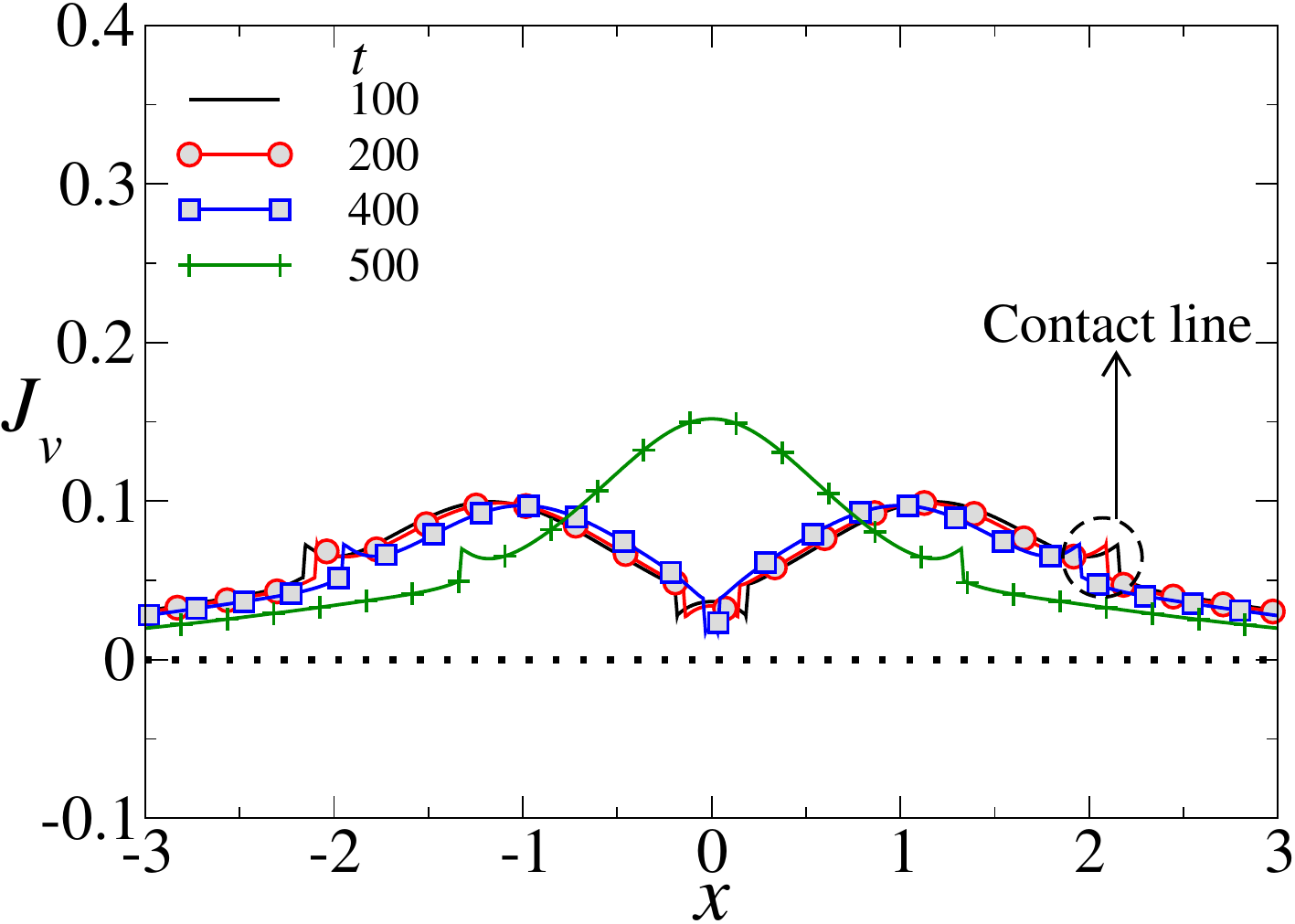}\\
\hspace{0.8cm} {\large (c)}\\
\includegraphics[width=0.45\textwidth]{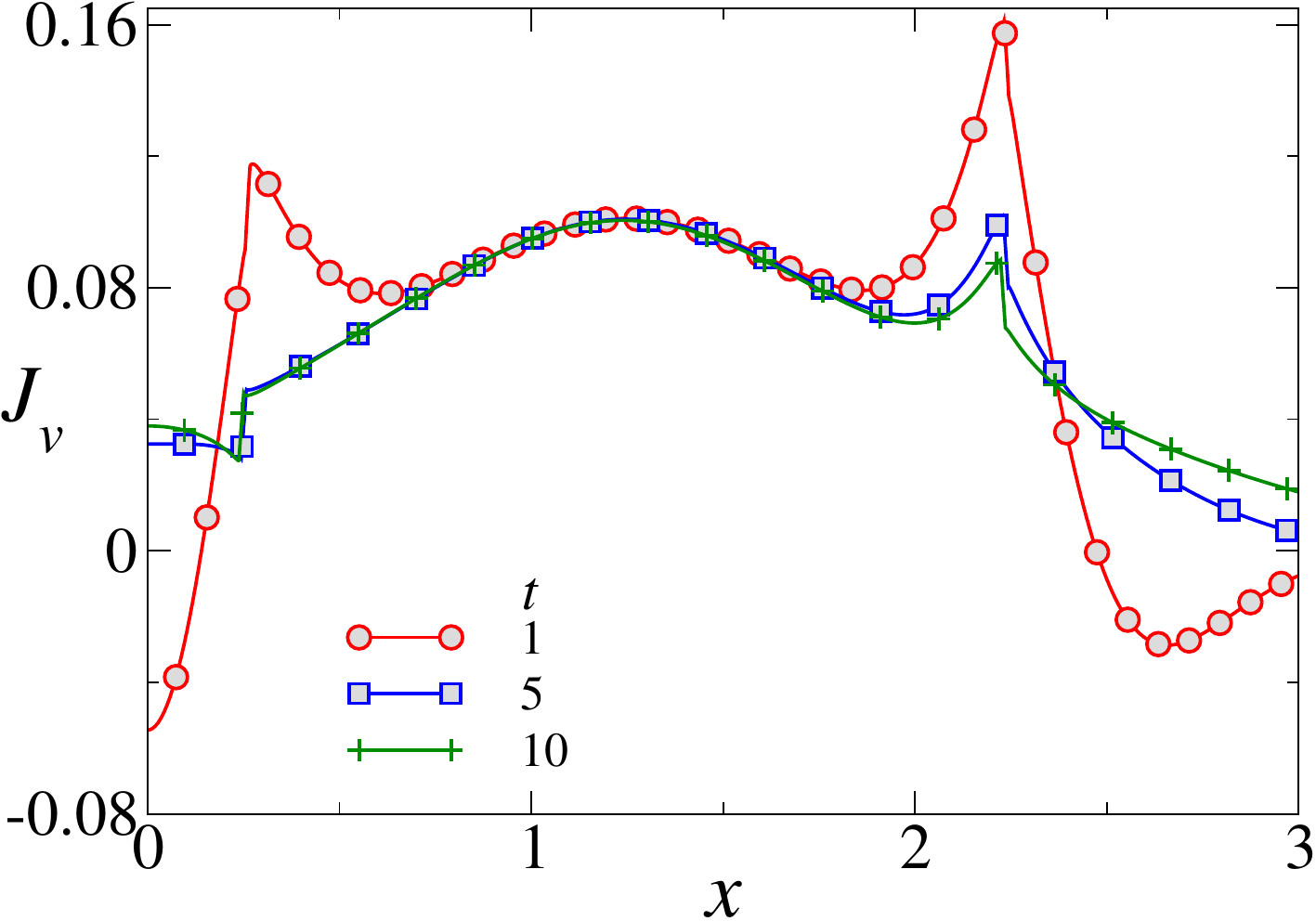}
\caption{The evolution of the evaporation flux ($J_v$) profile at (a) early and (b) later times. (c) The enlarged view of the evaporation flux ($J_v$) profile, shown in panel (a), at $t = 1$, $5$ and $10$ for the right drop. The rest of the dimensionless parameters are the same as figure \ref{fig:coalescence_demo}(b) (`base' parameters).}
\label{fig:evap_prof}
\end{figure}

\begin{figure}
\centering
\hspace{0.75cm}{\large (a)}   \hspace{5.5cm}  {\large (b)} \\
\includegraphics[width=0.45\textwidth]{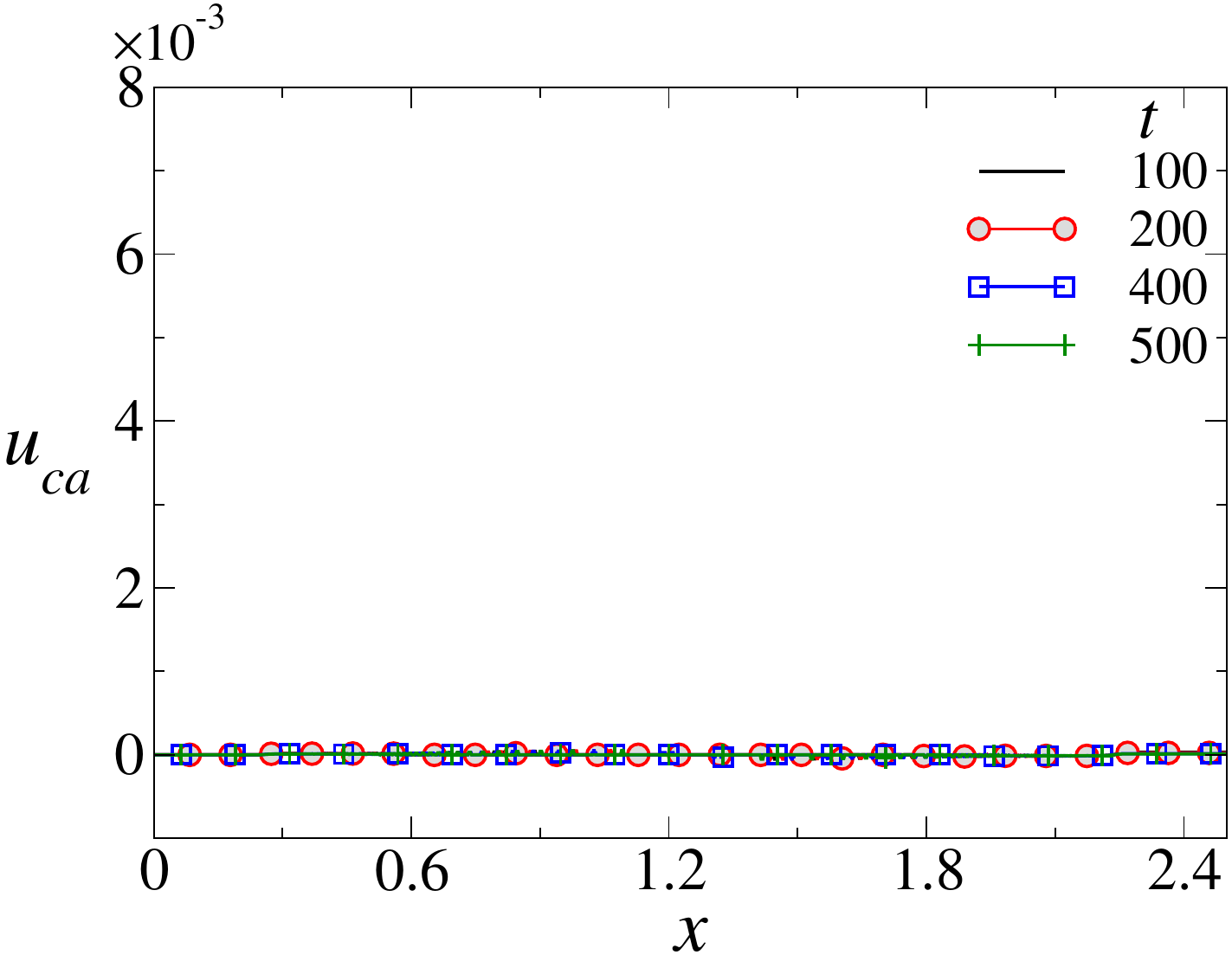} 
\hspace{0mm}
\includegraphics[width=0.45\textwidth]{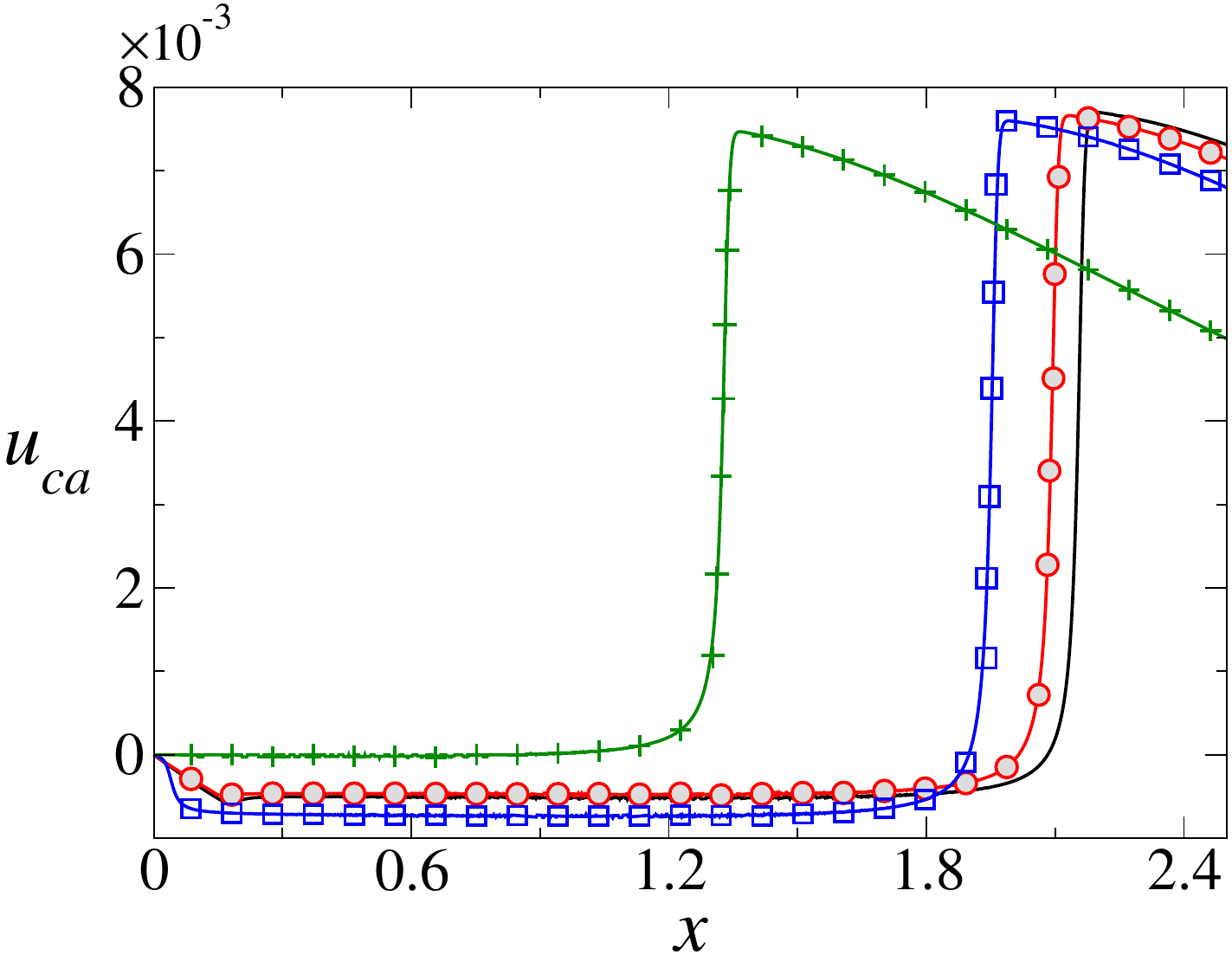}\\
\hspace{0.8cm} {\large (c)}\\
\includegraphics[width=0.45\textwidth]{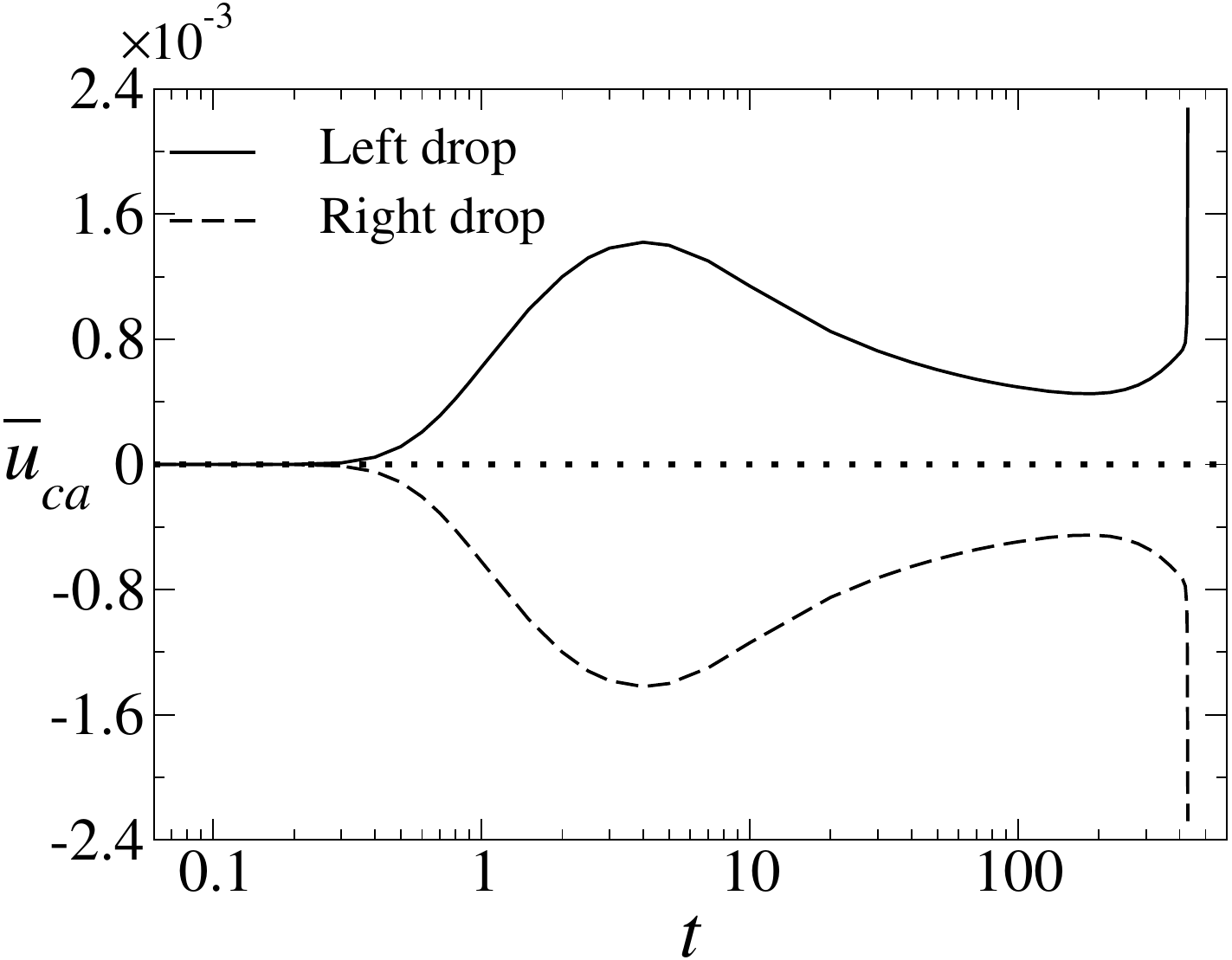}
\caption{Variation of the capillary velocity ($u_{ca}(x)$) along the substrate at different times in scenarios (a) without evaporation ($RH = 1.0$, $\chi = 0$, $\Delta = 0$, $\Psi = 0$, and $Pe_{v} = 0$) and (b) with evaporation ($RH = 0.9$, $\chi = 1.6$, $\Delta = 10^{-4}$, $\Psi = 0.02$, and $Pe_{v} = 1$). (c) Variation of the average capillary velocity ($\bar{u}_{ca}$) with time till coalescence for the left and right drops (with evaporation). The remaining dimensionless parameters in both the systems are $Ste = 2.53\times10^{-5}$, $T_{v} = 1.0$, $A_{n} = 17.0$, $D_v = 10^{-3}$, $D_{g} = 2.0$, $D_{s} = 0.9$, $\Lambda_{S} = 3.89$, $\Lambda_{W} = 0.33$, $\Lambda_{g} = 0.041$, $K = 8\times10^{-4}$, $D_{w} = 15.0$, $\epsilon=0.2$, and $\rho_{veR} = 1.0$.}
\label{fig:uca_plot}
\end{figure}

To understand the behaviour, in figure \ref{fig:evap_prof}(a) and (b), we analyze the evaporation flux $(J_v)$ during early and later stages for the system depicted in figure \ref{fig:coalescence_demo}(b). At $t = 0.1$, the droplets exhibit a symmetrical evaporation flux profile, as it takes some time for evaporation to initiate. As time progresses, the evaporation becomes asymmetrical, with higher $J_v$ values in the outer region and lower $J_v$ values in the inner area of the droplets due to the effect of vapour shielding, while the droplets migrate inwards. The asymmetry in the evaporation is clearly noticeable at times $t = 1$, $5$ and $10$ in figure \ref{fig:evap_prof}(a). An enlarged view of figure \ref{fig:evap_prof}(a) at $t = 1$, $5$ and $10$, highlighting the asymmetry in the evaporation flux $(J_v)$ at the inner and outer edges of the right drop, is shown in figure \ref{fig:evap_prof}(c). It is to be noted that when vapour from one droplet intersects with vapour from another droplet, it creates a more concentrated region, thereby diminishing the evaporation flux near the neighbouring droplet \citep{lee2023vapor}. This asymmetry in the evaporation flux persists until the droplets coalesce. After coalescence, the evaporation flux profile again becomes symmetrical, as observed at $t = 500$ in figure \ref{fig:evap_prof}(b). The slight spikes observed near the outer contact line of the drops (highlighted in figure \ref{fig:evap_prof}(b)) indicate a slightly higher evaporation flux in that region, a common characteristic is observed in thin droplets \citep{Karapetsas2016}. This observation suggests that increased evaporation at the outer edges of the droplets triggers a capillary flow towards this side to replenish the lost mass, which in turn pushes the droplet inwards where evaporation is low utilizing viscous friction from the substrate \citep{sadafi2019vapor}. This effect can be quantified by examining the capillary velocity ($u_{ca}(x)$) and the average capillary velocity ($\bar{u}_{ca}$) inside a drop, which are given by
\begin{equation}\label{ucaeq}
u_{ca} (x) = \frac{\int_{s}^{h}udz}{\int_{s}^{h}dz} = -\frac{1}{3}p_{x}(h-s)^{2},
\end{equation}
\begin{equation}\label{ucaavg}
\bar{u}_{ca} = \frac{\int_{x_{cl}}^{x_{cr}}\int_{s}^{h}udzdx}{\int_{x_{cl}}^{x_{cr}}\int_{s}^{h}dzdx} = -\frac{1}{3}\frac{\int_{x_{cl}}^{x_{cr}}p_{x}(h-s)^{3}dx}{\int_{x_{cl}}^{x_{cr}}(h-s)dx}.
\end{equation}
Here $x_{cl}$ and $x_{cr}$ represent the left and right contact lines of the drop. As the temperature of the unfrozen liquid remains near the melting point during the freezing process, we have not accounted in the Marangoni flow in our investigation. Consequently, the average capillary velocity of the droplet aligns with the velocity of its center of mass. Figures \ref{fig:uca_plot}(a) and \ref{fig:uca_plot}(b) illustrate the variation of capillary velocity ($u_{ca}(x)$) along the substrate at different times, without and with evaporation, for the set of parameters considered in figure \ref{fig:coalescence_demo}(a) and (b), respectively. It can be observed that while the scenario without evaporation displays negligible capillary velocity, the system undergoing evaporation exhibits a finite capillary velocity. The temporal evolution of the average capillary velocity $\bar{u}_{ca}$ for the left and right drops, as depicted in igure \ref{fig:uca_plot}(c), shows a positive value for the left drop and a negative value for the right drop. This indicates migration of the drops towards each other. The asymmetry in evaporation, shown in figure \ref{fig:evap_prof} during the early stages $(t < 5)$, results in a notable average capillary velocity within the drop. This velocity decreases as the freezing front evolves, followed by a rapid increase around $t \approx 430$ (figure \ref{fig:uca_plot}(c)), coinciding with the proximity of the two drops and subsequent coalescence. The following subsection provides a detailed examination of the interaction between the two drops through halos.

\subsection{Early dynamics: Interaction of frost halos} \label{sec:early}
\begin{figure}
\centering
\hspace{0.75cm}{\large (a)}   \hspace{5.5cm}  {\large (b)} \\
\includegraphics[width=0.45\textwidth]{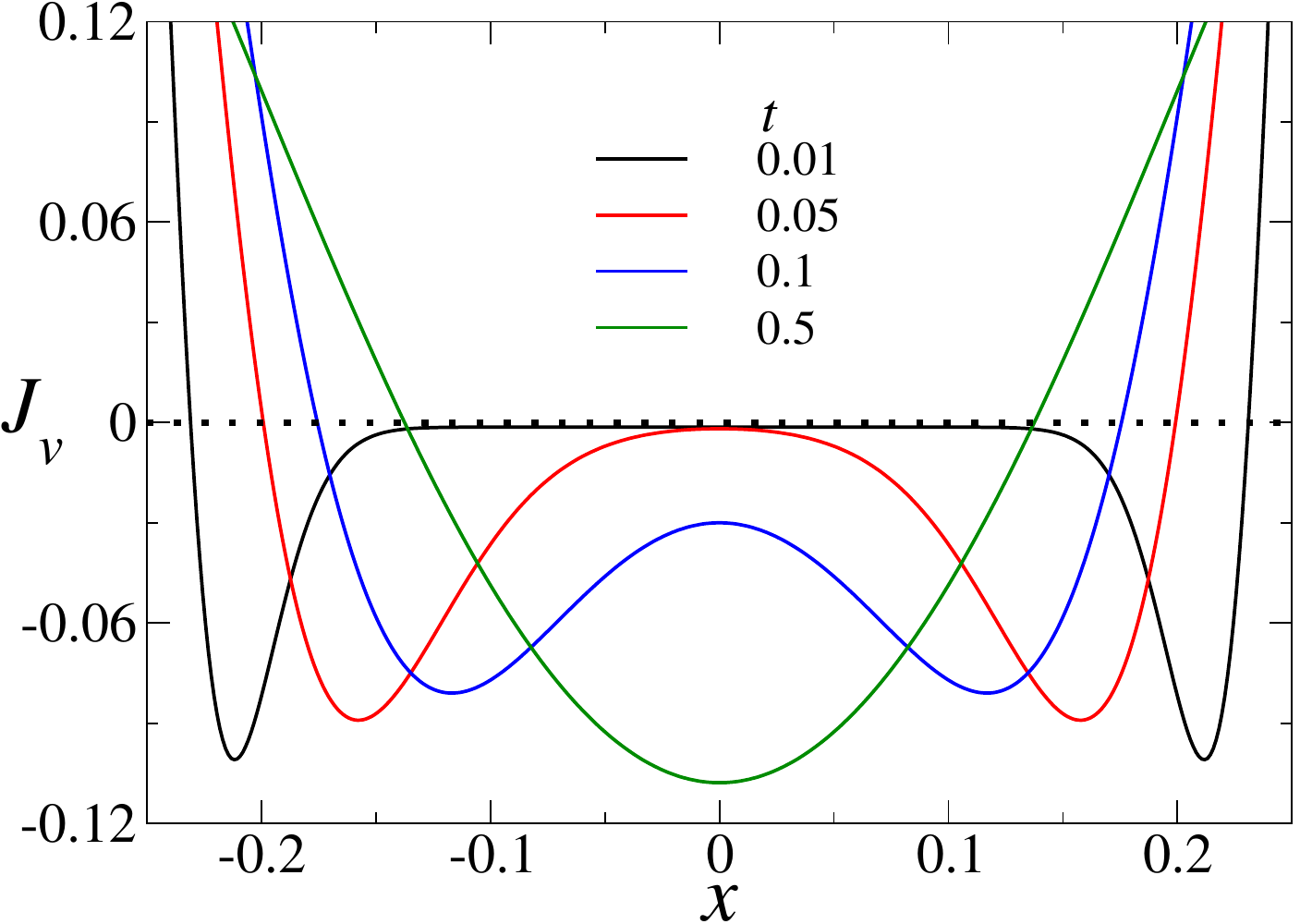} 
\hspace{0mm}
\includegraphics[width=0.45\textwidth]{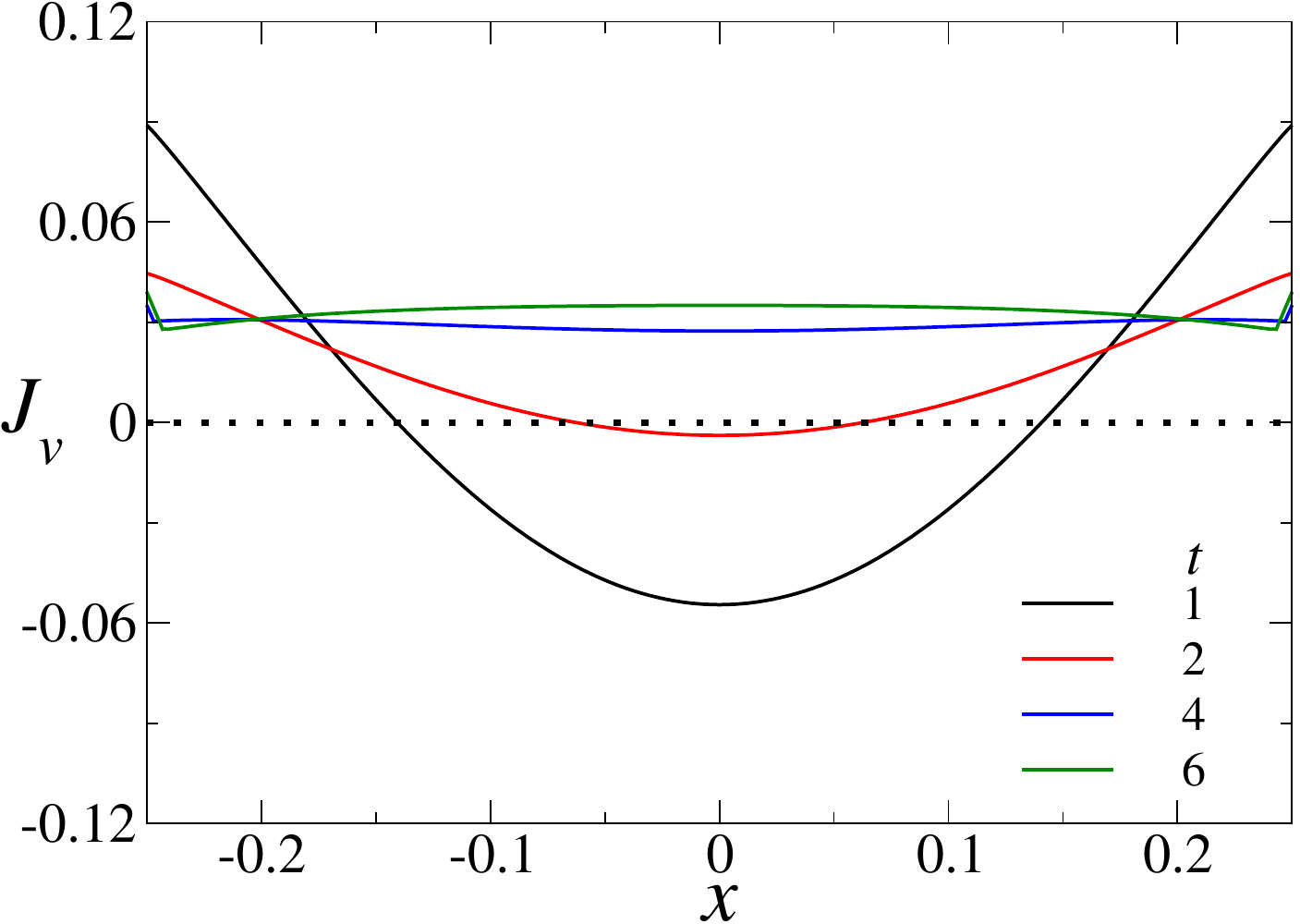}\\
\hspace{0.8cm} {\large (c)}\\
\includegraphics[width=0.45\textwidth]{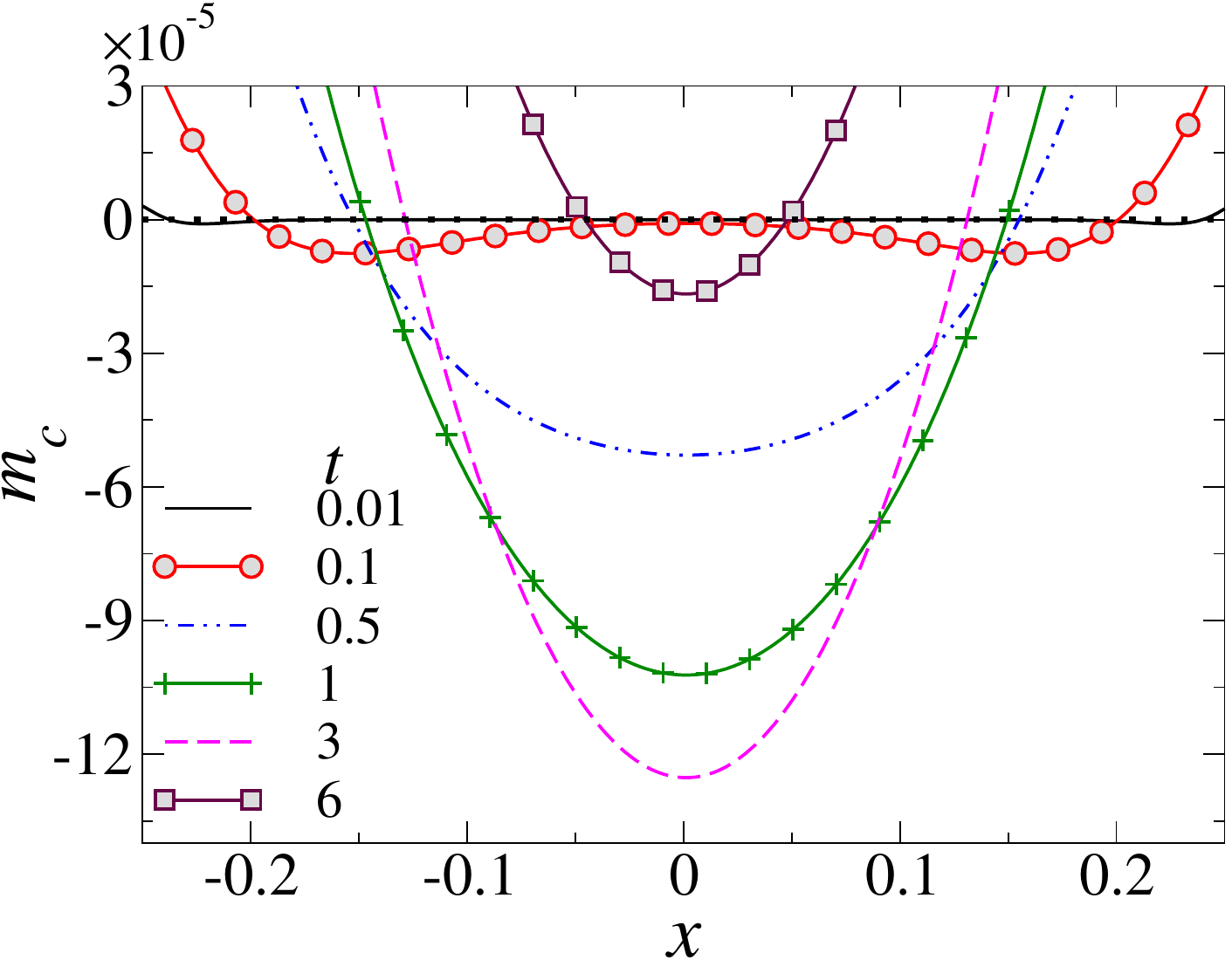}
\caption{The variation of evaporation flux ($J_v$) in the region between the two drops at the (a) early and (b) later times. (c) The variation of the total mass of the condensate deposited in between the two drops at different times. The remaining dimensionless parameters are $d_0 = 0.5$, $RH = 0.90$, $\chi = 1.6$, $\Delta = 10^{-4}$, $\Psi = 0.02$, $Pe_{v} = 1$, $Ste = 2.53\times10^{-5}$, $T_{v} = 1.0$, $A_{n} = 17.0$, $D_v = 10^{-3}$, $D_{g} = 2.0$, $D_{s} = 0.9$, $\Lambda_{S} = 3.89$, $\Lambda_{W} = 0.33$, $\Lambda_{g} = 0.041$, $K = 8\times10^{-4}$, $D_{w} = 15.0$, $\epsilon=0.2$, and $\rho_{veR} = 1.0$ (`base' parameters).}
\label{fig:two_drop_halo}
\end{figure}

In the initial stages, when evaporation is considered in the model, vapour generated by both drops due to the presence of unsaturated ambient conditions begins to condense on the substrate near their respective contact lines. The condensation in the vicinity of the drops can be identified by the negative values of evaporation ($J_v$), as depicted in figure \ref{fig:evap_prof}(a). Inspection of this figure reveals that the condensate is initially deposited closer to the droplet contact lines. As time progresses, the condensate accumulated closer to the contact lines of the drops re-evaporates, and the region where the net condensate is present moves away from the contact lines, as also reported by \cite{kavuri2023freezing} in the case of freezing of a single drop on a cold substrate. To understand the interactions between droplets, we examine the region close to the substrate between them $(-0.25 \le x \le 0.25)$. From figure \ref{fig:two_drop_halo}(a), it is clear that in this region, condensation occurs closer to the right contact line ($x = -0.25$) of the left drop and the left contact line ($x = 0.25$) of the right drop, which can be identified by the negative value of the evaporation flux ($J_v$). As time progresses, as shown in figures \ref{fig:two_drop_halo}(a) and \ref{fig:two_drop_halo}(b), the region where the evaporation flux is negative moves away from the respective contact lines of the droplets, and for $t>2$, condensation is no longer observed in the region between the two drops.

\begin{figure}
\centering
\hspace{0.50cm} {\large (a)}\\
\includegraphics[width=0.65\textwidth]{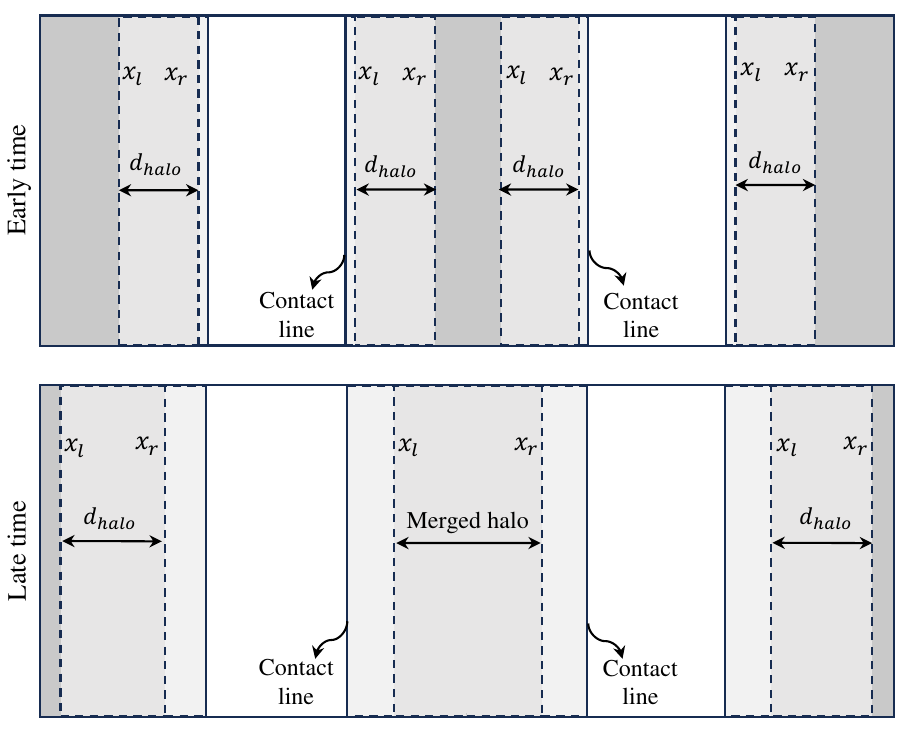}\\ 
\hspace{0.75cm}{\large (b)}   \hspace{5.5cm}  {\large (c)} \\
\includegraphics[width=0.45\textwidth]{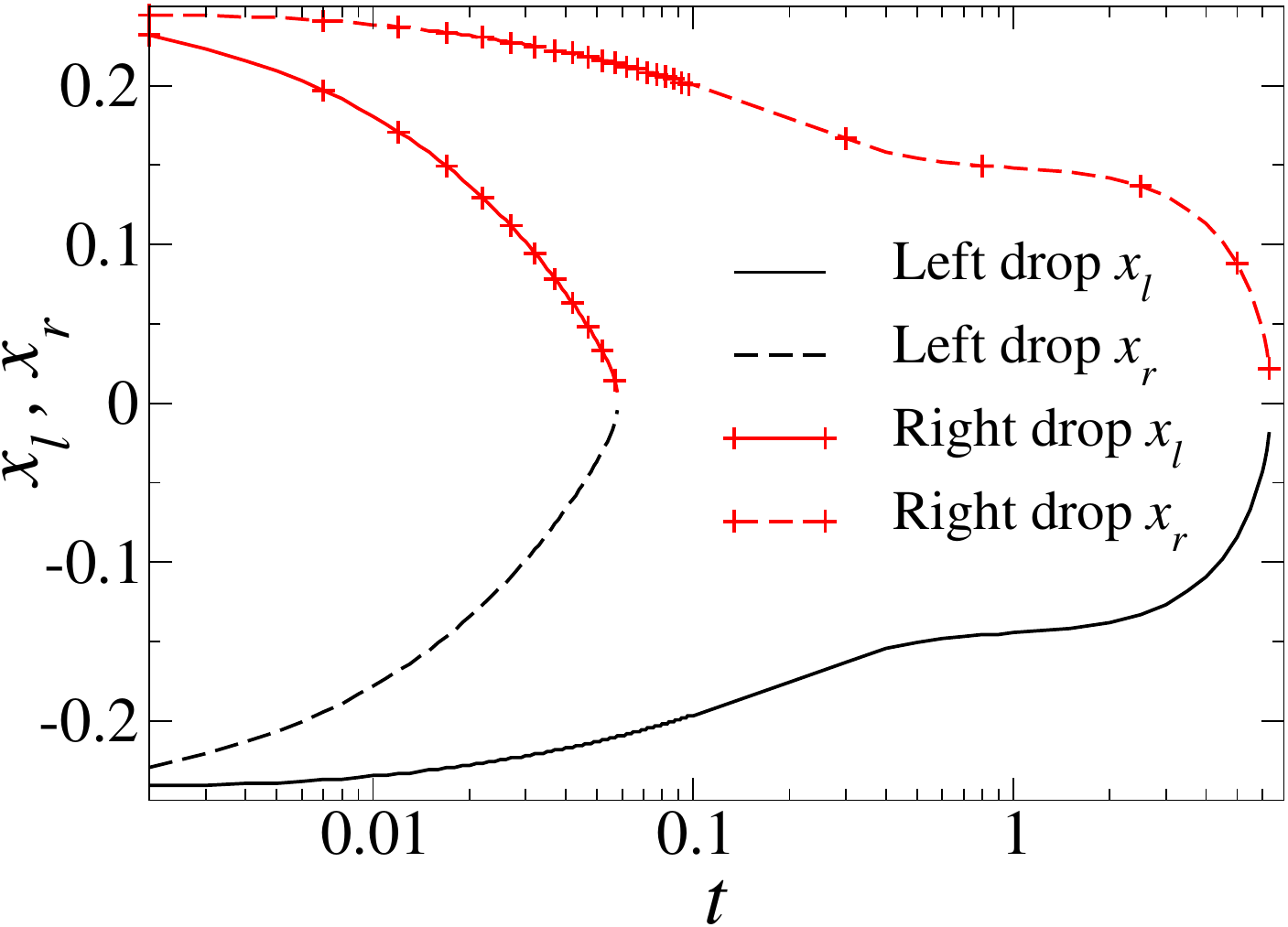}
\includegraphics[width=0.45\textwidth]{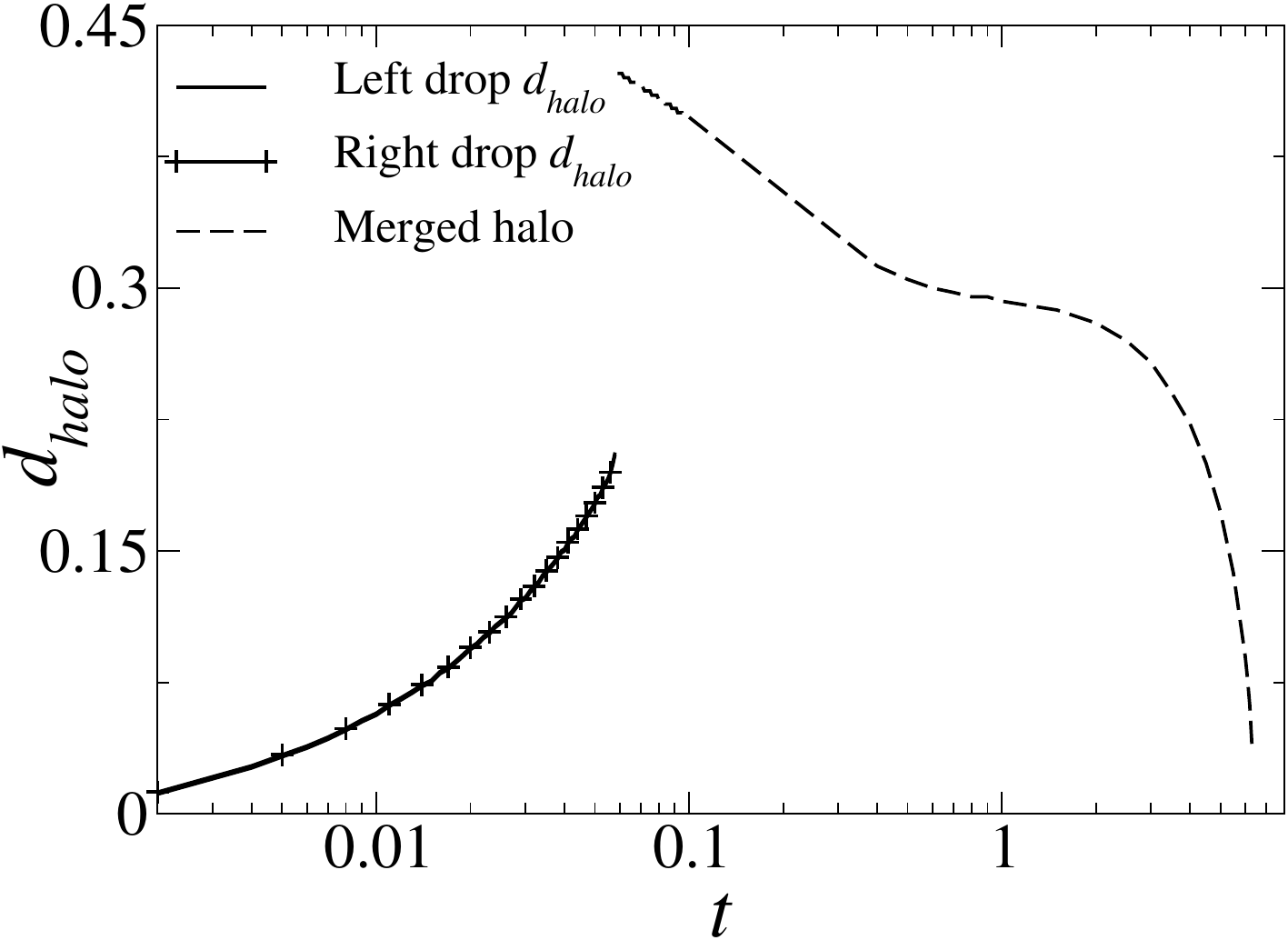}\\
\caption{Two dimensional schematic representation of the top view of the left ($x_{l}$) and right ($x_{r}$) edges of the condensate mass and the thickness of the halo ($d_{halo}=x_{r}-x_{l}$) in the vicinity of the drops. (b) The temporal variation of the left ($x_{l}$) and right ($x_{r}$) ends of the condensation halo in between two drops. (c) The temporal variation of the width of the condensation halo ($d_{halo}$). The remaining dimensionless parameters are $d_0 = 0.5$, $RH = 0.90$, $\chi = 1.6$, $\Delta = 10^{-4}$, $\Psi = 0.02$, $Pe_{v} = 1$, $Ste = 2.53\times10^{-5}$, $T_{v} = 1.0$, $A_{n} = 17.0$, $D_v = 10^{-3}$, $D_{g} = 2.0$, $D_{s} = 0.9$, $\Lambda_{S} = 3.89$, $\Lambda_{W} = 0.33$, $\Lambda_{g} = 0.041$, $K = 8\times10^{-4}$, $D_{w} = 15.0$, $\epsilon=0.2$, and $\rho_{veR} = 1.0$ (`base' parameters).}
\label{fig:xcl_xcr_plot}
\end{figure}

Next, we examine the interaction between the halos formed by the drops within the region separating them. We determine the net condensate mass ($m_c$) by numerically integrating the evaporation flux between the two drops. Initially, as depicted in figure \ref{fig:two_drop_halo}(c), condensate predominantly accumulates at the contact lines (around $x = -0.25$ and $x = 0.25$) of the respective droplets, approaching zero near the centre ($x = 0$) between the droplets until $t < 0.1$. To comprehend these results further, figure \ref{fig:xcl_xcr_plot}(a) shows a top-view schematic illustrating the interaction between the drops through their respective halos. In this figure, solid black lines denote the contact lines of the drops, while the dashed region represents the substrate area occupied by the halos. The boundaries of this halo region, denoted by $x_l$ and $x_r$, encompass the width of the halo, $d_{halo}$, defined as the separation between its left and right edges ($d_{halo} = x_r - x_l$). Shifting our attention to the region between the two drops, at early times in figure \ref{fig:xcl_xcr_plot}(a), we observe that the halos of both drops remain distinct and close to their respective contact lines. Over time, the accumulated condensate re-evaporates, causing the area containing net condensate to move away from the contact lines. This phenomenon results in the formation of a merged halo in the central region between the drops at later times, as depicted in figure \ref{fig:xcl_xcr_plot}(a). The interaction of vapour between the two drops causes the merging of the right edge ($x_r$) of the condensate originating from the left droplet with the left edge ($x_l$) of the condensate originating from the right droplet, as illustrated in figure \ref{fig:xcl_xcr_plot}(b) at $t = 0.06$. The extent of the condensate mass, indicated by negative $m_c$ values, defines the thickness of the halo ($d_{halo}$). Before $t \approx 0.06$, both drops exhibit halos of equal thickness, as depicted in igure \ref{fig:xcl_xcr_plot}(c). However, after $t \approx 0.06$, the merging of the halos occurs, resulting in a combined halo with nearly double the maximum thickness of the individual halos, as the right edge ($x_r$) of the halo of the left droplet combines with the left edge ($x_l$) of the halo of the right droplet. As condensate accumulates, re-evaporation begins, leading to a shift in the region of condensate presence. This is also evident in the shrinking of the left droplet's halo left edge (Left drop $x_l$) and the right droplet's halo right edge (Right drop $x_r$) of the combined halo shown in figure \ref{fig:xcl_xcr_plot}(b). Additionally, the width of the merged halo decreases due to re-evaporation, as shown in figure \ref{fig:xcl_xcr_plot}(c). Ultimately, for $t > 6$, the condensate between the droplets evaporates in the central region.

In the following section, we discuss the coalescence of the drops occurring at a later stage and examine the influence of the various parameters, such as relative humidity ($RH$), initial separation distance ($d_0$), and the temperature difference ($\Delta T$) between the melting temperature ($T_{m} = 0^{\circ}$C) and the temperature at the bottom of the substrate ($T_{c}$) on the coalescence behaviour.

\subsection{Late time dynamics - Coalescence of drops} \label{sec:late}

As discussed in \S \ref{sec:mechanism}, the asymmetry in evaporation flux drives capillary flow, causing the two drops to migrate toward each other, and they coalesce rapidly due to their inertia. The coalescence between the drops occurs when they make contact before the evolution of the freezing front, as the capillary velocity of the drops diminishes with the advancement of the freezing front. 

\begin{figure}
\centering
\hspace{0.75cm}{\large (a)}   \hspace{5.5cm}  {\large (b)} \\
\includegraphics[width=0.45\textwidth]{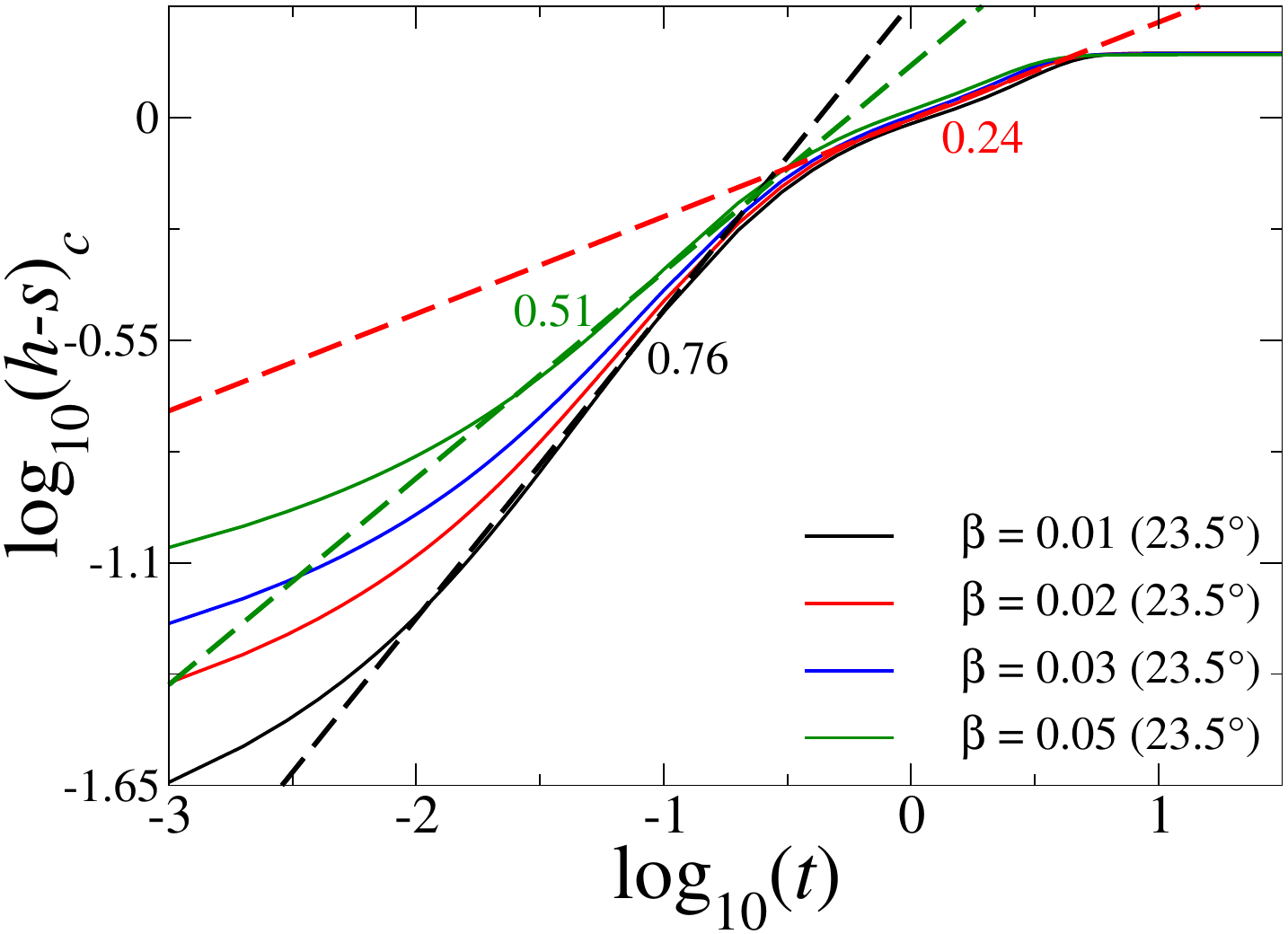}
\includegraphics[width=0.45\textwidth]{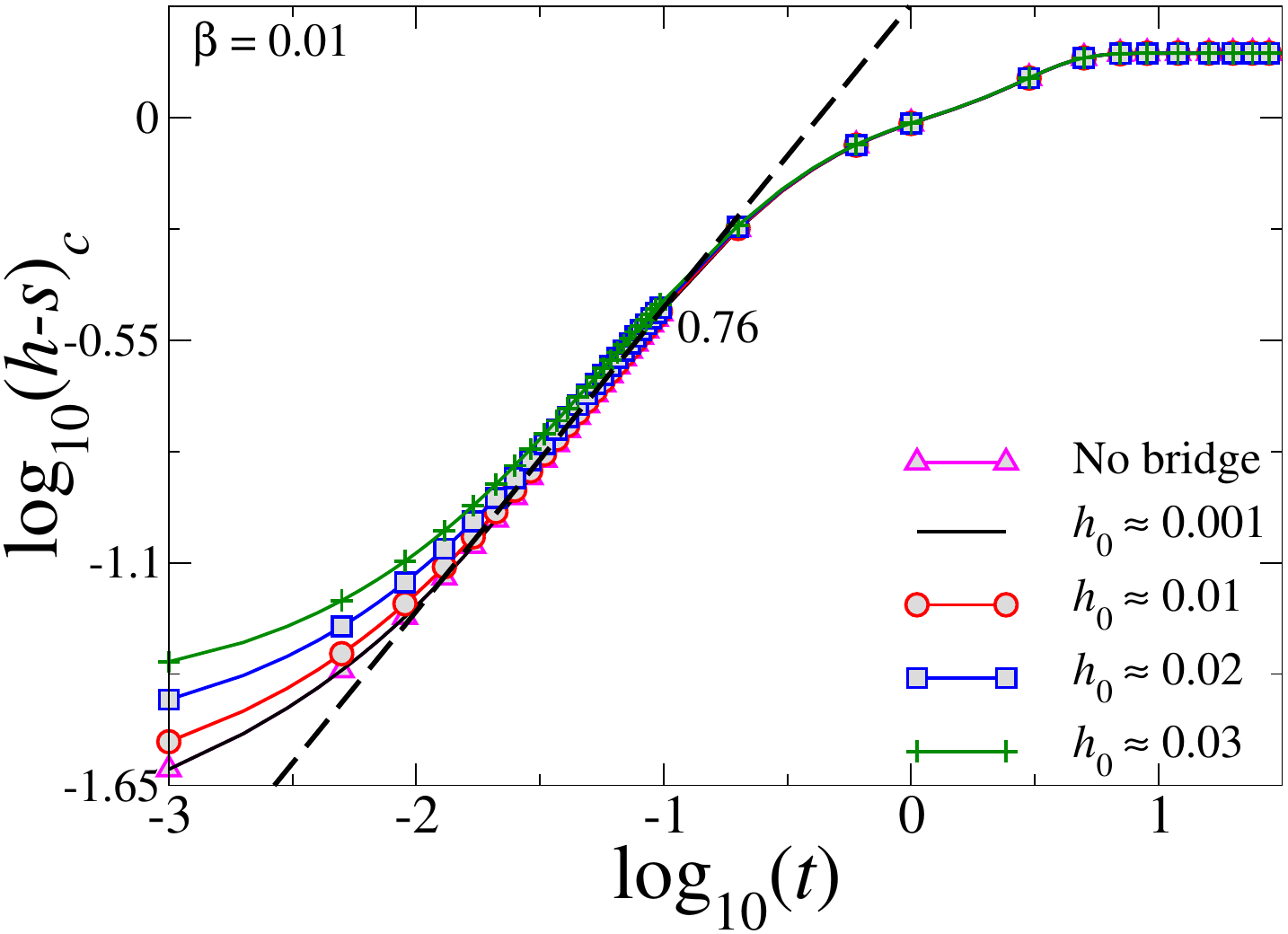}
\caption{Temporal variation of the height of the precursor layer at $x=0$ (denoted by $(h-s)_c$) for a system of two partially wetting drops without evaporation and freezing. (a) Effect of $\beta$ for $h_0=0$. (b) Effect of the initial bridge height ($h_0$) for $\beta=0.01$. The rest of the dimensionless parameters are $Ste = 0$, $d_0 = 0$, $T_{v} = 1.0$, $A_{n} = 17.0$, $D_v = 10^{-3}$, $D_{g} = 2.0$, $D_{s} = 0.9$, $\Lambda_{S} = 3.89$, $\Lambda_{W} = 0.33$, $\Lambda_{g} = 0.041$, $\chi = 0$, $RH = 1.0$, $K = 8\times10^{-4}$, $D_{w} = 15.0$, $\epsilon=0.2$, $\Delta = 0$, $\Psi = 0$, $\rho_{veR} = 1.0$ and $Pe_{v} = 0$.}
\label{noevap_pw}
\end{figure}

In the absence of evaporation and freezing, when two partially wetting drops are placed side by side with their contact lines touching $(d_0=0)$, the liquid bridge height increases with an exponent of $2/3$ over time \citep{sui2013inertial,eddi2013influence,varma2021coalescence}. Utilizing our current model for a two-dimensional (2D) system of two partially wetting drops with a precursor layer thickness ($\beta = 0.01$) and neglecting evaporation and freezing, we observe that $(h-s)_c$ at $x=0$ increases with time with an exponent of $0.76$. This growth rate decreases as the precursor layer thickness increases, as shown in figure \ref{noevap_pw}(a). Additionally, in figure \ref{noevap_pw}(b), we examine the influence of the initial bridge height on its growth rate while keeping the precursor layer thickness constant, revealing that the initial bridge height has a negligible effect on the coalescence process. We now shift our attention to the coalescence of two partially wetting sessile drops, taking into account both evaporation and freezing effects.

\begin{figure}
\centering
\hspace{0.75cm}{\large (a)}   \hspace{5.5cm}  {\large (b)} \\
\includegraphics[width=0.45\textwidth]{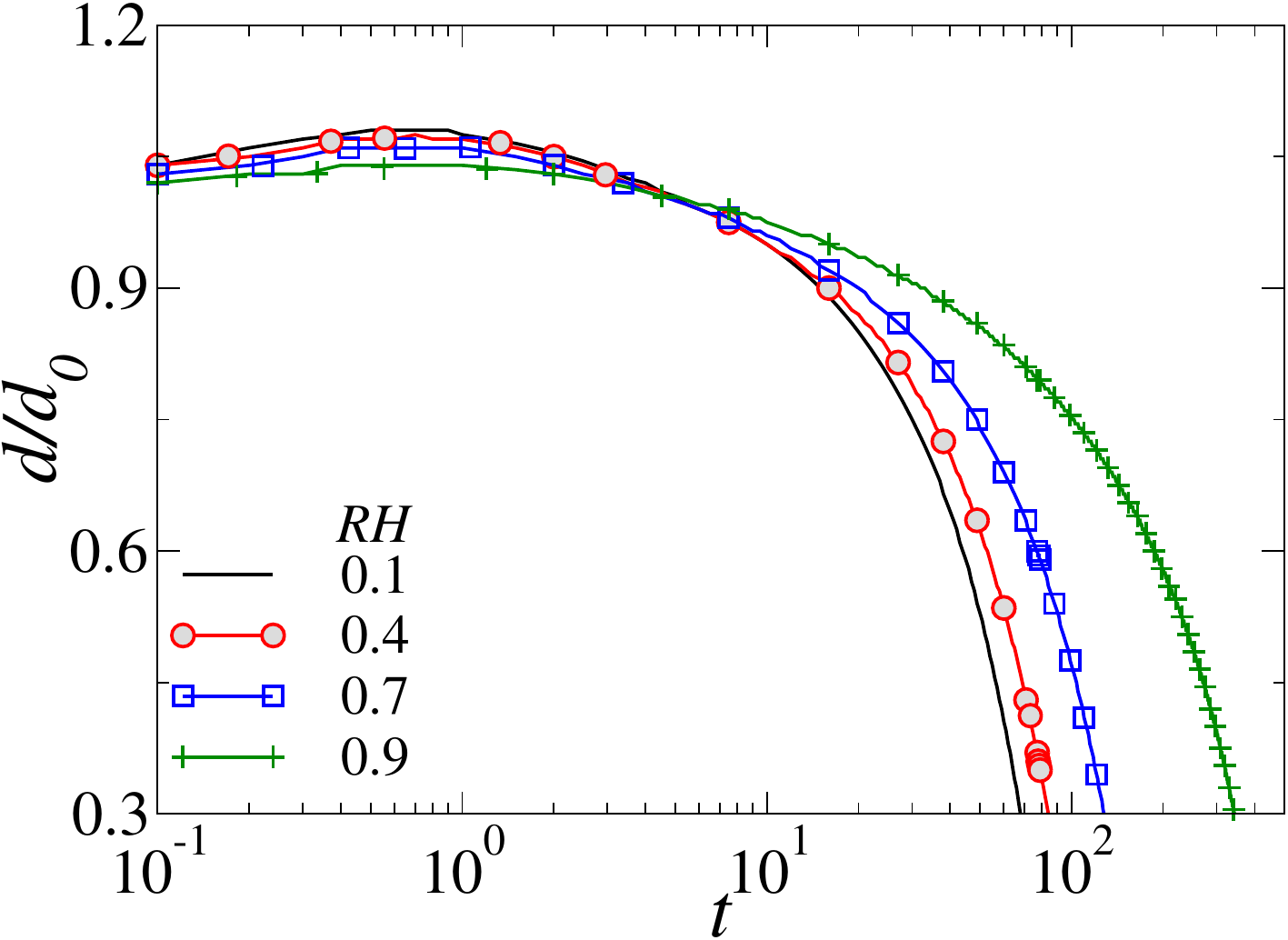}
\includegraphics[width=0.46\textwidth]{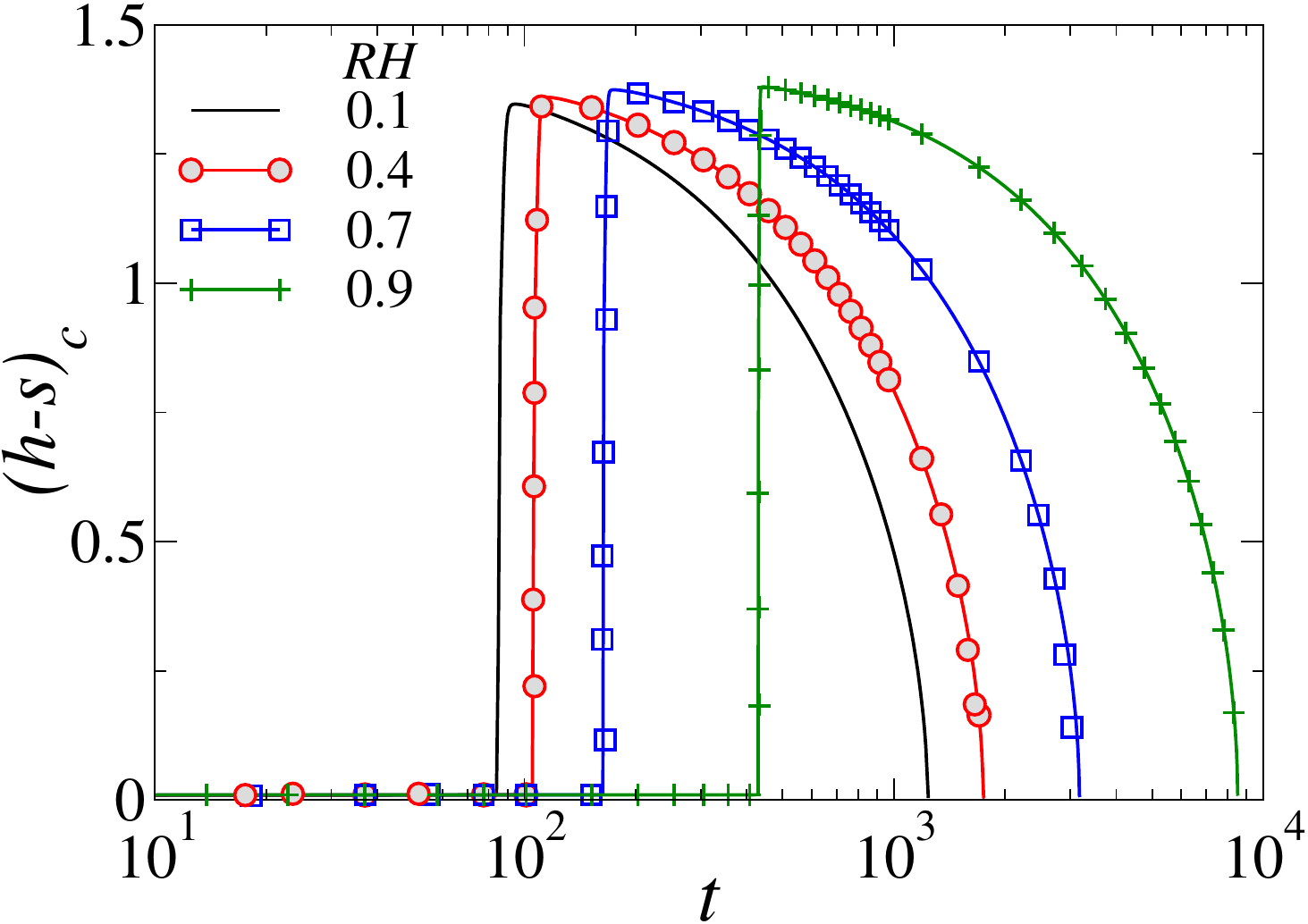}
\caption{Temporal evolution of (a) the normalised separation distance $(d/d_0)$ and (b) the liquid layer height at $x=0$ (denoted by $(h-s)_c$) for different values of $RH$. The remaining dimensionless parameters are $d_0 = 0.5$, $\chi = 1.6$, $\Delta = 10^{-4}$, $\Psi = 0.02$, $Pe_{v} = 1$, $Ste = 2.53\times10^{-5}$, $T_{v} = 1.0$, $A_{n} = 17.0$, $D_v = 10^{-3}$, $D_{g} = 2.0$, $D_{s} = 0.9$, $\Lambda_{S} = 3.89$, $\Lambda_{W} = 0.33$, $\Lambda_{g} = 0.041$, $K = 8\times10^{-4}$, $D_{w} = 15.0$, $\epsilon=0.2$, and $\rho_{veR} = 1.0$.}
\label{d_by_d0}
\end{figure}

Figure \ref{d_by_d0}(a) and \ref{d_by_d0}(b) depict the temporal evolution of the normalized separation distance $(d/d_0)$ and height of the liquid layer ($(h-s)_c$ at $x=0$) in the presence of evaporation and freezing for different values of $RH$. The remaining parameters are the same as the `base' parameters. Figure \ref{d_by_d0}(a) illustrates that an increase in relative humidity results in decreased evaporation, thereby reducing the capillary flow driven by the asymmetry in evaporation, which facilitates coalescence. Thus, it can be seen that higher relative humidity ($RH$) substantially extends the time for the droplets to approach each other. Conversely, the droplets migrate more rapidly at lower relative humidity, as observed in figure \ref{d_by_d0}(a). Consequently, the bridge evolves more rapidly than the coalescence of two partially wetting drops when evaporation and freezing are neglected in the two-dimensional (2D) model, as shown in figure \ref{d_by_d0}b. 

\begin{figure}
\centering
\hspace{0.75cm}{\large (a)} \hspace{5.5cm}  {\large (b)} \\
\includegraphics[width=0.45\textwidth]{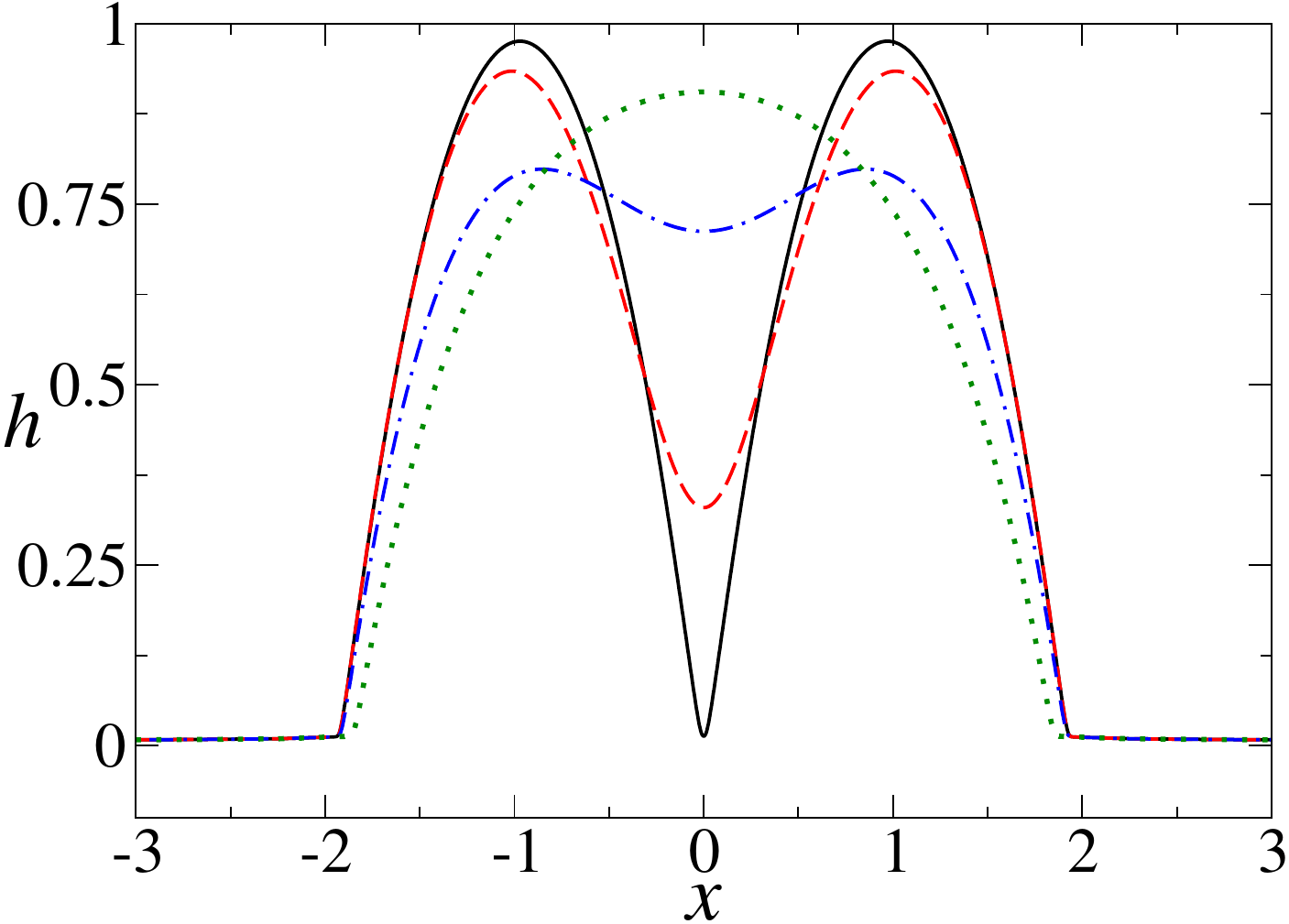} 
\hspace{0mm}
\includegraphics[width=0.45\textwidth]{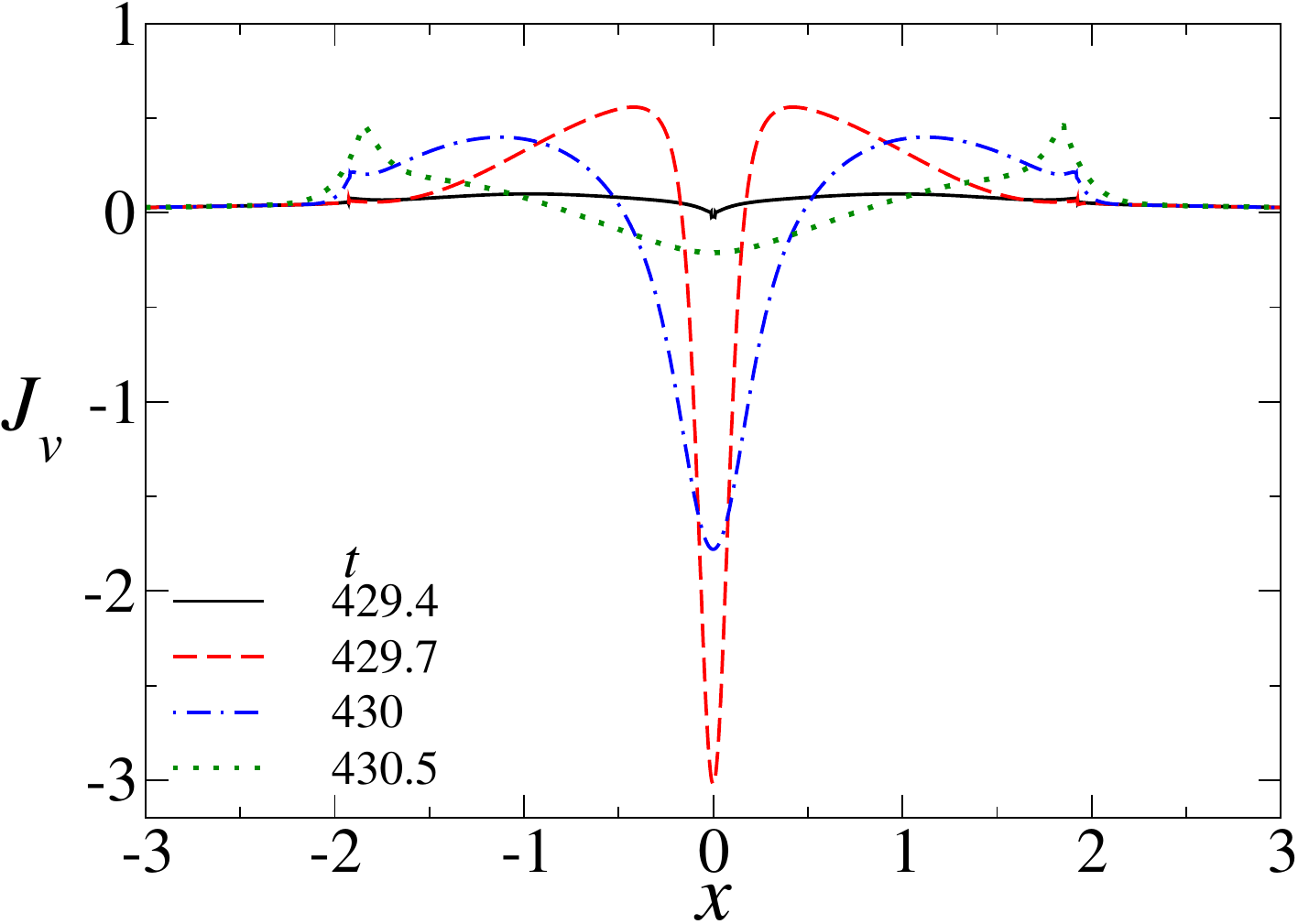}\\
\hspace{0.75cm}{\large (c)}   \hspace{5.5cm}  {\large (d)} \\
\includegraphics[width=0.45\textwidth]{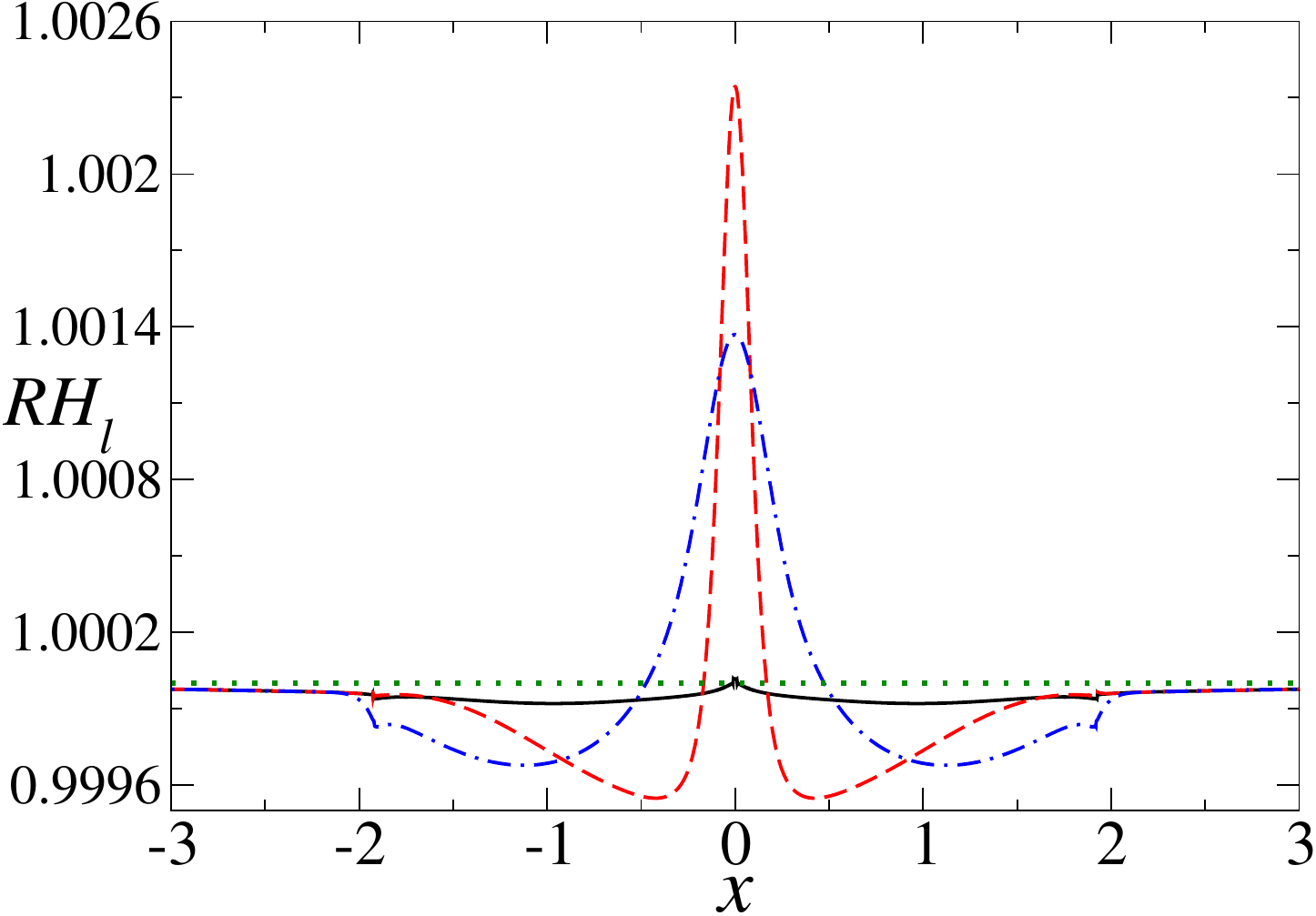}
\hspace{0mm}
\includegraphics[width=0.45\textwidth]{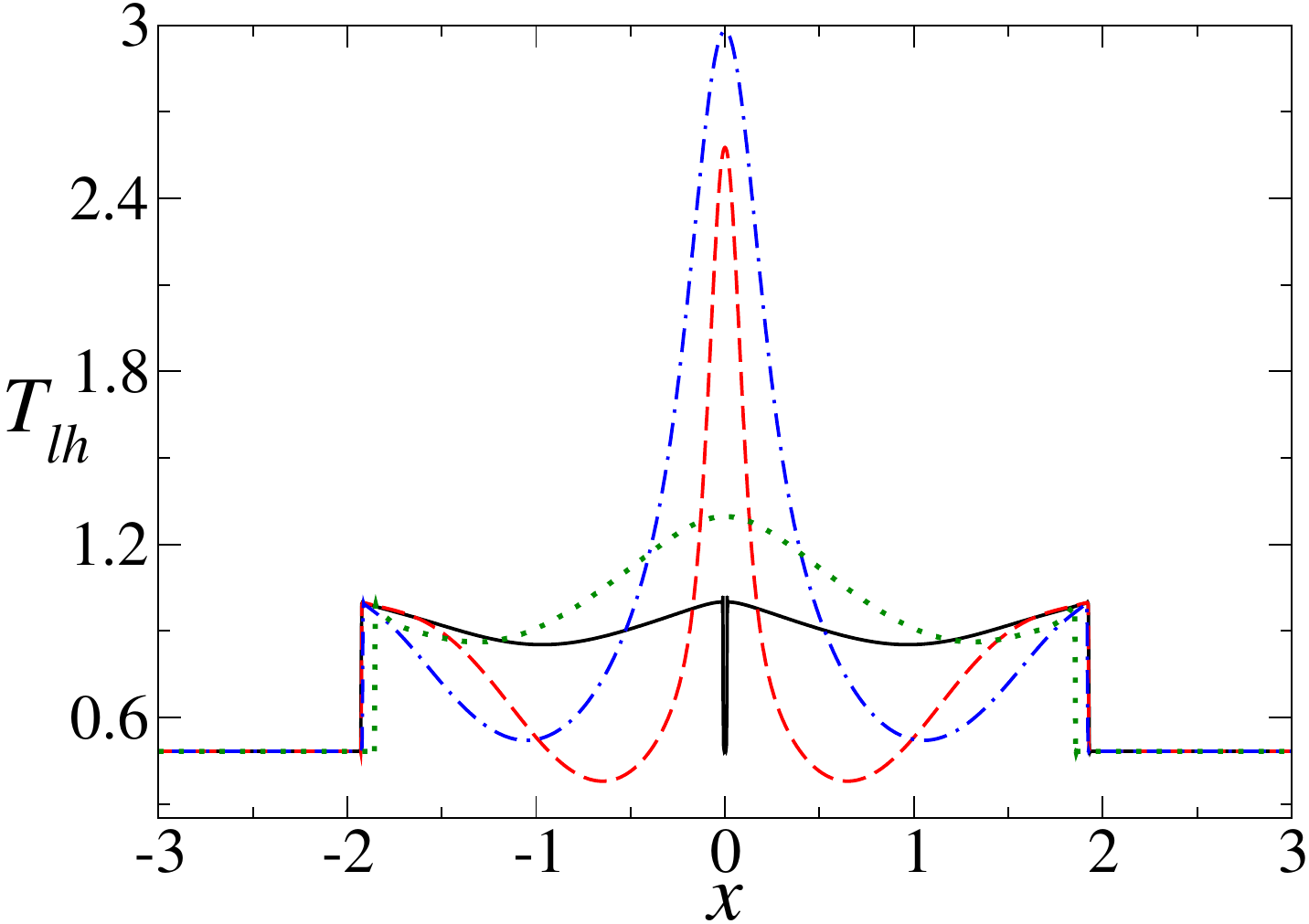}\\
\caption{Temporal evolution of (a) the liquid-gas interface ($h$), (b) the evaporation flux ($J_v$), (c) the local relative humidity ($RH_{l}$) and (d) temperature at the liquid-gas interface ($T_{lh}$) during coalescence for $RH = 0.90$. The remaining dimensionless parameters are $d_0 = 0.5$, $\chi = 1.6$, $\Delta = 10^{-4}$, $\Psi = 0.02$, $Pe_{v} = 1$, $Ste = 2.53\times10^{-5}$, $T_{v} = 1.0$, $A_{n} = 17.0$, $D_v = 10^{-3}$, $D_{g} = 2.0$, $D_{s} = 0.9$, $\Lambda_{S} = 3.89$, $\Lambda_{W} = 0.33$, $\Lambda_{g} = 0.041$, $K = 8\times10^{-4}$, $D_{w} = 15.0$, $\epsilon=0.2$, and $\rho_{veR} = 1.0$.}
\label{fig:zoomed_coal}
\end{figure}

Figure \ref{fig:zoomed_coal}(a) presents the shape of the drops ($h$) during the coalescence for relative humidity $RH = 0.9$, where the initial separation between the two drops is $d_0 = 0.5$, with all other parameters consistent with those considered in figure \ref{d_by_d0}. As the droplets approach each other, coalescence begins, and the local relative humidity in the coalescence region increases significantly, leading to substantial condensation, as demonstrated in figure \ref{fig:zoomed_coal}(b). The rise in local relative humidity ($RH_l$) during the coalescence process can be observed in figure \ref{fig:zoomed_coal}(c). Note that although the increase in local relative humidity may seem minor, given the substantially small value of the dimensionless number $K=8 \times 10^{-4}$ (eq. \ref{eq:nondimK}) associated with the equilibrium condition at the interface (equation \ref{bc:k}), even slight changes have a significant impact on the evaporation flux. As local relative humidity increases, vapour transitions from the gas phase to the liquid phase, releasing latent heat absorbed by the liquid-gas interface and the surroundings. This increase in temperature during coalescence due to condensation is depicted in figure \ref{fig:zoomed_coal}(d). In figure \ref{fig:zoomed_coal}(d), an increase in temperature at $x = 0$ is observed due to condensation, followed by a decrease when condensation significantly diminishes. The rise in local relative humidity and subsequent condensation result in a very rapid coalescence of the drops.

\begin{figure}
\centering
\includegraphics[width=0.95\textwidth]{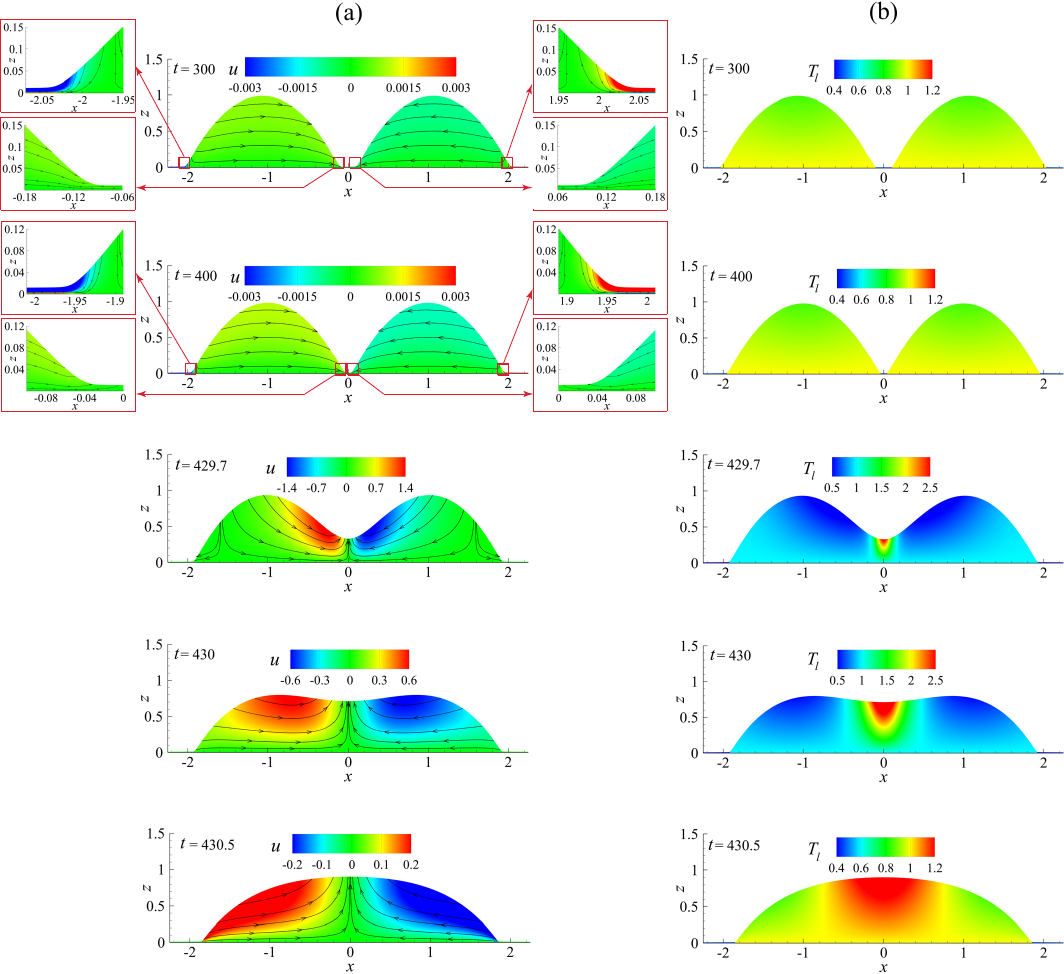}
     \caption{(a) The horizontal velocity field ($u$) along with streamlines and (b) the temperature distribution ($T_l$) at $t = 300$, $400$, $429.7$, $430$ and $430.5$ for $RH = 0.90$. The enlarged views of the velocity and streamline contours are presented for $t = 300$ and $t = 400$ to highlight the asymmetrical patterns near the inner and outer edges of the droplets. The remaining dimensionless parameters are $d_0 = 0.5$, $\chi = 1.6$, $\Delta = 10^{-4}$, $\Psi = 0.02$, $Pe_{v} = 1$, $Ste = 2.53\times10^{-5}$, $T_{v} = 1.0$, $A_{n} = 17.0$, $D_v = 10^{-3}$, $D_{g} = 2.0$, $D_{s} = 0.9$, $\Lambda_{S} = 3.89$, $\Lambda_{W} = 0.33$, $\Lambda_{g} = 0.041$, $K = 8\times10^{-4}$, $D_{w} = 15.0$, $\epsilon=0.2$, and $\rho_{veR} = 1.0$.}
\label{fig:u_contours}
\end{figure}

\begin{figure}
\centering
\hspace{0.75cm}{\large (a)}   \hspace{7.5cm}  {\large (b)} \\
\includegraphics[width=0.45\textwidth]{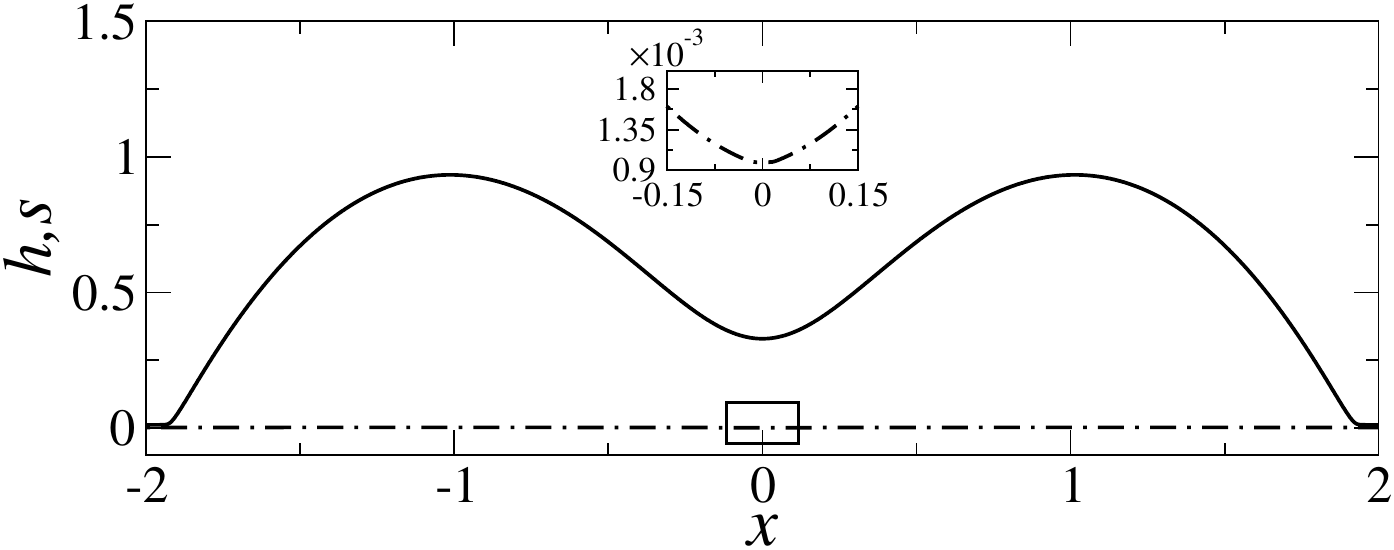}
\includegraphics[width=0.45\textwidth]{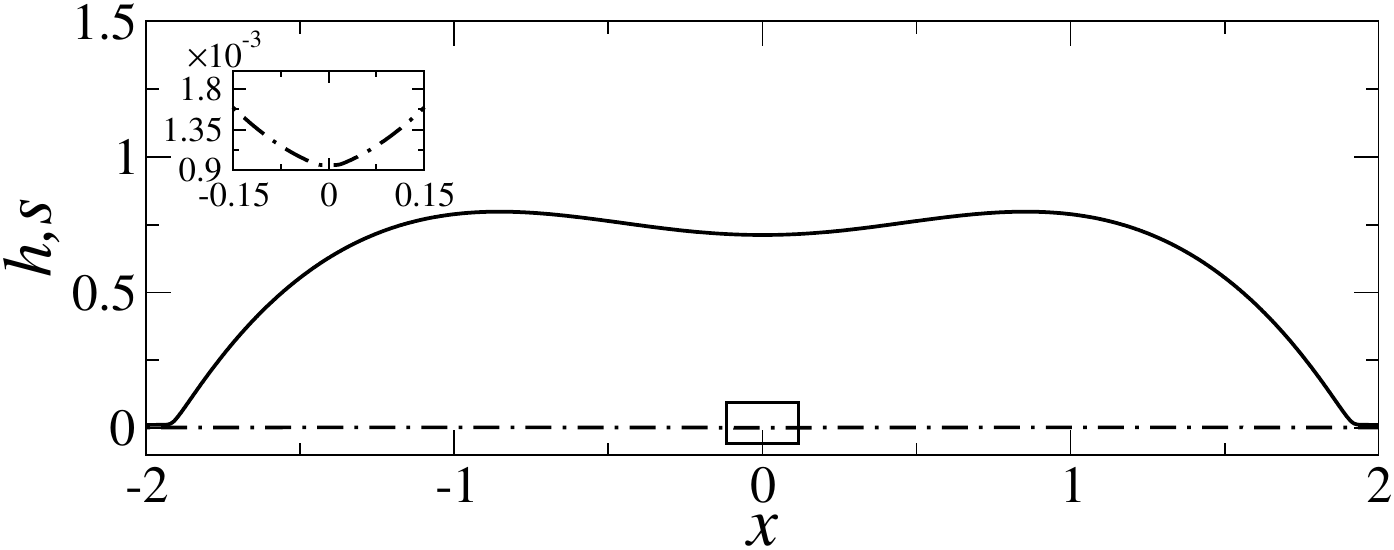}\\
\hspace{0.8cm} {\large (c)}\\
\includegraphics[width=0.45\textwidth]{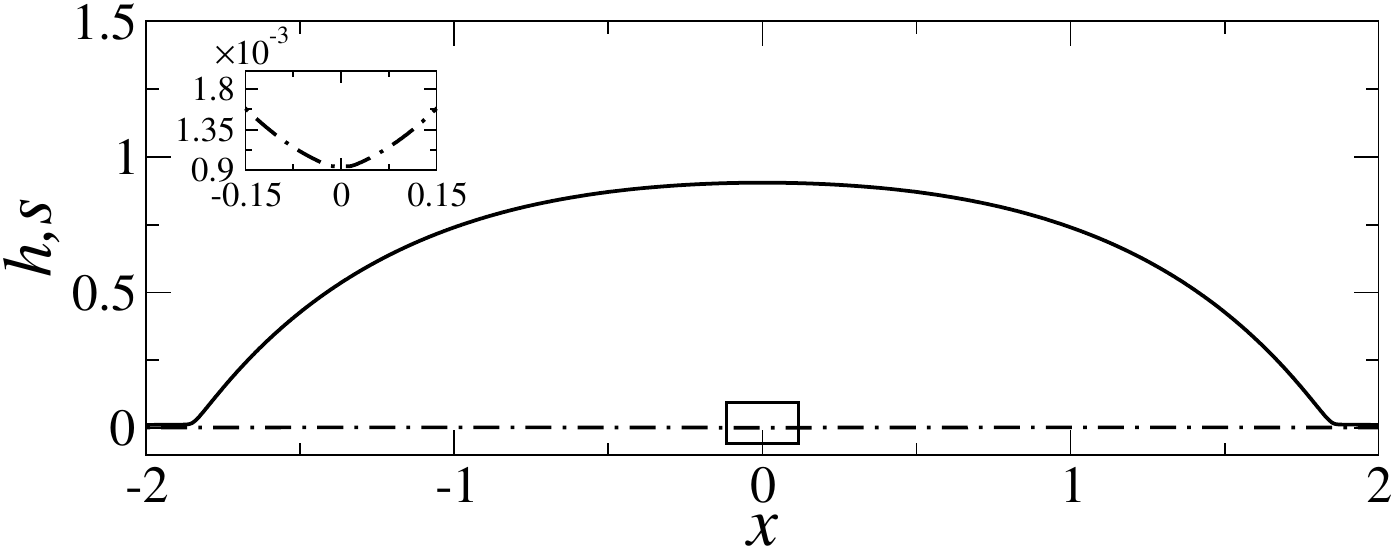}
     \caption{Temporal evolution of the liquid-gas interface (solid line) and freezing front (dot-dash line) at (a) $t = 429.7$, (b) $430$ and (c) $430.5$ during coalescence for $RH = 0.90$. The remaining dimensionless parameters are $d_0 = 0.5$, $\chi = 1.6$, $\Delta = 10^{-4}$, $\Psi = 0.02$, $Pe_{v} = 1$, $Ste = 2.53\times10^{-5}$, $T_{v} = 1.0$, $A_{n} = 17.0$, $D_v = 10^{-3}$, $D_{g} = 2.0$, $D_{s} = 0.9$, $\Lambda_{S} = 3.89$, $\Lambda_{W} = 0.33$, $\Lambda_{g} = 0.041$, $K = 8\times10^{-4}$, $D_{w} = 15.0$, $\epsilon=0.2$, and $\rho_{veR} = 1.0$.}
\label{fig:h_s_plot}
\end{figure}

The velocity and temperature fields before and during coalescence for the parameters considered in figure \ref{fig:zoomed_coal} are analyzed in figure \ref{fig:u_contours}(a) and \ref{fig:u_contours}(b), respectively. It depicts the flow dynamics at four stages of coalescence: before coalescence ($t = 300, 400$), early coalescence ($t = 429.7$), intermediate stage ($t = 430$), and final stage of coalescence ($t = 430.5$). In figure \ref{fig:u_contours}(a), it can be observed that before coalescence at $t = 300$ and $t = 400$, there is more evaporation occurring at the outer edges of the droplets than at the inner edges, as highlighted in the enlarged panels in figure \ref{fig:u_contours}(a). This creates a replenishing flow toward the outer edge, while the overall capillary flow remains directed toward the inner edges, driving the drops closer together. Additionally, the temperature contours at $t = 300$ and $t = 400$ shown in figure \ref{fig:u_contours}(b) indicate that before the coalescence of the two drops, their temperatures are close to their melting point. At the early stage of coalescence ($t = 429.7$), the flow is directed towards the coalescence bridge formed between the drops, leading to its rapid growth. The approach velocity of the drops is highest near this central region of the coalescence bridge. By the intermediate stage ($t = 430$), the bridge height increases considerably, and the flow distribution becomes more spread near the center, rather than being concentrated directly at the center as initially observed. As the bridge continues to grow, the droplet minimizes its surface energy by reducing its surface area, resulting in a higher $u$ velocity near the contact line. As droplets merge, the accumulating vapor near the coalescence region increases the local relative humidity, leading to condensation and a rise in temperature at the coalescence region, as shown in figure \ref{fig:u_contours}(b). This increase in local relative humidity is also evident in figure \ref{fig:zoomed_coal}(c). This rise in local relative humidity ($RH_l$) causes the vapor to condense from the gas phase to the liquid phase, releasing latent heat that is absorbed by the liquid-gas interface and the surrounding environment. This increase in temperature due to condensation is evident at the bridge between the droplets, as shown in figure \ref{fig:u_contours}(b). At $t = 429.7$, the temperature rise is primarily observed at the bridge due to the relatively low ambient and substrate temperatures. As coalescence progresses ($t = 430.5$), the temperature increases slowly throughout the drop, with the maximum temperature occurring at the center. The thermal conductivity of the substrate is similar to PMMA, and both the ambient and substrate temperatures are close to the freezing point of water, leading to a slower propagation of the freezing front compared to the evaporation rate. As detailed in Section 3.3, a significant evolution of the freezing front can restrict the movement of the contact line, hindering coalescence. This is reflected in figure \ref{fig:h_s_plot}, where the dot-dashed line representing the freezing front did not advance significantly, allowing the coalescence of the drops. The temperature increase at the bridge during coalescence due to condensation resulted in a slightly less advanced freezing front at the center.

\begin{figure}
\centering
\hspace{0.75cm}{\large (a)}   \hspace{5.5cm}  {\large (b)} \\
\includegraphics[width=0.45\textwidth]{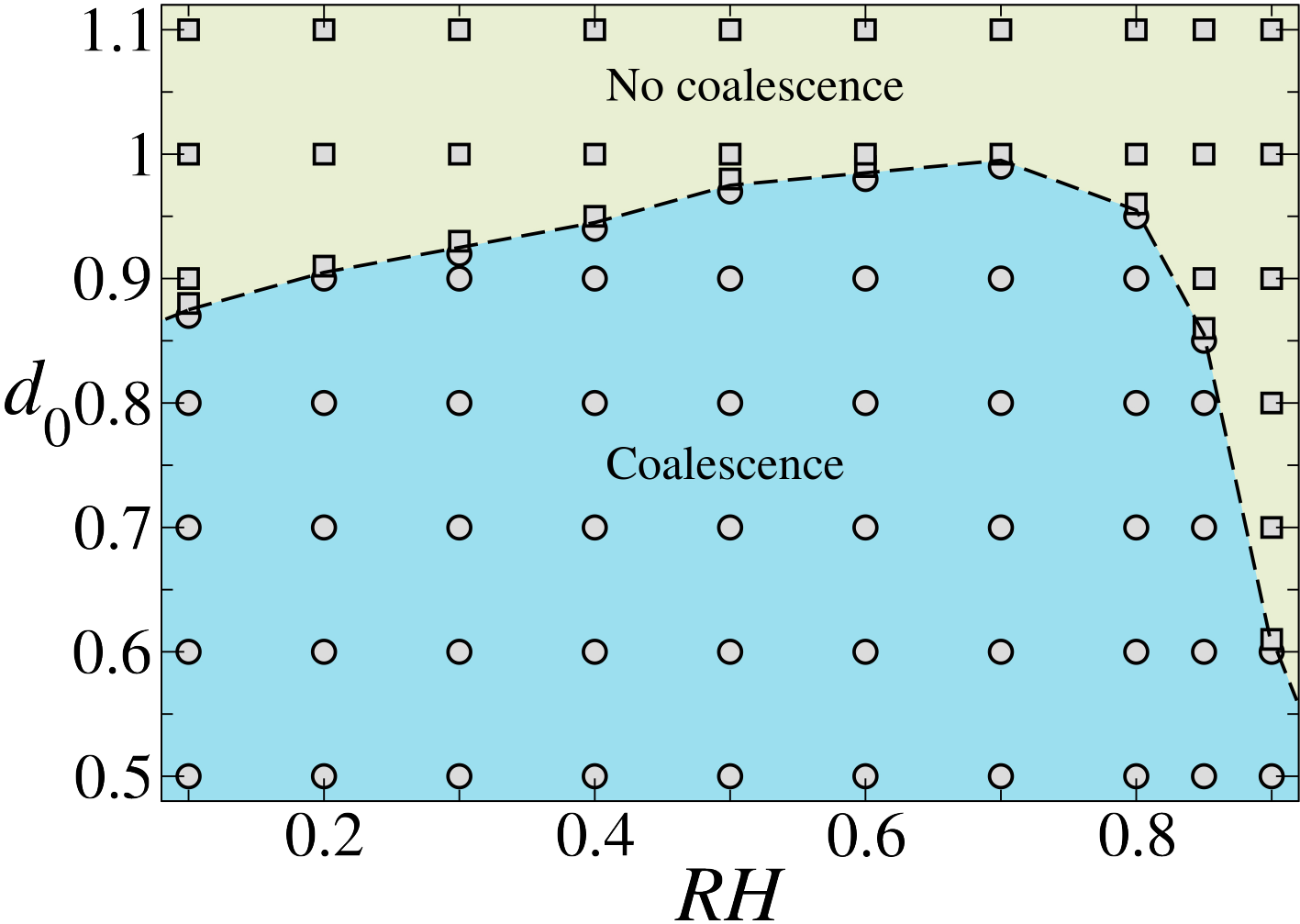}
\includegraphics[width=0.45\textwidth]{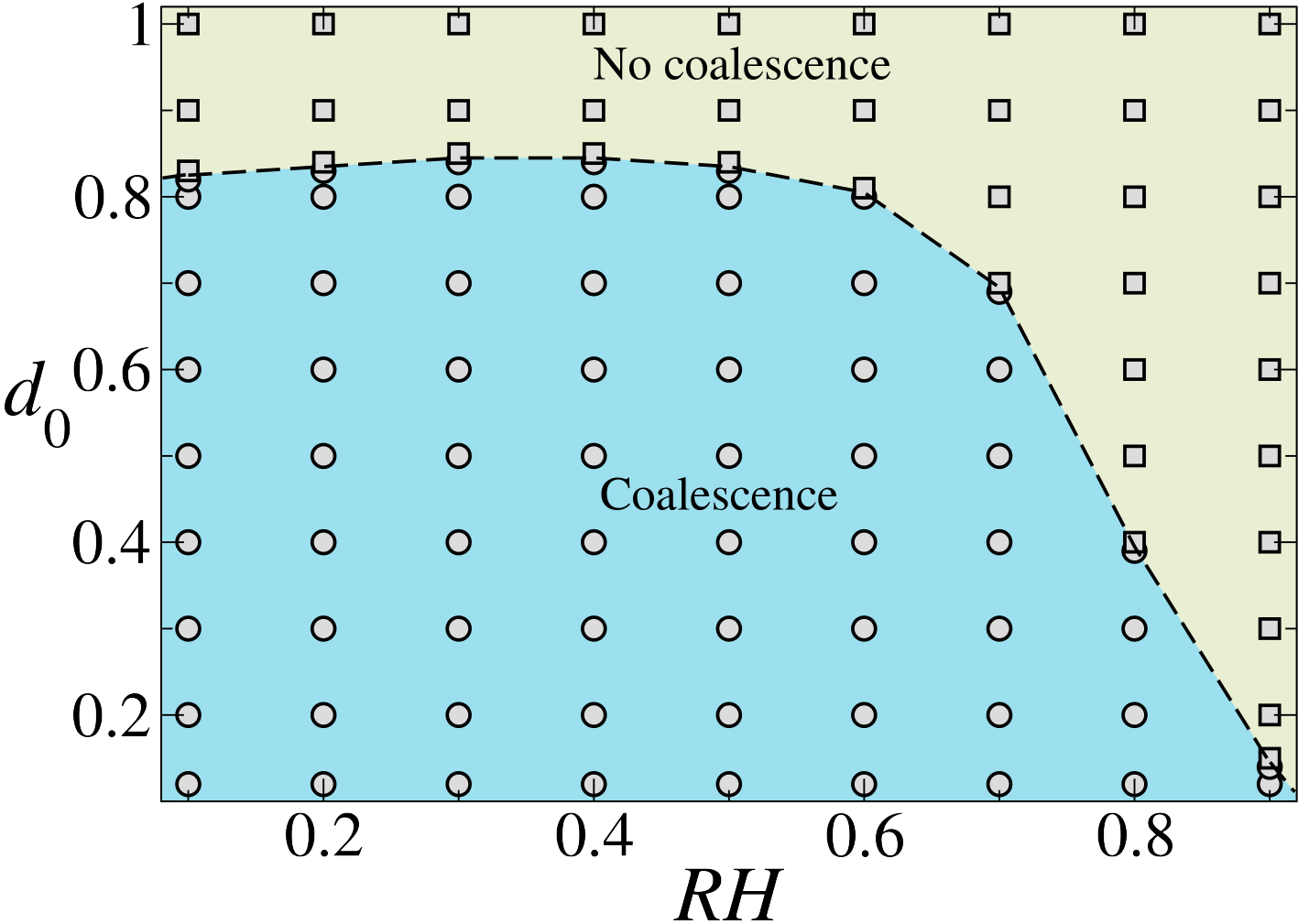}\\
\hspace{0.8cm} {\large (c)}\\
\includegraphics[width=0.45\textwidth]{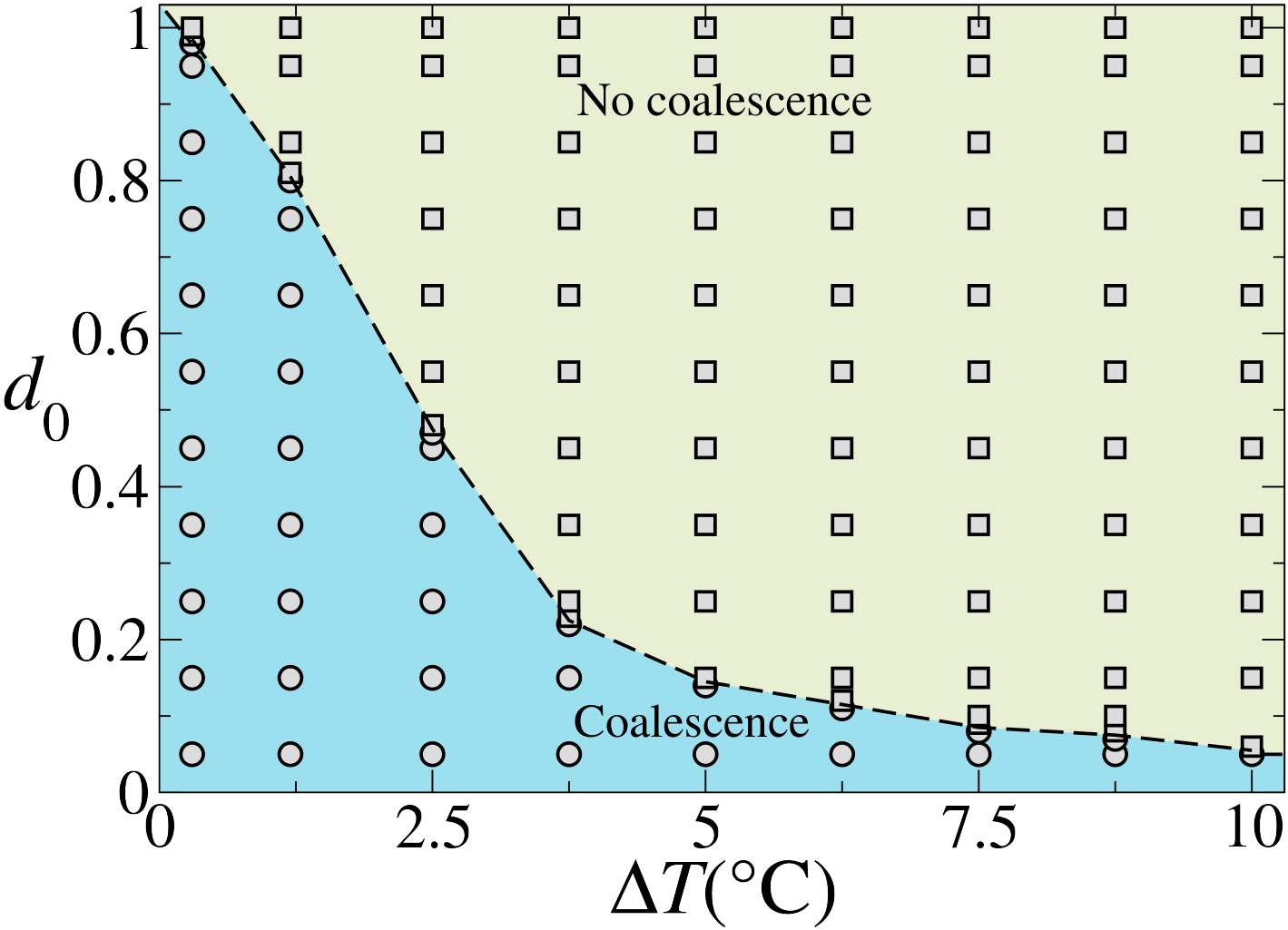}
\caption{Phase diagram demarcating the coalescence and no coalescence regimes in $d_0 - RH$ space for (a) $\Delta{T} = 0.3^{\circ}$C ($Ste = 2.53\times10^{-5}$, $\chi = 1.6$ and $\Psi = 0.02$) and (b) $\Delta{T} = 1.2^{\circ}$C ($Ste = 10^{-4}$, $\chi = 0.4$ and $\Psi = 0.085$). Panel (c) depicts the phase diagram in $d_0-\Delta{T}$ space for $RH = 0.6$. Here, $\Delta{T}$ denotes the temperature difference between the bottom of the substrate and the melting point temperature. The rest of the dimensionless parameters are the same as the `base' parameters.}
\label{RH_vs_d_phase}
\end{figure}

As discussed above, the coalescence of sessile drops on a cold substrate is influenced by several parameters. Thus, we finally examine the effect of the relative humidity ($RH$), initial separation between the two drops ($d_0$), and the temperature difference between the bottom of the substrate and the melting temperature ($\Delta T$) on the coalescence of the drops. We demarcate the regime of coalescence and non-coalescence by analyzing the combinations of these parameters in figure \ref{RH_vs_d_phase}(a-c).

In figure \ref{RH_vs_d_phase}(a) and \ref{RH_vs_d_phase}(b), we present regime maps in $d_0-RH$ space for $\Delta{T} = 0.3^{\circ}$C and $\Delta{T} = 1.2^{\circ}$C, respectively. In terms of dimensionless numbers, $\Delta{T} = 0.3^{\circ}$C and $\Delta{T} = 1.2^{\circ}$C correspond to ($Ste = 2.53\times10^{-5}$, $\chi = 1.6$, and $\Psi = 0.02$) and ($Ste = 10^{-4}$, $\chi = 0.4$, and $\Psi = 0.085$), respectively. The coalescence of two drops is influenced by the capillary flow within the drops and the height of the freezing front, as indicated by Eq. (\ref{ucaavg}). The drops are anticipated to migrate more rapidly at lower relative humidity ($RH$) due to increased evaporation rates. However, the evaporative cooling becomes more pronounced at lower relative humidity levels, leading to a competition between these effects. 
In figure \ref{RH_vs_d_phase}(a), when the relative humidity is $RH = 0.1$, it is evident that when the initial distance is $d_0 = 0.9$, coalescence does not occur. A substantial initial separation distance ($d_0$) can inhibit coalescence as the freezing front may progress significantly, halting the motion of the contact line before the drops can merge. At lower relative humidity levels, the time required for the droplets to approach each other increases with the initial separation distance $d_0$, and the freezing front undergoes considerable evolution due to evaporative cooling, impeding contact line motion and resulting in non-coalescence. Moreover, as relative humidity rises, evaporative cooling diminishes, but the time for droplets to converge increases, potentially providing ample opportunity for the freezing front to develop and impede contact line motion.

\begin{figure}
\centering
\hspace{0.75cm}{\large (a)}   \hspace{5.5cm}  {\large (b)} \\
\includegraphics[width=0.45\textwidth]{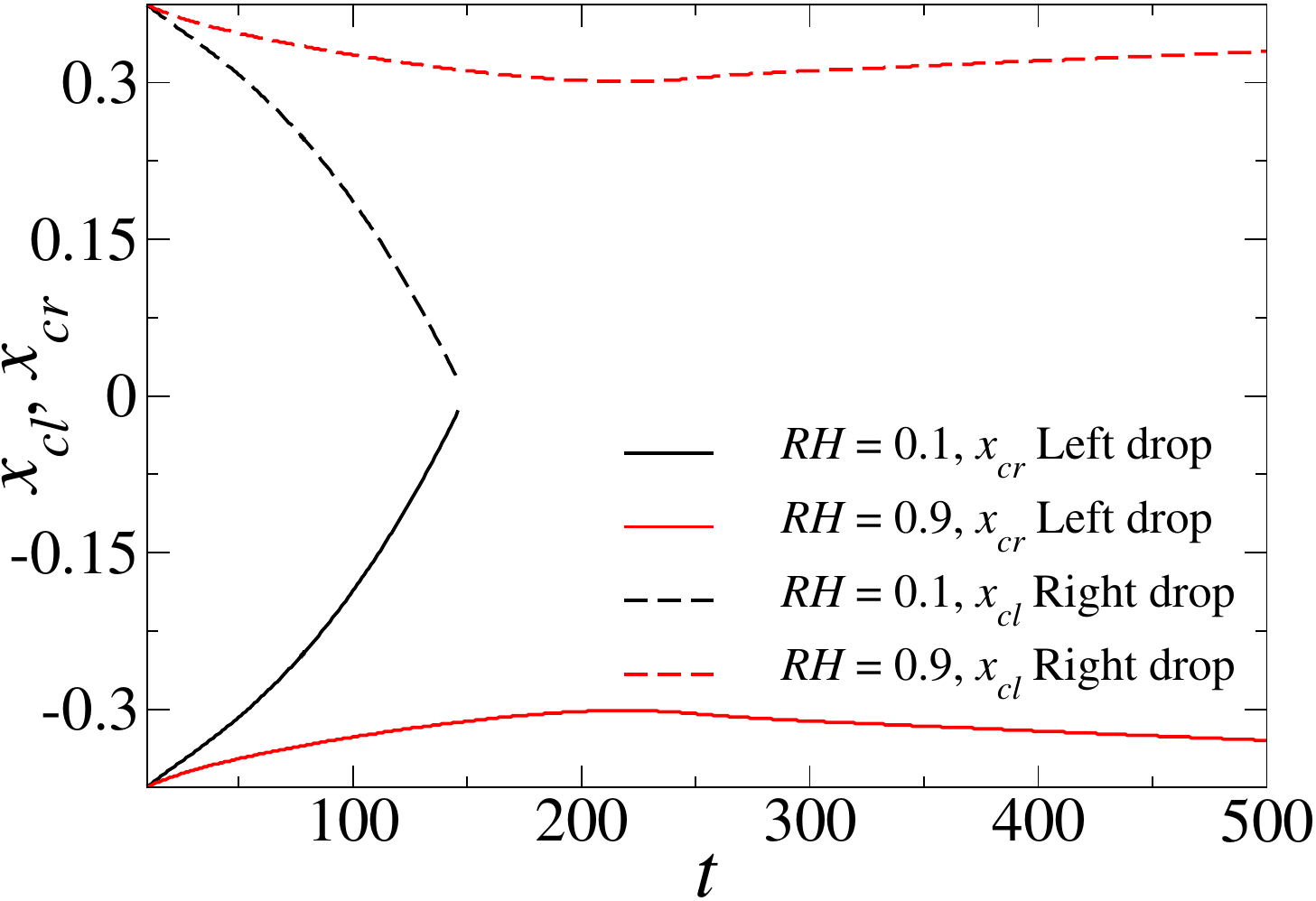} 
\hspace{0mm}
\includegraphics[width=0.45\textwidth]{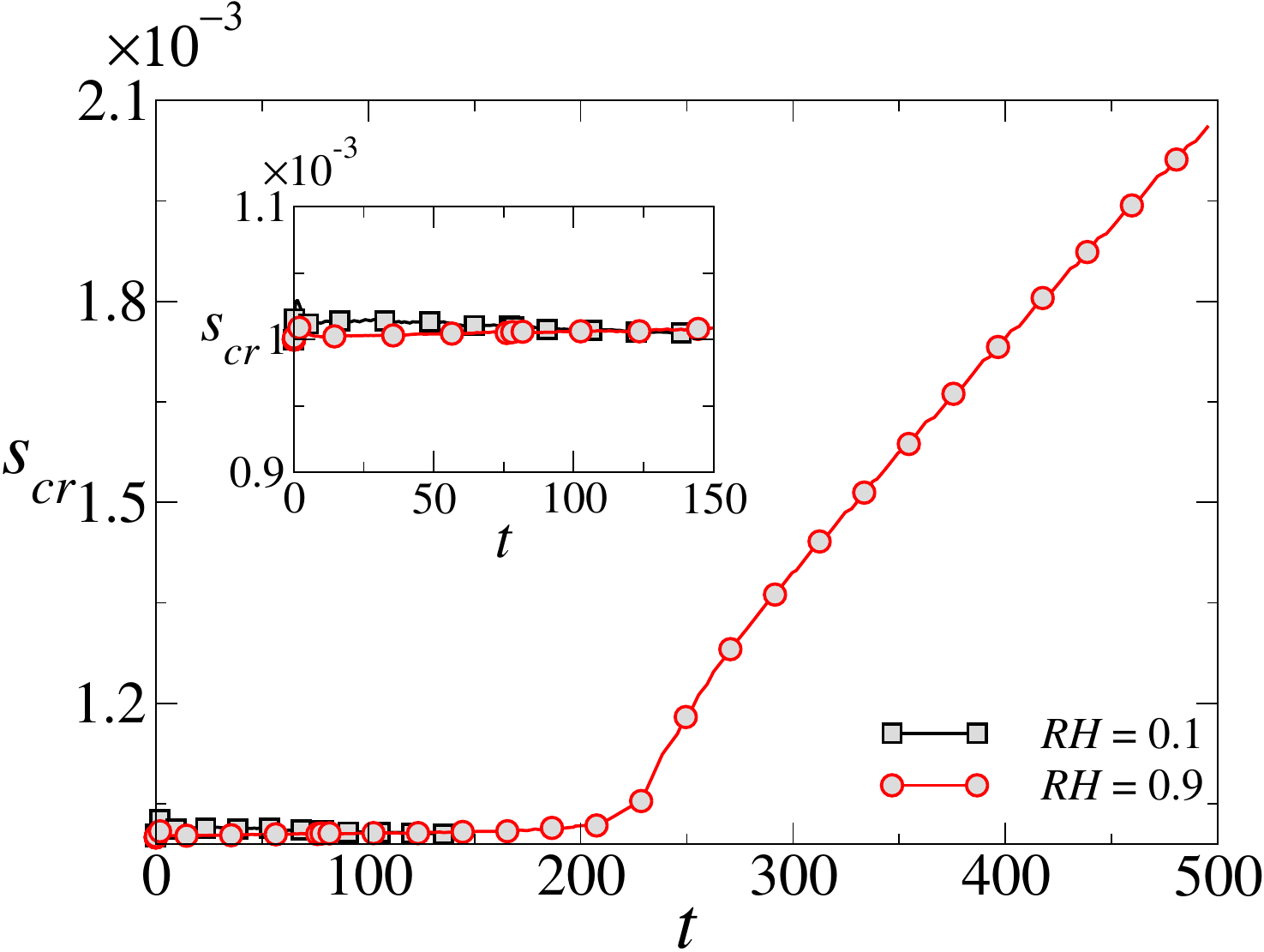}\\
\hspace{0.8cm} {\large (c)}\\
\includegraphics[width=0.45\textwidth]{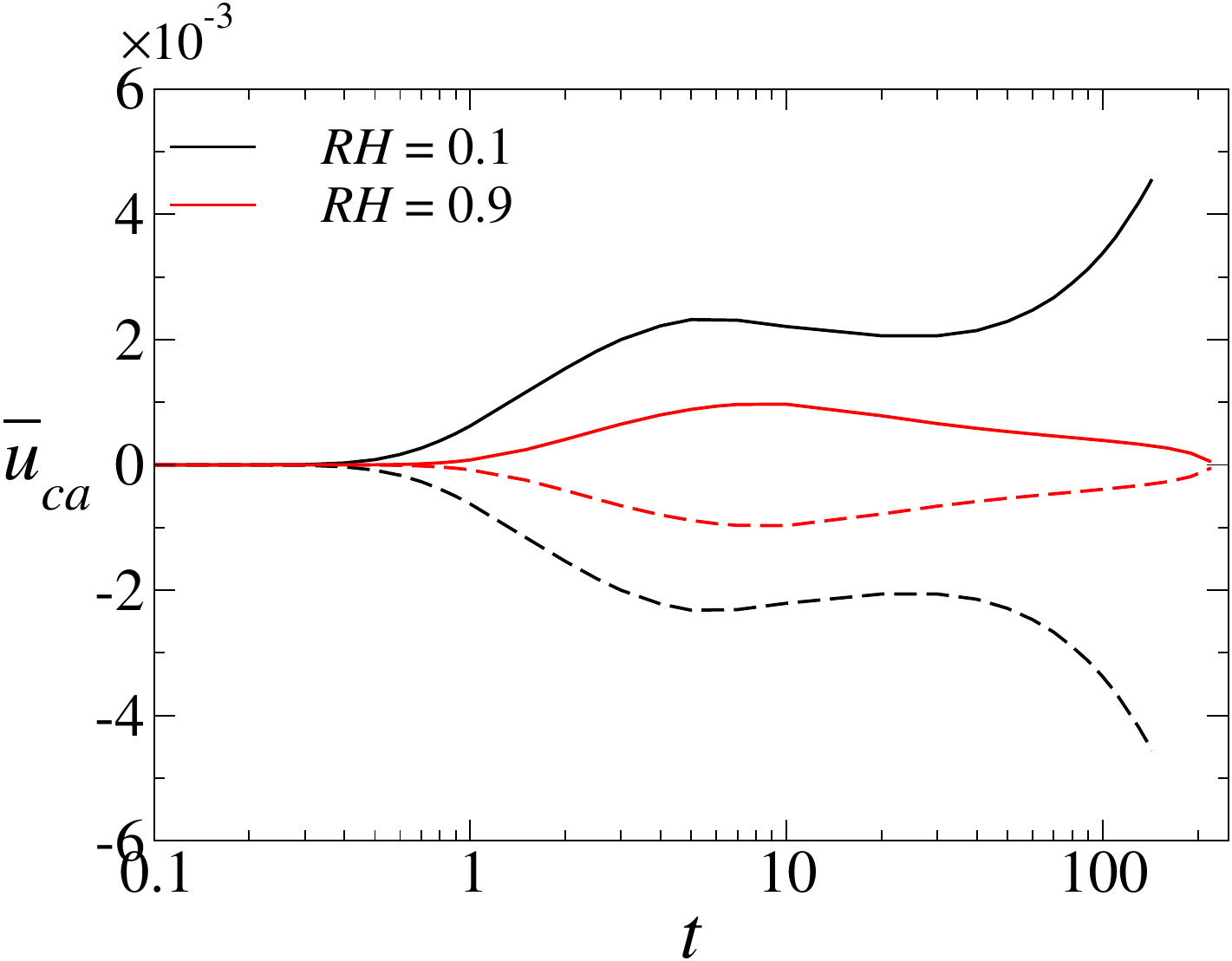}
\caption{Temporal variation of (a) the right contact line ($x_{cr}$) of the left drop and the left contact line ($x_{cl}$) of the right drop, (b) the freezing front height ($s_{cr}$) near the right contact line ($x_{cr}$), and (c) the average capillary velocity ($\bar{u}_{ca}$) for $RH = 0.1$ and $RH = 0.9$. The rest of the dimensionless parameters are the same as the `base' parameters.}
\label{fig:RH_0p1_RH_0p9}
\end{figure}

To gain more insights, in figure \ref{fig:RH_0p1_RH_0p9}(a), we examine a system of two drops with an initial distance of $d_0 = 0.75$ for two levels of relative humidity ($RH=0.1$ and $RH=0.9$) while keeping other parameters constant. Tracking the transition of the contact lines allows us to understand coalescence dynamics. It is evident that for $RH = 0.9$, the contact lines of adjacent drops approach each other until $t = 200$ and then move apart. This behavior can be explained by the evolution of the freezing front at the contact line of a drop ($s_{cr}$) over time ($t$). Figure \ref{fig:RH_0p1_RH_0p9}(b) illustrates that for $RH = 0.1$, the freezing front thickness at the contact line does not evolve significantly before coalescence occurs at around $t = 144$, indicated by the convergence of the contact lines. Conversely, for $RH = 0.9$, the freezing front evolves significantly before the contact lines approach each other. This can also be corroborated by plotting the average capillary velocity inside the drop, as shown in figure \ref{fig:RH_0p1_RH_0p9}(c). For $RH = 0.1$, the average capillary velocity initially increases, decreases slightly due to freezing front evolution, and peaks when the drops are close to each other, indicating coalescence. In contrast, for $RH = 0.9$, the average capillary velocity becomes very low at $t = 200$, corresponding to evolving freezing front thickness, leading to the slight separation of adjacent contact lines after $t = 200$. 

In figure \ref{RH_vs_d_phase}(a), as relative humidity increases to $RH = 0.7$, the initial separation distance ($d_0$) for coalescence slightly increases due to decreased evaporative cooling. However, a further increase in relative humidity reduces the average capillary velocity, leading to coalescence only at lower separation distances ($d_0$), attributed to significant freezing front evolution. Increasing the temperature difference between the substrate bottom and the melting temperature accelerates freezing front growth, reducing the initial distance ($d_0$) for coalescence. Lowering the substrate bottom temperature to $T_c = -1.2^{\circ}$C, corresponding to an increased temperature difference of ($\Delta T = 1.2^{\circ}$C), promotes freezing front growth due to a higher Stefan number ($Ste$) and reduced evaporative cooling as the scaled latent heat of vaporization ($\chi$) decreases. In figure~\ref{RH_vs_d_phase}(b), as relative humidity increases from $RH = 0.1$ to $RH = 0.4$, there is no significant increase in initial separation distance $d_0$ for coalescence due to decreased evaporative cooling. Subsequently, as relative humidity increases from $0.4$ to $0.9$, the average capillary velocity decreases significantly due to low evaporation at high humidity and freezing front evolution, leading to coalescence only at lower initial separation distances ($d_0$). In figure \ref{RH_vs_d_phase}(c), fixing relative humidity to $RH = 0.6$ and varying the temperature difference between the substrate bottom and the melting temperature demonstrates that increasing the temperature difference decreases the initial distance ($d_0$) for coalescence. It is observed that cooler temperatures facilitate faster freezing front propagation, completely restricting contact line movement. \\

\section{Conclusions} \label{sec:Conc}

We numerically investigate the evaporation-induced coalescence of two volatile sessile drops on a cold substrate undergoing freezing by employing the lubrication approximation in the finite element method framework. Our two-dimensional model incorporates mass conservation across the liquid, solid, and gas phases and momentum conservation within the liquid phase. Additionally, our model accounts for energy balance in the liquid, solid, and gas phases and considers heat flux across interfaces and substrate boundaries. 

We focus on the interactions between halos and the coalescence of two volatile sessile drops undergoing freezing, primarily driven by asymmetrical evaporation. For volatile drops, evaporation significantly dominates the halo formation. We observe that under freezing conditions on a cold substrate, two volatile drops can coalesce before the freezing front eventually confines their contact lines. Further, we find that this phenomenon cannot be accurately modelled without considering evaporation. In our simulations, neglecting evaporation results in stagnant sessile drops while considering it leads to coalescence due to asymmetrical evaporation rates. We establish a mechanism that drives this coalescence based on the average capillary velocity of the individual drop. It is also observed that when the two drops are in close vicinity, the halos of the individual droplets interact to form a merged halo. Furthermore, our observations indicate that drops approach each other more rapidly under lower relative humidity conditions, resulting in rapid coalescence, compared to partially wetting drops placed side by side without undergoing evaporation. While condensation and halo formation may not significantly contribute to bringing the droplets closer to each other, upon contact and merging, we observe a substantial increase in local relative humidity near the liquid-gas interface at the coalescence site. This surge triggers significant condensation, further expediting the process of coalescence. This condensation also raises the temperature of the liquid-gas interface as the surroundings absorb the latent heat released during gas-to-liquid transformation.

To gain further insights into the coalescence and non-coalescence behaviour of volatile sessile drops, we examine the influence of relative humidity, the initial separation distance between the drops, and the temperature difference between the bottom of the substrate and the melting temperature (with water as reference). Our findings reveal that the time required for the drops to approach each other at high relative humidity levels is significantly prolonged. This extended duration allows for greater propagation of the freezing front and reduces the average capillary velocity of the drops, leading to coalescence only at low initial separation distances. Moreover, when the bottom of the substrate is much colder, the freezing front propagates more rapidly, requiring minimal time to restrict the contact line. Consequently, to achieve coalescence, the droplets must be placed in close proximity to each other.\\
\\

\noindent{\bf Acknowledgement:} {K.C.S. thanks the Science \& Engineering Research Board and IIT Hyderabad, India for their financial supports through grants CRG/2020/000507 and IITH/CHE/F011/SOCH1, respectively. C.S.S. acknowledges IIT Ropar for financial support through ISIRD (Grant No. 9-388/2018/IITRPR/3335).} \\

\noindent{\bf Declaration of interests:} The authors report no conflict of interest.

\appendix
\numberwithin{equation}{section}
\makeatletter
\newcommand{\section@cntformat}{Appendix \thesection:\ }
\makeatother
\section{Validation of the numerical model} \label{sec:Val}

\subsection{Comparison with \citet{marin2014universality}} \label{A1}

\begin{figure}
\centering
\includegraphics[width=0.6\textwidth]{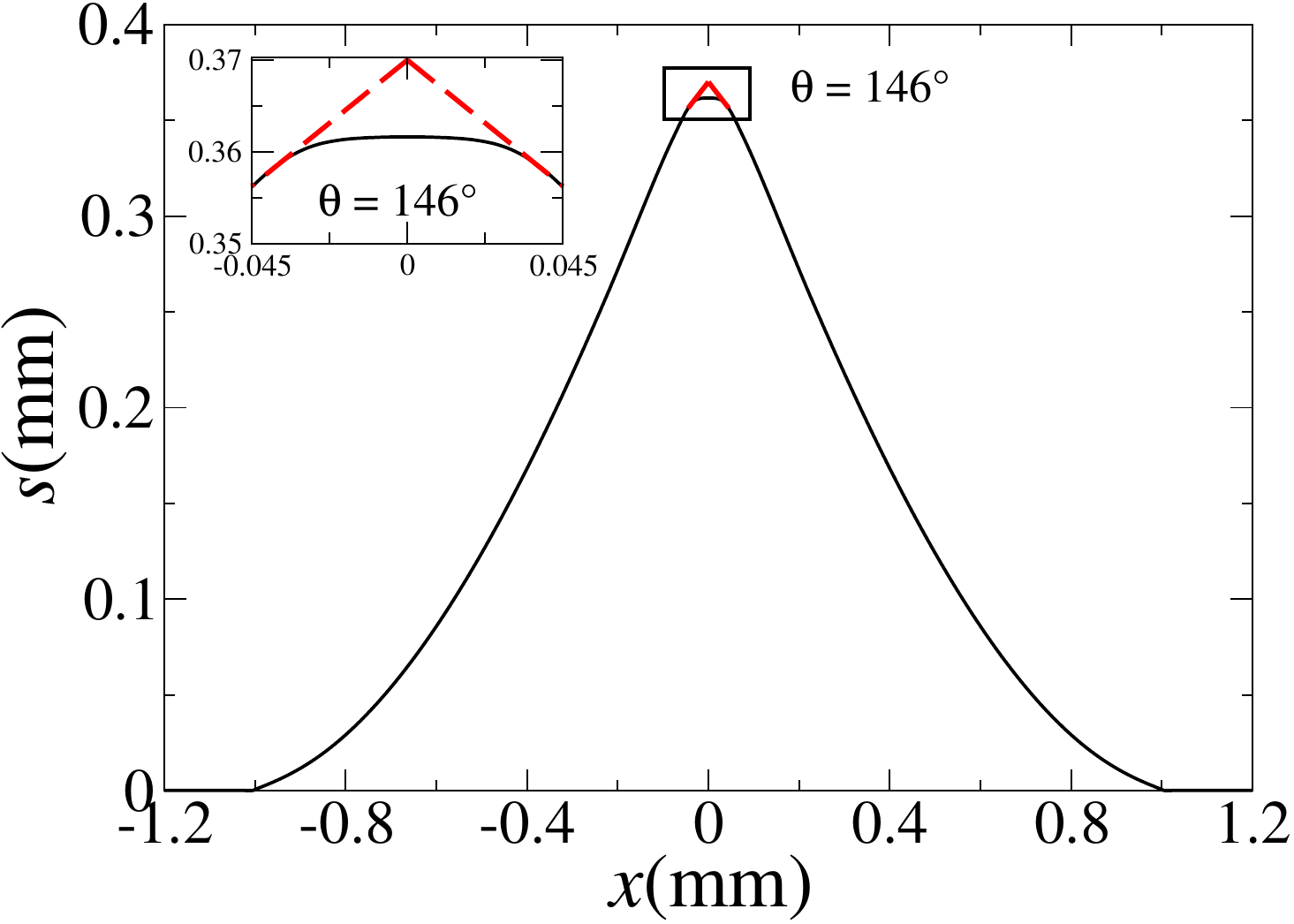}
\caption{The tip angle near the cusp obtained from our model considering both evaporation and freezing for $Ste = 1.22 \times 10^{-3}$, $T_{v} = 1$, $A_{n} = 17$, $D_{g} = 2$, $D_{s} = 0.9$, $\Lambda_{S} = 3.89$, $\Lambda_{W} = 698$, $\chi = 0.01$, $K = 8\times10^{-4}$, $\rho_{veR} = 1$, $\Lambda_{g} = 0.041$, $D_{w} = 15$, $RH = 0.70$, $\epsilon=0.2$, $D_{v} = 1.65 \times 10^{-6}$, $\Delta = 10^{-4}$, $\Psi = 0.30$ and $Pe_{v} = 1$. Note that \citet{marin2014universality} experimentally observed a tip angle of $\sim 139^{\circ}$.}
\label{fig:Tip_angle}
\end{figure}

We validated our model by comparing the tip angle at the end of freezing with the angle observed in the experimental study by \citet{marin2014universality} for a typical set of parameters (figure \ref{fig:Tip_angle}). To incorporate evaporation and condensation in our freezing model, we incorporated a precursor layer model to prevent shear stress singularities at the water-ice interface (see eq. \ref{pre_model}). To maintain a consistent precursor layer thickness and ensure accurate predictions for condensation and evaporation, we applied a penalty function as described by \citet{zadravzil2006droplet} (see eq. \ref{pre_model1})). However, this approach imposes a constraint on the freezing rate as the liquid layer thickness approaches that of the precursor layer. Consequently, our model may not fully capture the final tip formation at the end of freezing. Despite this limitation, by estimating the angle between the tangents near the cusp, we found that the tip angle is approximately $146^{\circ}$ for a typical set of parameters. This result closely matches the experimentally observed tip angle of about $\sim 139^{\circ}$ reported by \citet{marin2014universality}, who also noted that the tip angle is independent of substrate temperature, wettability, and solidification rate.

\subsection{Comparison with \citet{zadravzil2006droplet,kavuri2023freezing}} \label{A2}

\begin{figure}
\centering
\includegraphics[width=0.9\textwidth]{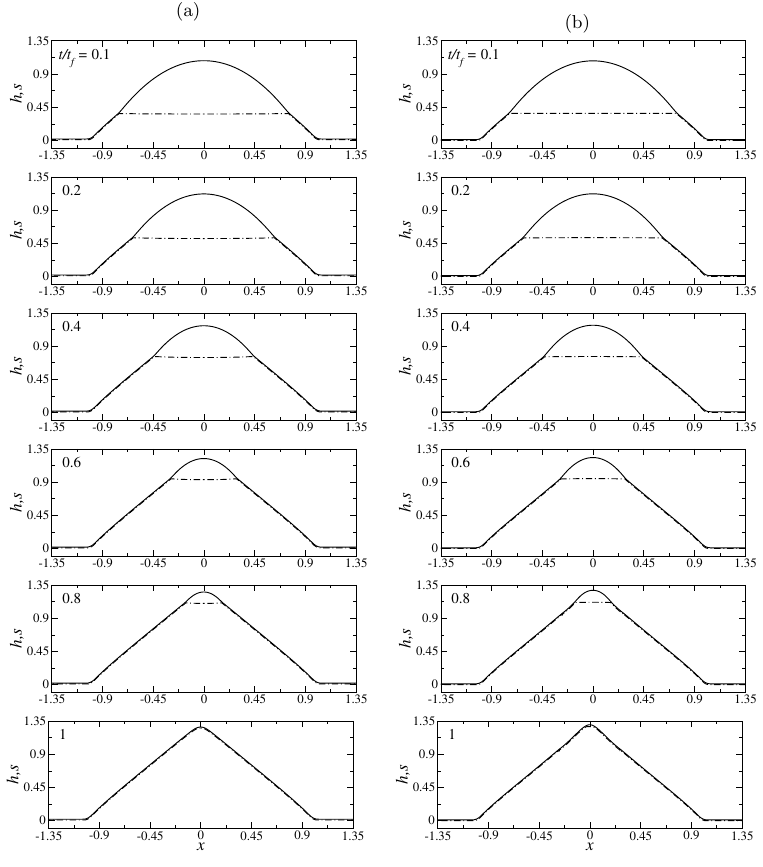}
        \caption{Temporal evolution of the droplet shape, $h$ (solid line), and the freezing front, $s$ (dot-dashed line) for a drop placed on a cold substrate. Panel (a) corresponds to data extracted from a scenario where only freezing is considered, without evaporation, as illustrated in Fig. 3 of \citet{kavuri2023freezing}, which mimics Fig. 15 of \citet{zadravzil2006droplet}. Panel (b) represents the same scenario but analyzed using our present formulation. The rest of the dimensionless parameters are $\epsilon=0.2$, $Ste = 0.04$, $T_{v} = 0.5$, $A_{n} = 6.25$, $D_{s} = \Lambda_{S} = \Lambda_{W} = RH = K = \rho_{veR} = 1$, $\Lambda_{g} = 0.6$, and $D_{w} = D_{v} = \Delta = \Psi = \chi = Pe_{v} = 0$. The value of dimensionless total freezing time $t_f$ in (a) and (b) are 15 and 16, respectively.}
        \label{fig:freez_comp}
\end{figure}

Further validation of our freezing model was conducted by simulating a scenario previously examined by \citet{kavuri2023freezing} and \citet{zadravzil2006droplet}. Figure \ref{fig:freez_comp} shows the temporal evolution of the droplet shape, $h$ (solid line), and the freezing front, $s$ (dot-dashed line), for a drop placed on a cold substrate. Figure \ref{fig:freez_comp}(a) corresponds to data extracted from a scenario where only freezing is considered, without evaporation, as depicted in figure 3 of \citet{kavuri2023freezing}, which mimics figure 15 of \citet{zadravzil2006droplet}. Figure \ref{fig:freez_comp}(b) represents the same scenario but is analyzed using our present formulation. The rest of the dimensionless parameters are $\epsilon=0.2$, $Ste = 0.04$, $T_{v} = 0.5$, $A_{n} = 6.25$, $D_{s} = \Lambda_{S} = \Lambda_{W} = RH = K = \rho_{veR} = 1$, $\Lambda_{g} = 0.6$, and $D_{w} = D_{v} = \Delta = \Psi = \chi = Pe_{v} = 0$. The dimensionless total freezing time in figure \ref{fig:freez_comp}(a) and figure \ref{fig:freez_comp}(b) are $t_f=15$ and $t_f=16$, respectively, which are reasonably in good agreement. It is noted that the set of parameters used in the present simulations corresponds to the typical set of parameters considered in figure 3 of \citet{kavuri2023freezing}, except the Biot number ($Bi$). In our model, ($Bi$) is zero as we do not consider heat transfer due to convection. To account for the higher heat transfer in our model, we exaggerated the thermal conductivity ratio of the gas phase to the liquid phase $\Lambda_g = \lambda_g/\lambda_l$ by comparing eqs. (\ref{eq262}) and (\ref{eq:Freezing_rate}) of the present model with eqs. (A29) and (A31) from \citet{kavuri2023freezing}. Additionally, we found that the evolution of the freezing front, $s$ (dot-dashed lines) and shape of the drop $h$ (solid line) placed on a cold substrate obtained using 4001, 9601, and 12001 grid points are indistinguishable (figure \ref{fig:Grid_study}).

\subsection{Comparison with \citet{wen2019vapor}} \label{A4}

\begin{figure}
\centering
\includegraphics[width=0.6\textwidth]{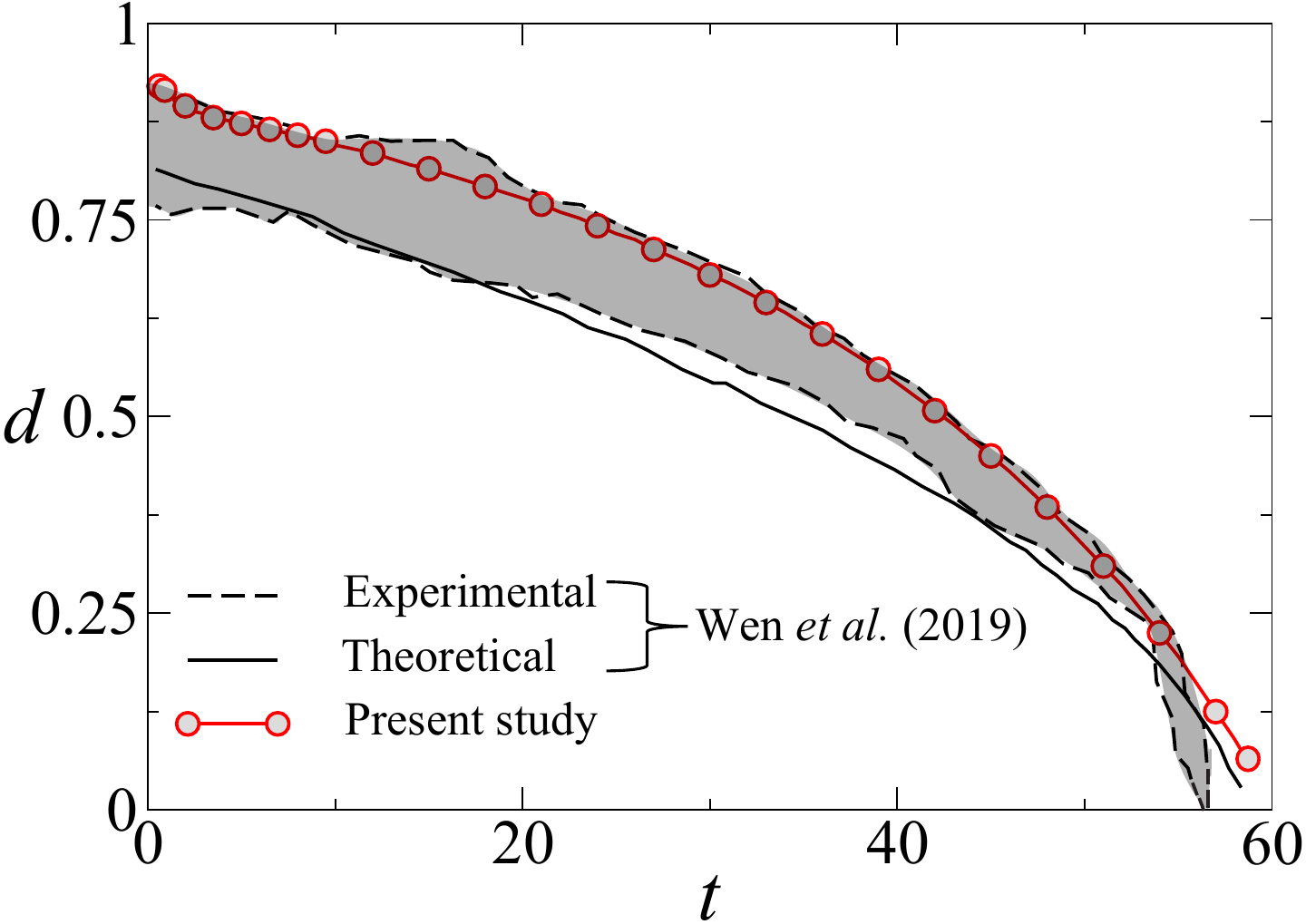}
\caption{Comparison with the experimental and theoretical results of \citet{wen2019vapor} to validate our evaporation model. The shaded region between the two dashed black lines represents experimental data obtained from various trials for the evaporation of single-component n-Hexane droplets at room temperature, while the solid black line shows the theoretical prediction, as reported in figure 2a of \citet{wen2019vapor}. The red solid line with circle symbols presents our simulation results, albeit without freezing. The dimensionless parameters used in our simulations are $d_0 = 0.9$, $\chi = 0.21$, $\Delta = 10^{-6}$, $\Psi = 0.03$, $Pe_{v} = 0.048$, $Ste = 0$, $T_{v} = 0$, $A_{n} = 1.28$, $D_v = 10^{-3}$, $D_{g} = 5.0$, $D_{s} = 0.9$, $\Lambda_{S} = 3.89$, $\Lambda_{W} = 11.5$, $\Lambda_{g} = 0.12$, $RH = 0$, $K = 1.1\times10^{-5}$, $D_{w} = 15.0$, $\epsilon = 0.06$, and $\rho_{veR} = 1.0$.}
\label{fig:evap_coal_val}
\end{figure}

\begin{figure}
\centering
\includegraphics[width=0.9\textwidth]{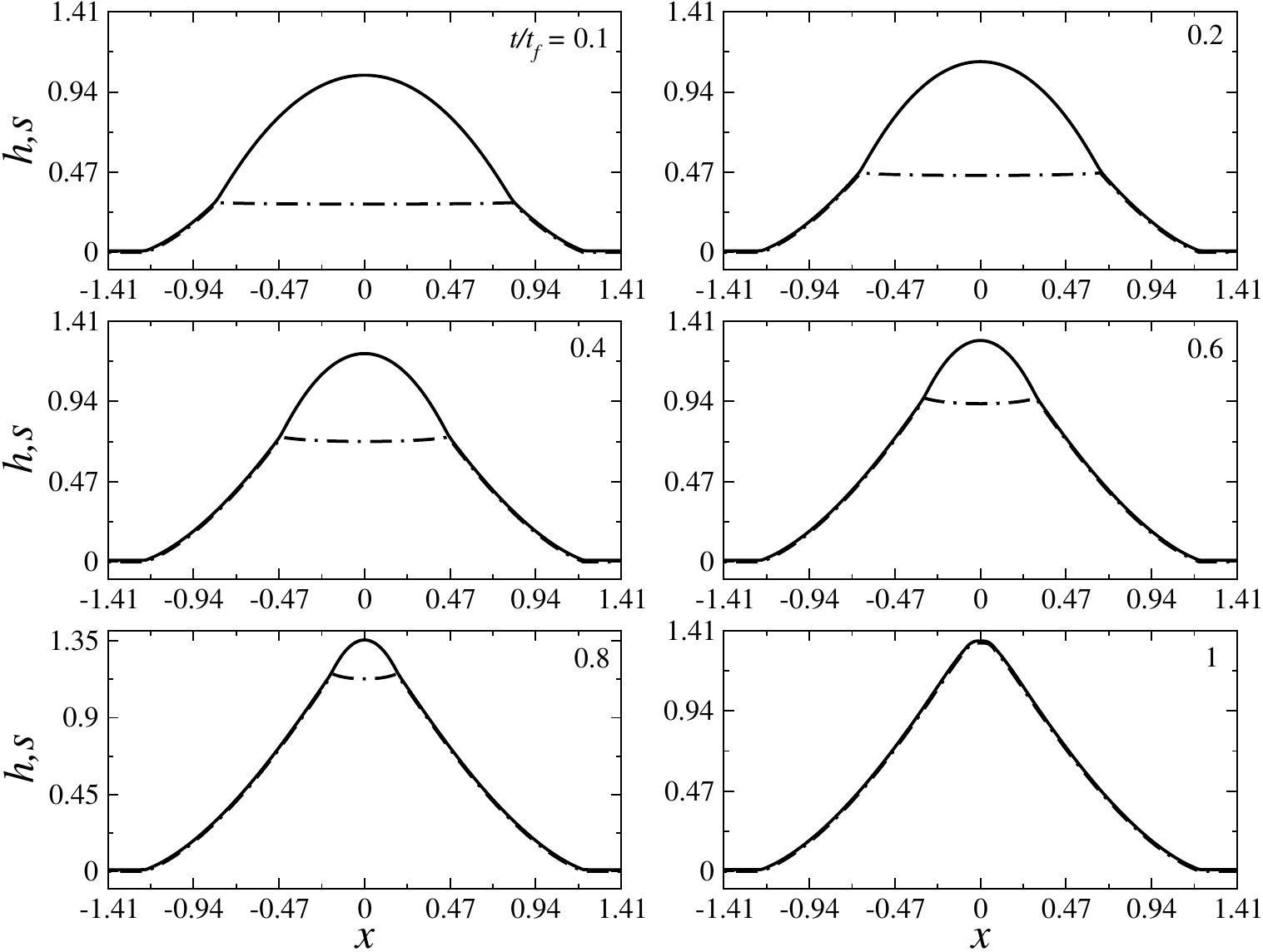}
\caption{Evolution of the freezing front, $s$ (dot-dashed lines) and shape of the drop $h$ (solid line) placed on a cold substrate. The results obtained using 4001, 9601, and 12001 grid points are found to be indistinguishable. The rest of the dimensionless parameters are $Ste = 1.7 \times 10^{-4}$, $T_{v} = 0.2$, $A_{n} = 6.25$, $D_{g} = 2$, $D_{s} = 0.9$, $\Lambda_{S} = 3.82$, $\Lambda_{W} = 191$, $\chi = 0.23$, $K = 8\times10^{-4}$, $\rho_{veR} = 1.09$, $\Lambda_{g} = 3.5$, $D_{w} = 7.5$, $RH = 0.90$, $\epsilon=0.2$, $D_{v} = 4.45 \times 10^{-6}$, $\Delta = 10^{-4}$, $\Psi = 0.14$ and $Pe_{v} = 1$. The values of the dimensionless total freezing time ($t_f$) of the droplet is 1210, respectively.}
\label{fig:Grid_study}
\end{figure}

Here, we validate our evaporation model by simulating a scenario examined by \citet{wen2019vapor}, who experimentally and theoretically investigated vapor-induced migration of two pure liquid droplets, without accounting for freezing. Figure \ref{fig:evap_coal_val} presents a comparison of the temporal evolution of the distance $d$ between two droplets prior to coalescence, as obtained from our simulations (neglecting freezing effects), with the results from \citet{wen2019vapor}. The shaded region between the two dashed black lines represents experimental data obtained from various trials for the evaporation of single-component n-Hexane droplets at room temperature, while the solid black line shows the theoretical prediction, as reported in figure 2a of \citet{wen2019vapor}. The red solid line with circle symbols in figure \ref{fig:evap_coal_val} corresponds to our simulation results using the current formulation, which also neglects freezing. The dimensionless parameters used in our simulations are $d_0 = 0.9$, $\chi = 0.21$, $\Delta = 10^{-6}$, $\Psi = 0.03$, $Pe_{v} = 0.048$, $Ste = 0$, $T_{v} = 0$, $A_{n} = 1.28$, $D_v = 10^{-3}$, $D_{g} = 5.0$, $D_{s} = 0.9$, $\Lambda_{S} = 3.89$, $\Lambda_{W} = 11.5$, $\Lambda_{g} = 0.12$, $RH = 0$, $K = 1.1\times10^{-5}$, $D_{w} = 15.0$, $\epsilon = 0.06$, and $\rho_{veR} = 1.0$. These parameters are consistent with those used in figure 2a of \citet{wen2019vapor}. Since evaporation experiments of \citet{wen2019vapor} were conducted at room temperature without freezing, we have adjusted the non-dimensionalization of the temperature to match their scenario within our freezing-inclusive formulation. Specifically, we modify $\Delta T = T_{m} - T_{c}$ to $\Delta T = \epsilon^2 T_c$ in Eq. (2.30) of our formulation. Additionally, we have considered a slightly greater gas layer thickness ($D_g$) than in our previous simulations due to the absence of specific information in \citet{wen2019vapor}, and we reasonably assumed that the chamber height significantly exceeds the droplet height. Figure \ref{fig:evap_coal_val} demonstrates that our numerical simulation closely matches the experimental data of \citet{wen2019vapor}, validating our evaporation model.

\clearpage

\subsection{Grid convergence test} \label{A3}

\clearpage

\end{document}